%% file: main.tex
\documentclass[lettersize,journal]{IEEEtran}
\IEEEoverridecommandlockouts
\usepackage{cite}
\usepackage{amsmath,amssymb,amsfonts}
\usepackage{dsfont}
\usepackage{algorithmic}
\usepackage{graphicx}
\usepackage{textcomp}
\usepackage{xcolor}
\usepackage{hyperref}
\usepackage{algorithm}
\usepackage{tabularx,booktabs}

\newcolumntype{Y}{>{\centering\arraybackslash}X}
\usepackage{multirow}
\usepackage[font=small,labelfont=bf]{caption}

\usepackage{amsthm}\theoremstyle{plain}

\newtheorem{definition}{Definition}

\def\BibTeX{{\rm B\kern-.05em{\sc i\kern-.025em b}\kern-.08em
    T\kern-.1667em\lower.7ex\hbox{E}\kern-.125emX}}

\allowdisplaybreaks

\begin{document}



\title{Mitigating Evasion Attacks in Federated Learning Based Signal Classifiers}

\author{
    Su Wang*\thanks{*S. Wang and R. Sahay contributed equally to this work. S Wang conducted this work while at Purdue University, IN, USA. He is currently with Princeton University, NJ, USA, email: hw5731@princeton.edu.}, 
    Rajeev Sahay*\thanks{R. Sahay is with UC San Diego, CA, USA email: r2sahay@ucsd.edu.}, 
    Adam Piaseczny,~\IEEEmembership{Student Member,~IEEE,}\\
    Christopher G. Brinton~\IEEEmembership{Senior Member,~IEEE}\thanks{A. Piaseczny and C. Brinton are with Purdue University, IN, USA email: \{apiasecz, cgb\}@purdue.edu.}
    \thanks{A preliminary version of this material appeared in the Proceedings of the 2023 IEEE International Conference on Communications (ICC)~\cite{wang2023potent}.}
    \thanks{This work was supported in part by the Office of Naval Research (ONR) under grants N00014-21-1-2472 and N00014-22-1-2305, and by the National Science Foundation (NSF) under grant CNS-2146171.}
}

\maketitle

\begin{abstract} \space \hspace{-1.5mm}
{\color{black}Recent interest in leveraging federated learning (FL) for radio signal classification (SC) tasks has shown promise but FL-based SC remains susceptible to model poisoning adversarial attacks. 
These adversarial attacks mislead the ML model training process, damaging ML models across the network and leading to lower SC performance. 
In this work, we seek to mitigate model poisoning adversarial attacks on FL-based SC by proposing the Underlying Server Defense of Federated Learning (USD-FL).
Unlike existing server-driven defenses, USD-FL does not rely on perfect network information, i.e., knowing the quantity of adversaries, the adversarial attack architecture, or the start time of the adversarial attacks. 
Our proposed USD-FL methodology consists of deriving logits for devices' ML models on a reserve dataset, comparing pair-wise logits via 1-Wasserstein distance and then determining a time-varying threshold for adversarial detection. 
As a result, USD-FL effectively mitigates model poisoning attacks introduced in the FL network. 
Specifically, when baseline server-driven defenses do have perfect network information, USD-FL outperforms them by (i) improving final ML classification accuracies by at least $6\%$, (ii) reducing false positive adversary detection rates by at least $10\%$, and (iii) decreasing the total number of misclassified signals by over $8\%$. 
Moreover, when baseline defenses do not have perfect network information, we show that USD-FL achieves accuracies of approximately $74.1\%$ and $62.5\%$ in i.i.d. and non-i.i.d. settings, outperforming existing server-driven baselines, which achieve $52.1\%$ and $39.2\%$ in i.i.d. and non-i.i.d. settings, respectively.}
\end{abstract}


\section{Introduction}
\label{sec:intro}
\input{introduction}

\section{Related Work}
\label{sec:related_work}
\input{related_work}

\begin{figure}[t]
    \centering
    \includegraphics[width=0.48\textwidth]{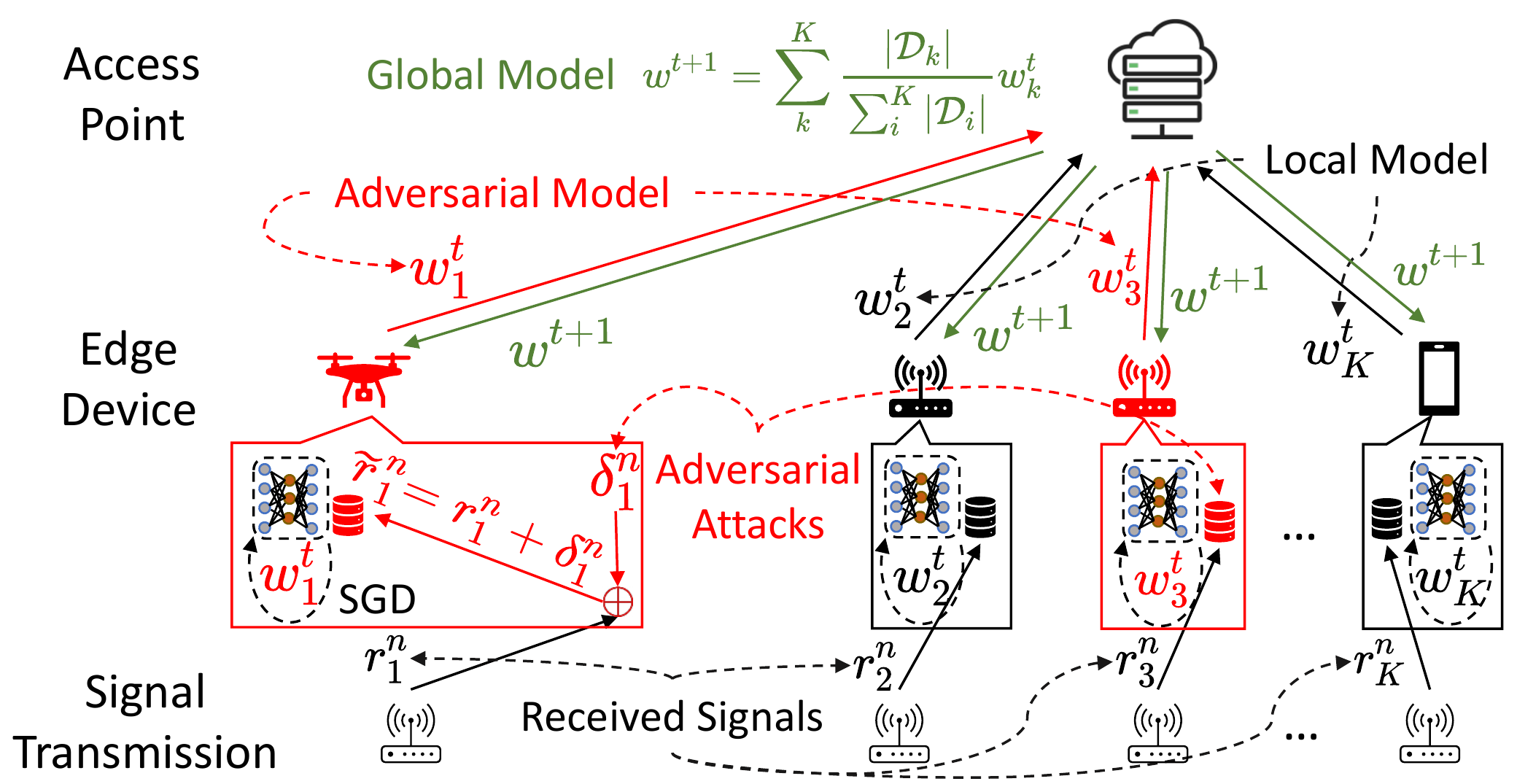}
    \caption{FL-based SC framework in which select devices train their local ML models on datasets perturbed by adversarial evasion attacks. 
    As a result, the server conducts global aggregations with both poisoned and unpoisoned model parameters, and subsequently distributes the now poisoned global ML model throughout the network.} 
    \label{fig:sys_model}
    \vspace{-4mm}
\end{figure}

\section{Evasion Attacks on FL-based SC}
\label{sec:atk_ovr_method}
\input{atk_ovr_method}


\input{atk_eval}

\section{Methodology of USD-FL} 
\label{sec:def_ovr_method}

\input{def_ovr_method}

\section{Evaluation of Defensive Framework}
\label{sec:def_eval}

\input{def_eval}



\section{Conclusion and Future Work}
\label{sec:conclusion}
\input{conclusion}

\bibliography{references}

\bibliographystyle{IEEEtran}
\begin{IEEEbiography}
[{\includegraphics[width=0.9\textwidth,clip]{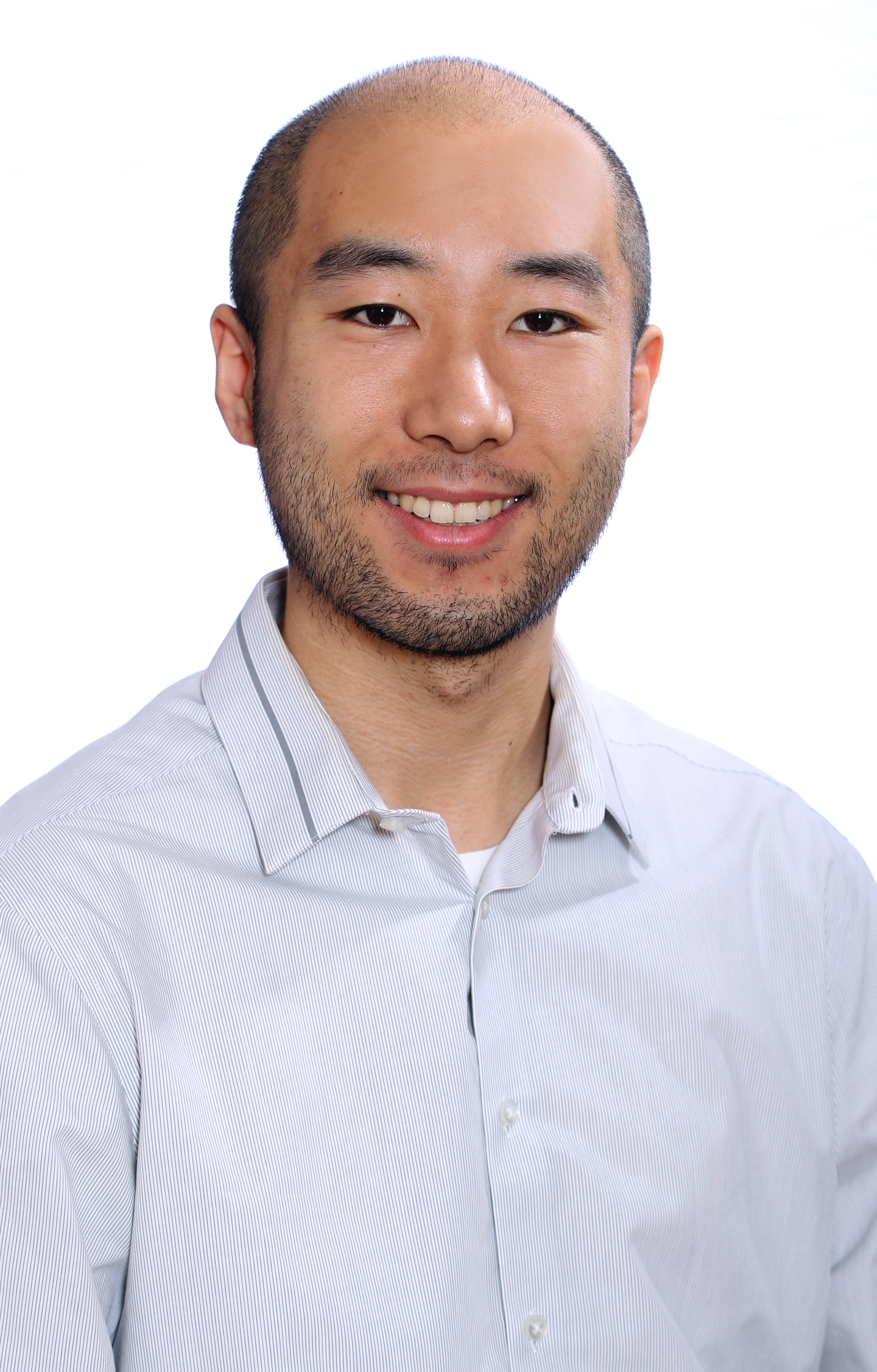}}]
{(Henry) Su Wang} received the B.S. (with Distinction) and Ph.D. degrees in Electrical and Computer Engineering from Purdue University, West Lafayette in 2018 and 2023, respectively. 
He is currently a Postdoctoral Research Associate in the Department of Electrical and Computer Engineering at Princeton University. 
His current research explores the intersection of edge/fog networking and distributed machine learning, particularly federated learning. In these settings, he focuses on network relationships to enhance both system efficiency and machine learning development.
\end{IEEEbiography}

\begin{IEEEbiography}
[{\includegraphics[width=0.9\textwidth,clip]{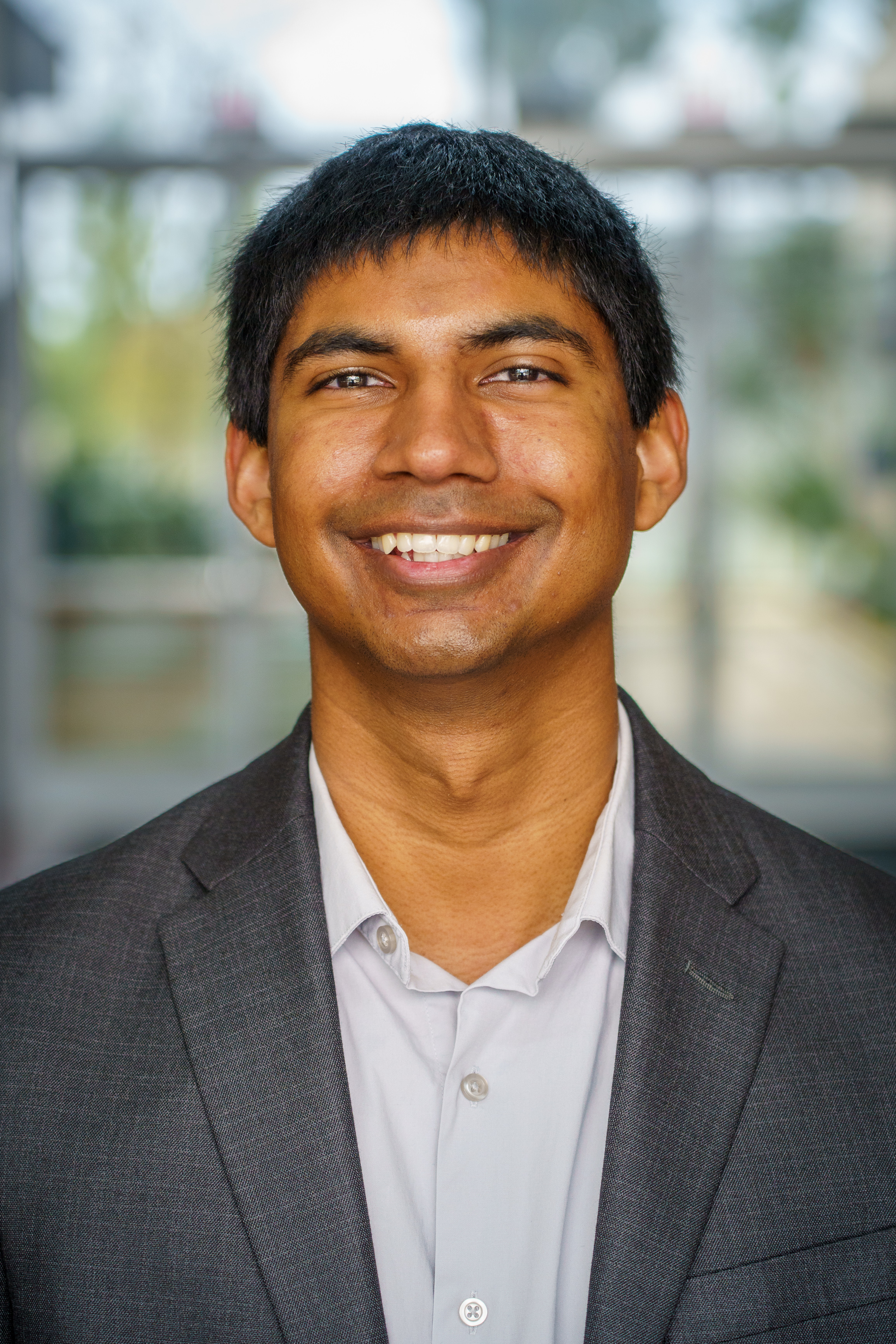}}]
{Rajeev Sahay} received the B.S. degree in electrical engineering from The University of Utah, Salt Lake City, UT, USA, in 2018, and the M.S. and Ph.D. degrees in electrical and computer engineering from Purdue University, West Lafayette, IN, USA, in 2021 and 2022, respectively. Currently, he is a faculty member in the Department of Electrical and Computer Engineering at UC San Deigo. He was the recipient of the Purdue Engineering Dean’s Teaching Fellowship and was named an Exemplary Reviewer by the IEEE Wireless Communications Letters. His research interests lie in the intersection of networking and machine learning, especially in their applications to wireless communications and engineering education. 
\end{IEEEbiography}

\begin{IEEEbiography}
[{\includegraphics[width=\textwidth,clip]{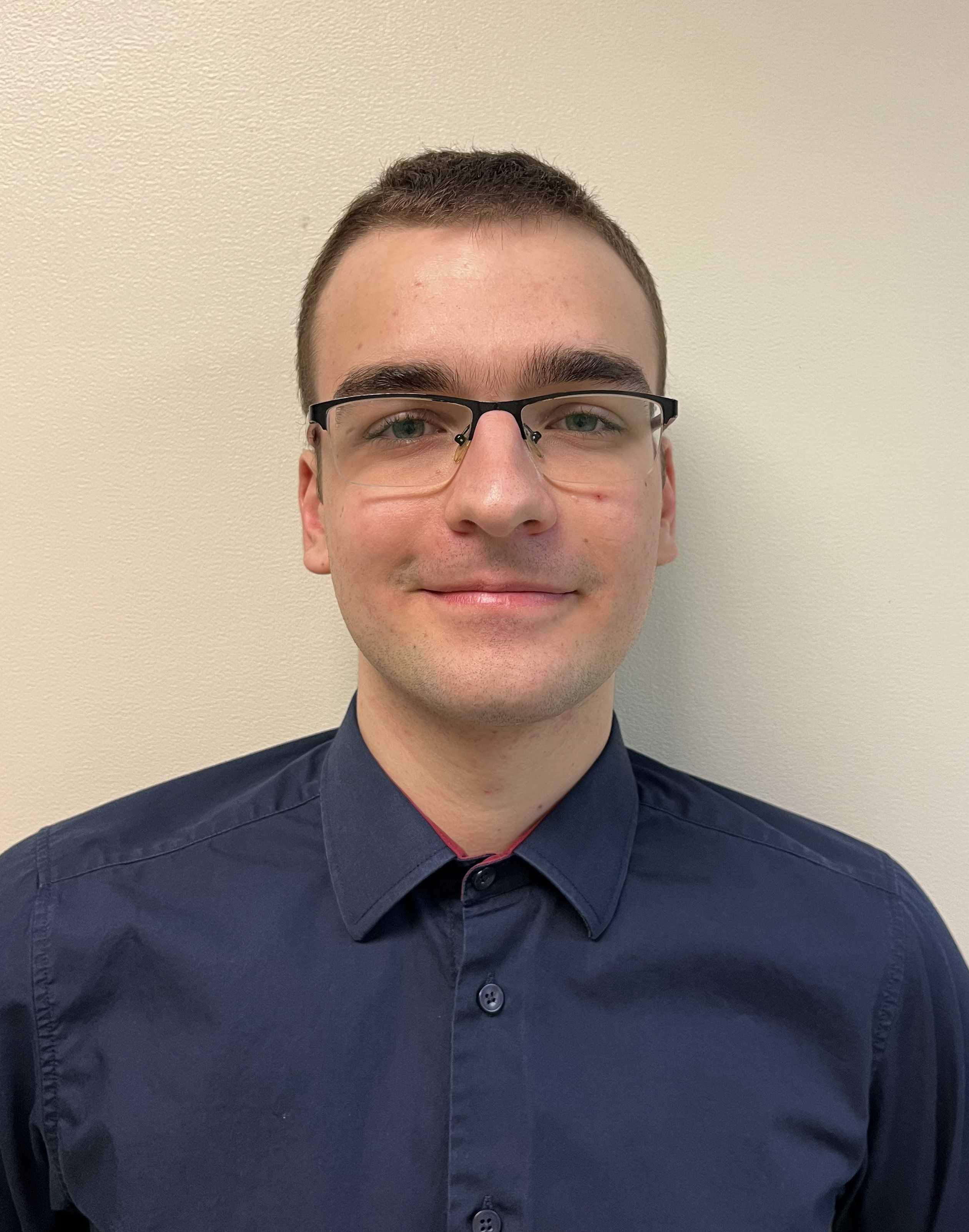}}]
{Adam Piaseczny} is currently pursuing his B.S. in Electrical and Computer Engineering at Purdue University, West Lafayette. His interests lie in Cognitive AI, AI Security and Distributed Machine Learning.
\end{IEEEbiography}

\begin{IEEEbiography}
[{\includegraphics[width=\textwidth,clip]{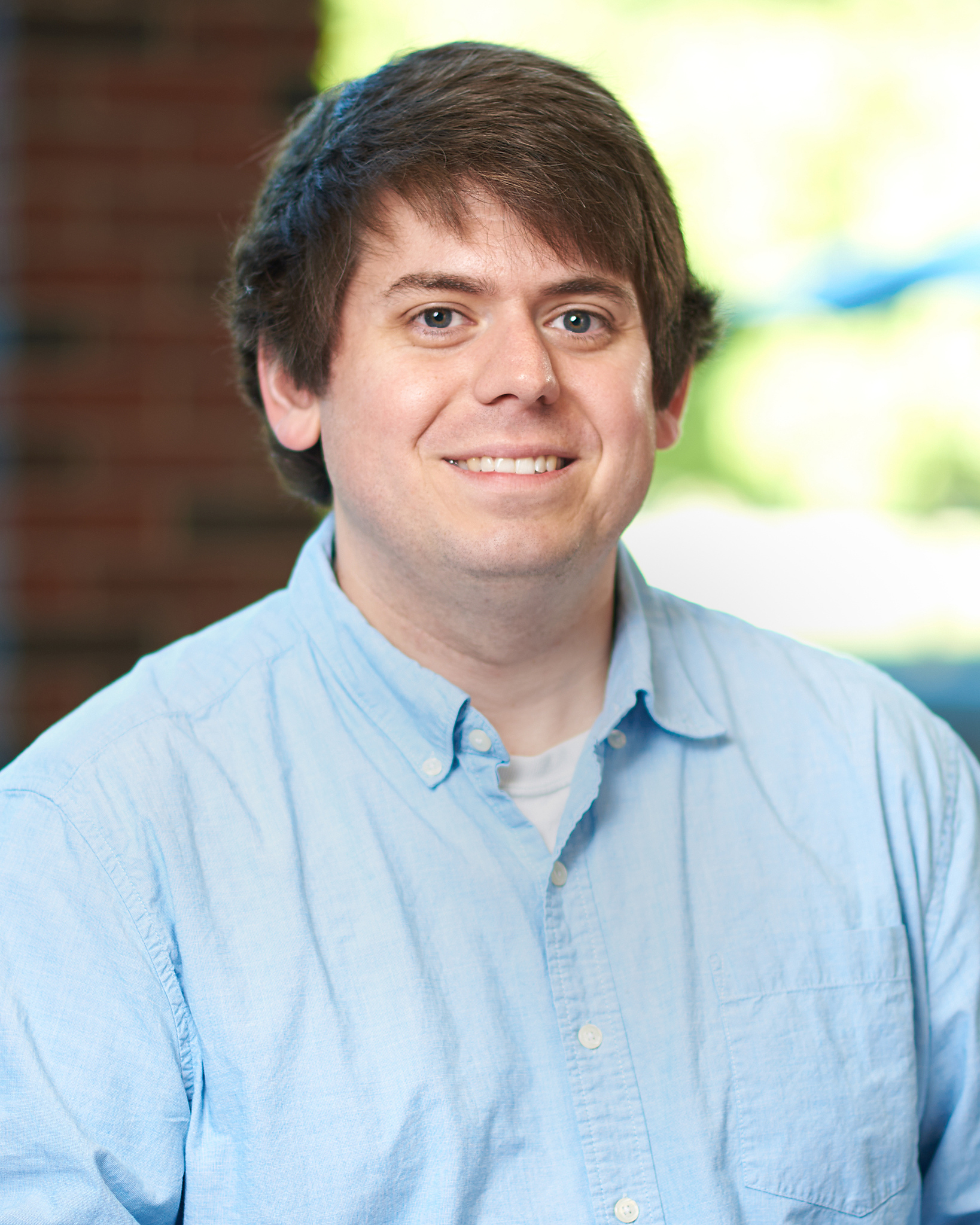}}]
{Christopher G. Brinton (S’08, M’16, SM’20)} is the Elmore Associate Professor of Electrical and Computer Engineering (ECE) at Purdue University. His research interest is at the intersection of networking, communications, and machine learning, specifically in fog/edge network intelligence, distributed machine learning, and AI/ML-inspired wireless network optimization. Dr. Brinton is a recipient of four of the US top early career awards, from the National Science Foundation (CAREER), Office of Naval Research (YIP), Defense Advanced Research Projects Agency (YFA), and Air Force Office of Scientific Research (YIP). He is also a recipient of the IEEE Communication Society William Bennett Prize Best Paper Award, the Intel Rising Star Faculty Award, the Qualcomm Faculty Award, and Purdue College of Engineering Faculty Excellence Awards in Early Career Research, Early Career Teaching, and Online Learning. Dr. Brinton currently serves as an Associate Editor for IEEE/ACM Transactions on Networking, and previously was an Associate Editor for IEEE Transactions on Wireless Communications. Prior to joining Purdue, Dr. Brinton was the Associate Director of the EDGE Lab and a Lecturer of Electrical Engineering at Princeton University. He also co-founded Zoomi Inc., a big data startup company that has provided learning optimization to more than one million users worldwide and holds US Patents in machine learning for education. His book \textit{The Power of Networks: 6 Principles That Connect our Lives} and associated Massive Open Online Courses (MOOCs) have reached over 400,000 students. Dr. Brinton received the PhD (with honors) and MS Degrees from Princeton in 2016 and 2013, respectively, both in Electrical Engineering.
\end{IEEEbiography}

\newpage
\clearpage
\begingroup
\let\clearpage\relax 
\onecolumn
\appendices

\newpage
\input{more_exps/conf_mats} 
\newpage

\end{document}

%% file: introduction.tex

As the Internet of Things (IoT) expands, efficient management of the wireless spectrum is critical for next-generation wireless networks. Intelligent signal classification (SC) techniques, such as automatic modulation classification (AMC), are a key technology for enabling such efficiency in the increasingly crowded radio spectrum. Such methods dynamically predict signal characteristics, such as its modulation scheme, direction of arrival, and channel state information (CSI), using the in-phase and quadrature (IQ) time samples of received signals. Deep learning is known to be highly effective for SC, outperforming likelihood-based classifiers without requiring specific feature engineering of the IQ samples \cite{dl_amc}. 

Federated learning (FL) \cite{fl,wang2023towards}, a technique for distributing model training, and its advancements~\cite{dong2022federated,dong2023federated,hu2019decentralized,hegedHus2019gossip,wang2023multi} have recently been considered for DL-based SC~\cite{amc_fl1}. 
In FL-based SC, each participating device trains a model on their locally collected dataset of received signals. Periodically, each local device transmits their model parameters to a global server, which aggregates all the received model parameters. The global server then communicates the updated aggregated model to all participating devices. The participating FL devices resume training from the received model parameters returned from the global server.


{\color{black}As a result of this design, locally received/collected signals are never transmitted over the network, as required by centralized SC, thus mitigating the potential of data leakage. 
While FL does not directly transmit datasets, it remains susceptible to model poisoning adversarial attacks, which reduce the shared model's performance by perturbing either the model parameters directly (during transmission or at a device) or the on-device data used to train a particular local model.}

{\color{black}In this work, we focus on the latter and aim to mitigate FL-based adversarial attacks that are induced by perturbing local datasets. 
Here, we consider attack frameworks in which adversarial evasion perturbations~\cite{evasion_atks} are used to conduct model poisoning attacks~\cite{model_poison} in FL-based SC. Specifically, we consider the common adversarial FL setting in which a subset of participating clients are adversaries in the FL network and thus intentionally perturb their local datasets in an effort to poison the globally learned model.}


In response, we develop a server-driven defense called Underlying Server Defense of Federated Learning (\textbf{USD-FL}) to mitigate the effects of poisoning attacks on FL-based SC. USD-FL is designed to analyze the distribution of model parameters returned by each participating client. Through this analysis, our proposed defense can accurately identify devices that were trained on poisoned data and distinguish them from devices trained on noisy (i.e., low signal-to-noise ratio) data, which often introduces false alarms in previously proposed adversarial defense frameworks \cite{cao2022mpaf,yin2018byzantine,blanchard2017machine,byz_atk}. 

\begin{figure*}[t!]
    \centering
    \includegraphics[width=\textwidth]{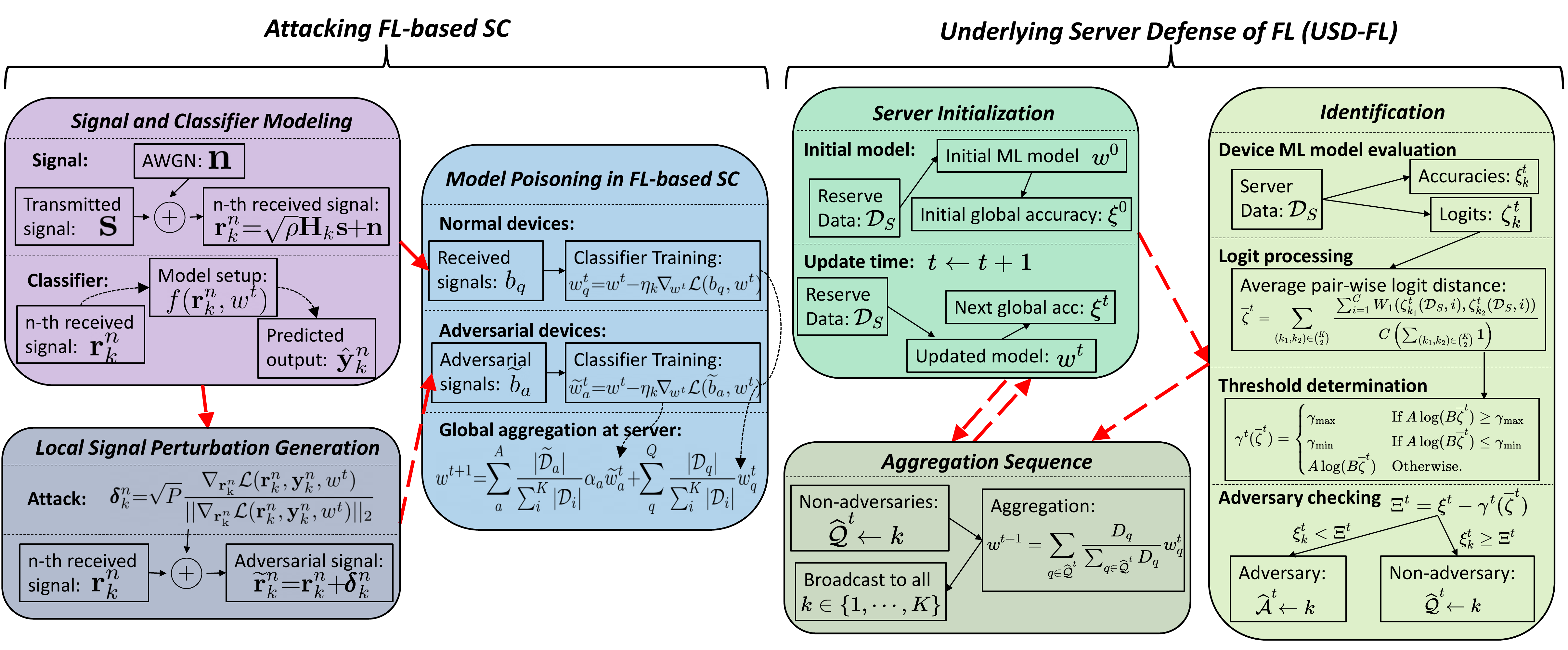}
    \caption{Overall flow of our adversarial framework for FL-based SC and our proposed defense methodology, USD-FL. Both methodologies rely on signal and classifier modeling decisions presented in Sec.~\ref{sec:sig_mod}. } 
    \label{fig:ovr_flow}
    \vspace{-4mm}
\end{figure*}

Our proposed USD-FL methodology examines the distribution of logits via the 1-Wasserstein distance in FL-based SC, and subsequently leverages that information to develop a threshold function for each device's model parameters, which separates devices that are extremely heterogeneous from those that are likely compromised by adversarial perturbations. 
{\color{black}Logits refer to raw, un-normalized predictions from a machine learning (ML) model, and our methodology determines devices' logits by applying their ML models on the server's reserve dataset, a small cache of data distributed in a non-i.i.d. way with respect to the devices' local datasets. 
USD-FL leverages logits to detect adversaries and defend FL because, as established via~\cite{carlini2017towards,kannan2018adversarial}, attack potency and detectability are strongly linked with logit distributions in centralized settings.}

{\color{black}Subsequently, we compare logits in a pairwise manner via the 1-Wasserstein distance (i.e., a permutation distance), which enables for more precise estimates of the underlying degree of heterogeneity, including potential adversarial perturbations, throughout the network without requiring   perfect knowledge of both the quantity of adversaries and the start of adversarial attacks as in existing server-driven defenses.}
This, in turn, allows USD-FL to setup an adaptive threshold function that automatically conforms to different network conditions, enabling the server to filter away devices with adversarial perturbations and yielding a modified FL model aggregation rule.



\subsection{Outline}
{\color{black}
We now explain the structure of our paper. 
First, we review relevant literature relevant to adversarial attacks on and defenses for FL-based SC in Sec.~\ref{sec:related_work}. 
Next, we present the system model for standard FL-based SC in Sec.~\ref{sec:atk_ovr_method}.  
Subsequently, we present the development and experimental evaluation of the proposed USD-FL defense methodology in Sec.~\ref{sec:def_ovr_method} and~\ref{sec:def_eval}, with a demonstration of the potency of popular evasion attacks to further motivate USD-FL in Sec.~\ref{sec:perf_eval}.
For clarity, we provide a high-level visualization of our workflow in Fig.~\ref{fig:ovr_flow}. 
}

\subsection{Summary of Contributions:}


\begin{enumerate}
    
    \item \textbf{Development of USD-FL:} (Sec.~\ref{sec:def_ovr_method}) 
    Our proposed USD-FL methodology is among the first to show that ML models compromised by evasion attacks on unperturbed data still yields compromised logits. 
    Simultaneously, USD-FL shows that information can be extracted from logits via the 1-Wasserstein distance to automatically adjust a threshold function, and thereby defend FL-based SC against adversarial perturbations.
    
    \item \textbf{Resilience against Poisoning Attacks:} (Sec.~\ref{sec:def_comp_exp} and Sec.~\ref{sec:conf_mat})
    We evaluate the performance of USD-FL relative to four popular defenses from literature through experiments conducted on a real-world AMC dataset. These experiments illustrate that USD-FL 
    offers the greatest performance improvements against a range of data poisoning attacks.

    \item \textbf{Minimize False Positive Rates:} (Sec.~\ref{sec:false_positive} and Supplementary Materials)
    USD-FL reduces false positive device classifications for both i.i.d. and non-i.i.d. settings in FL-based SC.
    This improvement enhances convergence speed and contributes to USD-FL's superior ML model performance, highlighting USD-FL's efficiency in defending FL-based SC. 

\end{enumerate}

%% file: related_work.tex
\noindent 
\textbf{Adversarial attacks in FL-based SC:} 
Centralized DL-based SC has been shown to be susceptible to adversarial evasion attacks \cite{adv_in_rf,amc_adv_atk1,amc_adv_atk2,amc_adv_atk3}. 
In these settings, the SC DL classifier is first trained using a collection of labeled radio signals, then, during test time, the adversary perturbs inputs to induce the trained classifier to output erroneous predictions, thereby attacking the SC DL classifier during the inference phase. 
Several defenses have been proposed to mitigate such attacks \cite{amc_def1,amc_def2}, but these methods are designed specifically for test-time attacks in the centralized SC scenario. Our focus, on the other hand, is on defending against adversarial attacks that poison the model training process and lead to a compromised post-training model rather than test-time attacks. 

One very effective technique for mitigating evasion attacks on centralized SC systems is adversarial training \cite{adv_trn1,adv_trn2,tian2022exploring}, where the training set is augmented with adversarial examples in order to increase test-time performance in the presence of such attacks. However, adversarial training on samples with high-bounded perturbations results in the model overfitting to adversarial examples, thus reducing classification performance on unperturbed samples \cite{adv_trn_overfitting}. 
We investigate this property in FL-based SC, showing that augmenting the local training sets of particular FL devices with imperceptible adversarial evasion attacks can instead poison the global model during training and reduce its classification performance.

{\color{black}In terms of FL-based systems, existing works~\cite{atks_in_fl,label_flip,byz_atk} have investigated how to corrupt trained classifier performances. 
In this context, model poisoning attacks, which aim to corrupt the training process, have been proposed for image processing tasks \cite{atks_in_fl}. Such attacks consist of label flipping \cite{label_flip} and model parameter perturbations \cite{byz_atk}. In the former case, the resulting attack potency is low and can be mitigated through global averaging of all model parameters. The latter case relies on perturbing weights after training, which can be detected using existing distributed SC algorithms \cite{fl_atk_det}.}

\begin{table}[t!]
\caption{{\color{black}Comparison of Relevant Literature}}
\vspace{-1.5mm}
\label{tab:lit_review} 
{\footnotesize
{\color{black}
\begin{tabularx}{0.49\textwidth}{m{12mm} c c c c}
\toprule[.2em] 
\multirow{2}{*}{Method} & Extensive & Unknown & Unknown & Low\\ 
& Heterogeneity & Adversaries & Attack Time & Overhead \\
\midrule
Median~\cite{yin2018byzantine,cao2022mpaf} & X & \checkmark & X & \checkmark \\
Trimmed~\cite{yin2018byzantine,cao2022mpaf} & X & X & X & \checkmark \\
UnionM~\cite{byz_atk} & \checkmark & \checkmark & X & \checkmark\\
UnionS~\cite{byz_atk} & X & X & X & \checkmark \\
MODEL~\cite{wu2024model} & X & \checkmark & \checkmark & X \\
Dual~\cite{xu2024dual} & X & \checkmark & \checkmark & X \\
FCD~\cite{kumar2024revamping} & \checkmark & X & \checkmark & X \\
USD-FL & \checkmark & \checkmark & \checkmark & \checkmark\\ 
\bottomrule
\end{tabularx} } }
\vspace{-4mm}
\end{table}

{\color{black}
Contrary to these works, we characterize attack frameworks that do not rely on perturbing the model parameters after local training, which bypass detection mechanisms from previous SC frameworks~\cite{wang2023potent} and motivate the need for our proposed USD-FL defense methodology for FL-based SC. 
}

{\color{black}We also want to emphasize that there exist theoretic works~\cite{lyu2024adversarial,kumar2023impact} which have established that such adversarial attack frameworks are very damaging for general FL-based classification tasks. Therefore, USD-FL's goal to mitigate the damage of adversarial evasion attacks on FL-based SC is an impactful one.}

\textbf{Defenses against adversarial devices in FL-based SC:}
{\color{black}
To preserve ML model performance against adversarial attacks in FL settings, existing works have focused on either comparing local device training data~\cite{shen2016auror,li2021lomar,zhang2024flpurifier} or designing effective server-driven FL defenses~\cite{tao2023byzantine,yin2018byzantine,jiang2023data,xie2019zeno,yang2024roseagg}, which typically consist of modifying the global aggregation rule by discarding device ML models or components.} 
However, both lines of research have difficulty adapting to the extensive device and training data heterogeneity in wireless networks. 
In particular,~\cite{shen2016auror} relies on homogeneous training data distributions, while~\cite{li2021lomar} requires the server to obtain knowledge of the data distributions across network devices, which a central server in FL-based SC may have difficulty obtaining. 


{\color{black}
Similarly, current server-driven defenses for FL also suffer from heterogeneity concerns induced by wireless networks~\cite{chen2022uav,jia2022energy}. 
For example,~\cite{yin2018byzantine} and~\cite{tao2023byzantine} rely on the server to determine and subsequently discard the model parameters that deviate too far from the global average ML model. 
Meanwhile,~\cite{xie2019zeno} and~\cite{byz_atk} rely on a reserve dataset at the server to assess the output of various device ML models, subsequently excluding ML models with inferior performance in terms of classification error and/or ML model loss.} 

{\color{black} While existing server-driven defense methodologies are effective when network devices exhibit homogeneity in their underlying data distributions, they have difficulty adapting to environments with pervasive heterogeneity, as often observed in wireless networks.
In particular, wireless networks contain devices that are heterogeneous with respect to (i) the local quality of wireless equipment~\cite{wang2022uav,vilajosana2023challenges,lin20215g,wild2021joint} and (ii) the types of modulated signals received, which includes various modulation schemes and signal-to-noise ratios~\cite{mohamed2020strategies,cai2017modulation,liu2022integrated}.

Naturally, these factors lead to highly heterogeneous local training data, which subsequently results in highly heterogeneous yet non-adversarial local model parameters. 
Consequently, existing defenses~\cite{byz_atk,yin2018byzantine,cao2022mpaf} have difficulty distinguishing between model parameters trained by wireless devices with non-adversarial but noisy signal data and those poisoned by a genuine adversarial attack. 
This is especially problematic in FL-based SC as filtering away model parameters trained by devices with honest but noisy training data can lead to global aggregations that are further biased towards the model parameters from adversarial devices.}

{\color{black}Using the concept of reserve server datasets from~\cite{byz_atk}, we aim to address this problem by first examining the logits derived from the server's reserve dataset, and then adjusting a threshold function to distinguish between heterogeneity and adversary. In this manner, our methodology aims to provide a defense with low false positive rates to detect adversarial devices in FL-based SC. Within our experimental evaluation, we then confirm that the proposed USD-FL methodology better adapts to extensive network heterogeneity.}

{\color{black}Moreover, we want to emphasize that existing works such as~\cite{byz_atk,kumar2024revamping,tao2023byzantine,alharbi2024collusive} rely on perfect network information at the server, i.e., the server knows the number of adversaries, the start time of adversarial attacks, and more. 
In practical network settings, none of this information would be available at the server, and thus, these existing defenses for FL-based SC may encounter some difficulty.

By contrast, the proposed USD-FL methodology does not require the exact number of adversarial devices or the starting attack time. 
Furthermore, the USD-FL defense incurs low computational overhead, especially compared to existing server-driven defenses such as~\cite{wu2024model,xu2024dual}, which requires re-performing k-means clustering until convergence for each global aggregation, or~\cite{kumar2024revamping}, which further requires that devices are all initially trustworthy for an extended period of time.}

{\color{black} We provide a summary of comparison to other defenses for FL-based SC in Table~\ref{tab:lit_review}. In particular, we can see that, relative to the existing server-driven baselines, the proposed USD-FL methodology is designed to adapt to extensive heterogeneity, unknown adversaries, unknown adversarial attack start time, and features low computational overhead for the network.}



%% file: atk_ovr_method.tex
This section will first discuss our notations and system model for evasion attacks on FL-based SC in Sec.~\ref{sec:sig_mod}-\ref{sec:mod_pois}.
An overview of the system described throughout this section is provided in Fig.~\ref{fig:sys_model}.


\subsection{Signal and Classifier Modeling} \label{sec:sig_mod}

We consider an FL framework consisting of $k = 1, 2, \ldots, K$ participating training devices, where each device contains a local dataset denoted by $\mathcal{D}_{k}$ consisting of $|\mathcal{D}_{k}|$ samples. At each device, $\mathcal{D}_{k}$ is comprised of a set of received signals, which were each transmitted to device $k$ through the channel $\mathbf{h}_{k} = [h_{k}[0],\ldots,h_{k}[\ell - 1]]^{T}$, where $\ell$ is the length of the received signal's observation window. We assume that the channel distribution between the transmitter and each device is independent and identically distributed (i.i.d.). Formally, the $n^{\text{th}}$ signal received at device $k$ is modeled by
\begin{equation}
    \mathbf{r}^{n}_{k} = \sqrt{\rho}\mathbf{H}_{k}\mathbf{s}^{n}_{k} + \mathbf{n},
\end{equation}
where $\mathbf{s}^{n}_{k} = [s[0], \ldots, s[\ell - 1]]$ is the transmitted signal, $\mathbf{H}_{k} = \text{diag}\{h_{k}[0],\ldots,h_{k}[\ell - 1]\} \in \mathbb{C}^{\ell \times \ell}$, $\mathbf{n} \in \mathbb{C}^{\ell}$ is complex additive white Gaussian noise (AWGN), and $\rho$ denotes the signal to noise ratio (SNR), which is known at the receiver of each device. 
Each realization of $ \mathbf{r}^{n}_{k}$ comes from various constellations, and the FL objective is to learn a global signal classifier by training all local models to classify the signal as one of $C$ possible signal constellations. 

{\color{black}
While all received signals are complex, $\mathbf{r}^{n}_{k} \in \mathbb{C}^{\ell}$, we represent each signal in terms of its real and imaginary components, $\mathbf{r}^{n}_{k} \in \mathbb{R}^{\ell \times 2}$, where the two columns correspond to the real and imaginary parts of $\mathbf{r}^{n}_{k}$. This representation allows us to (i) leverage all signal features during training and (ii) use real-valued DL architectures, which are predominantly employed in DL- and FL-based SC.}

{\color{black}At each training round, $t$, the global model transmits its parameters, $w^{t}$, to each FL device. Each device then trains, using $w^{t}$ as the starting point, its own local model, denoted by $f(\cdot, w^{t}): \mathbb{R}^{\ell \times 2} \rightarrow \mathbb{R}^{C}$, where $f()$ denotes the deep learning classifier (identical architecture at each device) and $(\cdot)$ represents the input.
When the training round ends, each device returns $w^{t}_{k}$, which are the model parameters of device $k$ after the completion of training round $t$ on $\mathcal{D}_{k}$, to the global server for aggregation (further discussed in Sec. \ref{sec:mod_pois}).}

{\color{black}
After aggregation, the global server transmits the updated model parameters, $w^{t+1}$, to all devices for the next round of training. The model prediction, after training round $t$, is given by $\hat{\mathbf{y}}^{n}_{k} = f(\mathbf{r}^{n}_{k}, w^{t})$, where $\hat{\mathbf{y}}^{n}_{k} \in \mathbb{R}^{C}$ denotes the predicted output vector of $\mathbf{r}^{n}_{k}$ from $f()$. 
Moreover, the predicted signal constellation is given by $\text{argmax}_{j} \hspace{1mm} \hat{\mathbf{y}}^{n}_{k, j}$, where $\hat{\mathbf{y}}^{n}_{k, j} \in \mathbb{R}$ is the $j^{\text{th}}$ element of $\hat{\mathbf{y}}^{n}_{k}$.}

\subsection{Local Data Perturbation Generation} \label{sec:atk_fw}

Here, we describe the process followed by adversarial devices, which are a subset of network devices that train on perturbed input data, while the remaining devices train on their original, unperturbed datasets. 
At the beginning of each training iteration, after the local model has received an updated global model, adversarial devices will craft adversarial evasion perturbations on each instance of $\mathbf{r}^{n}_{k}$. The $n^{\text{th}}$ resulting sample is denoted by 
\begin{equation} \label{adv_sig}
    \widetilde{\mathbf{r}}^{n}_{k} = \mathbf{r}^{n}_{k} + \pmb{\delta}^{n}_{k},
\end{equation}
where $\pmb{\delta}^{n}_{k}$ is the adversarial perturbation crafted for the $n^{\text{th}}$ signal on device $k$. 

The adversarial perturbation, $\pmb{\delta}^{n}_{k}$, could be crafted at each local device by utilizing common perturbation models such as AWGN or changing the local data completely by e.g., using zero-vectors as training samples or training on signals received from an out-of-distribution channel.
However, the injection of AWGN results in less potent attacks to the global model in comparison to other perturbation methodologies (as we will show in Sec. \ref{sec:perf_eval}). On the other hand, although changing the local training data may result in more potent attacks, the global model can simply query training samples from each local device to identify the adversarial device. Using adversarial evasion attacks, as we propose, induces a higher attack potency while simultaneously being imperceptible and, thus, is able to withstand existing FL adversarial attack detectors. 


To craft an effective and imperceptible perturbation, adversaries will aim to satisfy
\begin{subequations} \label{adv_opt:all-lines}
\begin{align}
    \underset{\pmb{\delta}}{\text{min}} \quad &  ||\pmb{\delta}^{n}_{k}||_{2} \label{adv_opt:line_1} \\
    \text{s. t.} \quad &  \hspace{0.5mm} f(\mathbf{r}^{n}_{k}, w^{t}) \neq f(\mathbf{r}^{n}_{k} + \pmb{\delta}^{n}_{k}, w^{t}), \label{adv_opt:line_2} \\ 
    \quad & ||\pmb{\delta}^{n}_{k}||_{2}^{2} \leq P, \label{adv_opt:line_3} \\
     \quad & \mathbf{r}^{n}_{k} + \pmb{\delta}^{n}_{k} \in \mathbb{R}^{\ell \times 2} \label{adv_opt:line4}, 
\end{align}
\end{subequations} 
where we use $\Vert \cdot \Vert_{2}$ for the $l_2$ norm and $P$ for the max perturbation power. 
\eqref{adv_opt:line_1} minimizes power to keep the perturbation imperceptible, \eqref{adv_opt:line_2} changes the prediction of the perturbed sample for a given model, \eqref{adv_opt:line_3} sets a maximum power, and \eqref{adv_opt:line4} keeps $\widetilde{\mathbf{r}}^{n}_{k}$ and $\mathbf{r}^{n}_{k}$ in the same dimensional space. 

\begin{algorithm}[h!] 
   \caption{FL model poisoning in SC at training iteration $t > t_{0}$, where the adversarial devices train on perturbed inputs.}
   \label{adv_alg}
   \begin{algorithmic}[1] 
        \STATE \textbf{input:} $w^{t}$: global parameter values at training round $t$ \\ 
        

            \FOR{$q = 1, \ldots, Q$} 
                \STATE $\mathcal{B}_q \gets$ split $\mathcal{D}_{q}$ into batches
                \STATE $w^{t}_q \gets w^{t}$
                \FOR{batch $b_q \in \mathcal{B}_q$}
                    \STATE $w^{t}_{q} \gets w^{t}_q - \eta \nabla_{w^{t}_q}\mathcal{L}(b_q, w^{t}_q)$
                \ENDFOR
            \ENDFOR

            \FOR{$a = 1, \ldots, A$}
                \STATE $\widetilde{\mathcal{D}}_{a} \gets \{\}$
                \FOR{$\mathbf{r}^{n}_{a} \in \mathcal{D}_{a}$}
                    \STATE $\pmb{\delta}^{n}_{a} = \sqrt{P} \frac { \nabla_{\mathbf{r}_{\text{k}}^{n}} \mathcal{L}(\mathbf{r}^{n}_{k}, \mathbf{y}^{n}_{k}, w^{t})} {|| \nabla_{\mathbf{r}_{\text{k}}^{n}} \mathcal{L}(\mathbf{r}^{n}_{k}, \mathbf{y}^{n}_{k}, w^{t})||_{2}}$
                    \STATE $\widetilde{\mathbf{r}}^{n}_{a} = \mathbf{r}^{n}_{a} + \pmb{\delta}^{n}_{a}$
                    \STATE add $\widetilde{\mathbf{r}}^{n}_{a}$ to $\widetilde{\mathcal{D}}_{a}$
                    
                \ENDFOR
                
                \STATE $\widetilde{\mathcal{B}}_a \gets$ split $\widetilde{\mathcal{D}}_{a}$ into batches
                \STATE $\widetilde{w}^{t}_a \gets w^{t}$
                \FOR{batch $\widetilde{b}_a \in \widetilde{\mathcal{B}}_a$}
                    \STATE $\widetilde{w}^{t}_{a} \gets \widetilde{w}^{t}_a - \eta \nabla_{\widetilde{w}^{t}_a}\mathcal{L}(\widetilde{b}_a, \widetilde{w}^{t}_a)$
                    
                \ENDFOR

            \ENDFOR

            \STATE $w^{t+1} = \sum_{a}^{A} \frac{|\widetilde{\mathcal{D}}_{a}|}{\sum_{i}^{K} |\mathcal{D}_{i}|} \alpha_{a} \widetilde{w}^{t}_{a} + \sum_{q}^{Q} \frac{|\mathcal{D}_{q}|}{\sum_{i}^{K} |\mathcal{D}_{i}|} w^{t}_{q}$

        \RETURN $w^{t+1}$
  
  \end{algorithmic}
\end{algorithm}

Due to its excessive nonlinearity, however, (\ref{adv_opt:all-lines}) is difficult to solve using traditional optimization methods. 
Thus, we approximate its solution using the fast gradient sign method (FGSM) \cite{fgsm}, and projected gradient descent (PGD)~\cite{madry2018towards}. 

\textbf{FGSM:} The FGSM perturbation for our proposed FL-based SC model is given by
\begin{equation} \label{fgsm}
    \pmb{\delta}^{n}_{k} = \sqrt{P} \frac { \nabla_{\mathbf{r}^{n}_{\text{k}}} \mathcal{L}(\mathbf{r}^{n}_{k}, \mathbf{y}^{n}_{k}, w^{t})} {|| \nabla_{\mathbf{r}^{n}_{\text{k}}} \mathcal{L}(\mathbf{r}^{n}_{k}, \mathbf{y}^{n}_{k}, w^{t})||_{2}},
\end{equation}
where 
\begin{equation}
    \mathcal{L}(\mathbf{r}^{n}_{k}, \mathbf{y}^{n}_{k}, w^{t}) = \sum_{j=1}^{C} \mathbf{y}^{n}_{k, j} \text{log}(\hat{\mathbf{y}}^{n}_{k})
\end{equation}
is the cross entropy loss with $\mathbf{y}^{n}_{k, j}$ denoting the $j^{\text{th}}$ element of the true label vector corresponding to $n^{\text{th}}$ sample on the $k^{\text{th}}$ device and $\nabla_{\mathbf{r}_{\text{k}}} \mathcal{L}(\mathbf{r}^{n}_{k}, \mathbf{y}^{n}_{k}, w^{t})$ denotes the gradient of $\mathcal{L}(\mathbf{r}^{n}_{k}, \mathbf{y}^{n}_{k}, w^{t})$ w.r.t. $\mathbf{r}^{n}_{k}$. Finally, $\sqrt{P} / || \nabla_{\mathbf{r}_{\text{k}}} \mathcal{L}(\mathbf{r}^{n}_{k}, \mathbf{y}^{n}_{k}, w^{t})||_{2}$ is the scaling factor used to satisfy the power constraint in (\ref{adv_opt:line_3}).

\textbf{PGD:} The PGD perturbation is an iterative extension of the FGSM attack, where a smaller power perturbation, $\alpha = \sqrt{P}/Q$, is added for $Q$ iterations. 
At the $q$-th iteration, 
\begin{equation} \label{eq:PGD_iteration}
    \boldsymbol{\Delta}^{n,(q)}_{k} = \boldsymbol{\Delta}^{n,(q-1)}_{k} + \alpha \frac { \nabla_{{\mathbf{r}}^{n}_{\text{k}}} \mathcal{L}(\widetilde{\mathbf{r}}^{n,(q-1)}_{k}, \mathbf{y}^{n}_{k}, w^{t})} {|| \nabla_{{\mathbf{r}}^{n}_{\text{k}}} \mathcal{L}(\widetilde{\mathbf{r}}^{n,(q-1)}_{k}, \mathbf{y}^{n}_{k}, w^{t})||_{2}},
\end{equation}
where the superscript $(q)$ denotes the $q$-th iteration, $\widetilde{\mathbf{r}}^{n,(q)}_k = \widetilde{\mathbf{r}}^{n,(q-1)}_k + \boldsymbol{\Delta}^{n,(q-1)}_{k}$, $\boldsymbol{\Delta}^{n,(0)}_{k} = 0$, and $\widetilde{\mathbf{r}}^{n,(0)}_k = \mathbf{r}^{n}_k$. This yields an effective perturbation $\pmb{\delta}^{n}_{k} = \boldsymbol{\Delta}^{n,(Q)}_k$, with the final signal, $\widetilde{\mathbf{r}}^{n}_k$, obtained via~\eqref{adv_sig}.

The objective of each adversarial device is to overfit their local model to the perturbed dataset generated using~\eqref{adv_sig} combined with either the FGSM attack~\eqref{fgsm} or the PGD attack~\eqref{eq:PGD_iteration} for each training sample. 
{\color{black}This attack framework of leveraging the ML model gradient to generate adversarial training data, outlined in Algorithm~\ref{adv_alg}, is very similar to that used in mimicry attacks~\cite{biggio2013evasion,zhang2020voiceprint}.}
We denote the batch of $N$ perturbed samples at an adversarial device $a$ as $\widetilde{b}_a = \{(\widetilde{\mathbf{r}}^{n}_{a}, \mathbf{y}^{n}_{a})\}_{n=1}^{N}$ and the weights at the end of training round, $t$ as $\widetilde{w}^{t}_{a}$. 
Similarly, the batch of $N$ unperturbed inputs as well as the weights at the end of training round $t$ at a non-adversarial device $q$ will be denoted as $b_q = \{(\mathbf{r}^{n}_{q}, \mathbf{y}^{n}_{q})\}_{n=1}^{N}$ and $w^{t}_{q}$, respectively. 
For the each devices' complete dataset (i.e., the superset of all possible local batches), we use $\widetilde{\mathcal{D}}_a$ and $\mathcal{D}_q$ for adversarial device $a$ and non-adversarial device $q$ respectively. 


\subsection{Model Poisoning in FL-Based Signal Classification} \label{sec:mod_pois}

To begin each training round, $t$, in the FL AMC training process, the global model will transmit $w^{t}$ to each participating FL device. Note that when $t=0$ (i.e., the first round of training), $w^{0}$ is initialized at the server, and subsequently synchronized across the network. After receiving $w^{t}$, each FL device $k$ will train $f(\cdot, w^{t})$ on $\mathcal{D}_{k}$. The model parameters of the $a^{\text{th}}$ adversarial device will be updated, beginning on training round $t_{0}$, according to 
\begin{equation} \label{eq:adv_sgd}
    \widetilde{w}^{t}_{a} = w^{t} - \eta \hspace{1mm} \nabla_{w^{t}}\mathcal{L}(\widetilde{b}_a, {w}^{t}), 
\end{equation}
while the model parameters of the $q^{\text{th}}$ non-adversarial device, along with adversarial devices prior to training round $t_{0}$, will be updated according to 
\begin{equation} \label{eq:non_adv_sgd}
    w^{t}_{q} = w^{t} - \eta \hspace{1mm} \nabla_{w^{t}}\mathcal{L}(b_q, w^{t}), 
\end{equation}
where $\eta$ is the learning rate.  
At the termination of training round $t$, each FL device will transmit its updated model parameters back to the global server. Although non-adversarial devices will transmit $w^{t}_{q}$ to the global model, adversarial devices will transmit $\alpha_{a} \hspace{0.5mm} \widetilde{w}^{t}_{a}$, where $\alpha_{a} > 0$ is a scaling factor used at adversarial device $a$ that can be used to make the effect of the perturbed weights more potent at the global model. Note that $\alpha_{a} = 1$ corresponds to not scaling the trained weights. In addition, each FL device will also transmit $|\mathcal{D}_{k}|$ to the global model for appropriate parameter scaling from each participating device during global aggregation. 

The global model will then perform a global aggregation using the received weights. From the perspective of the server, the aggregation scheme used to generate the model parameters for the next device training iteration has form: 
\begin{equation}
    w^{t+1} = \sum_{k}^{K} \frac{|\mathcal{D}_{k}|}{\sum_{i}^{K} |\mathcal{D}_{i}|} w^{t}_{k},
\end{equation}
where $K$ is the total number of FL devices. However, the true aggregation process, taking the effect of the adversarial devices into account, is given by
\begin{equation} \label{eq:true_avg}
    w^{t+1} = \sum_{a}^{A} \frac{|\widetilde{\mathcal{D}}_{a}|}{\sum_{i}^{K} |\mathcal{D}_{i}|} \alpha_{a} \widetilde{w}^{t}_{a} + \sum_{q}^{Q} \frac{|\mathcal{D}_{q}|}{\sum_{i}^{K} |\mathcal{D}_{i}|} w^{t}_{q}, 
\end{equation}
where $A$ and $Q$ are the total number of adversarial and non-adversarial devices, respectively, and $K = A + Q$. 
The complete overview of this model poisoning framework is given in Algorithm \ref{adv_alg}.


%% file: atk_eval.tex

%% file: def_ovr_method.tex
Our proposed USD-FL defense methodology investigates the ability of logits, which have been leveraged to improve convergence rates of standard FL~\cite{yan2023rethinking,zhang2022federated,itahara2021distillation}, to quantify the degree of heterogeneity among network devices. 
To do so, we assume the existence of a reserve dataset at the server, a common feature in many FL methodologies~\cite{chen2023importance,tian2022fedbert,chen2022pre,guo2022deep,byz_atk}, 
and acquire device logits by evaluating their ML models on the server's reserve dataset.

{\color{black}
In some works~\cite{byz_atk,guo2022deep}, the reserve dataset is equivalent to the validation dataset, but, for a more significant challenge, we assume that the reserve dataset is a much more limited dataset (e.g., fewer data, non-i.i.d., and unique labels) than the validation dataset - see Sec.~\ref{sec:def_eval}.} 

\begin{figure}[t]
    \centering
    \includegraphics[width=0.48\textwidth]{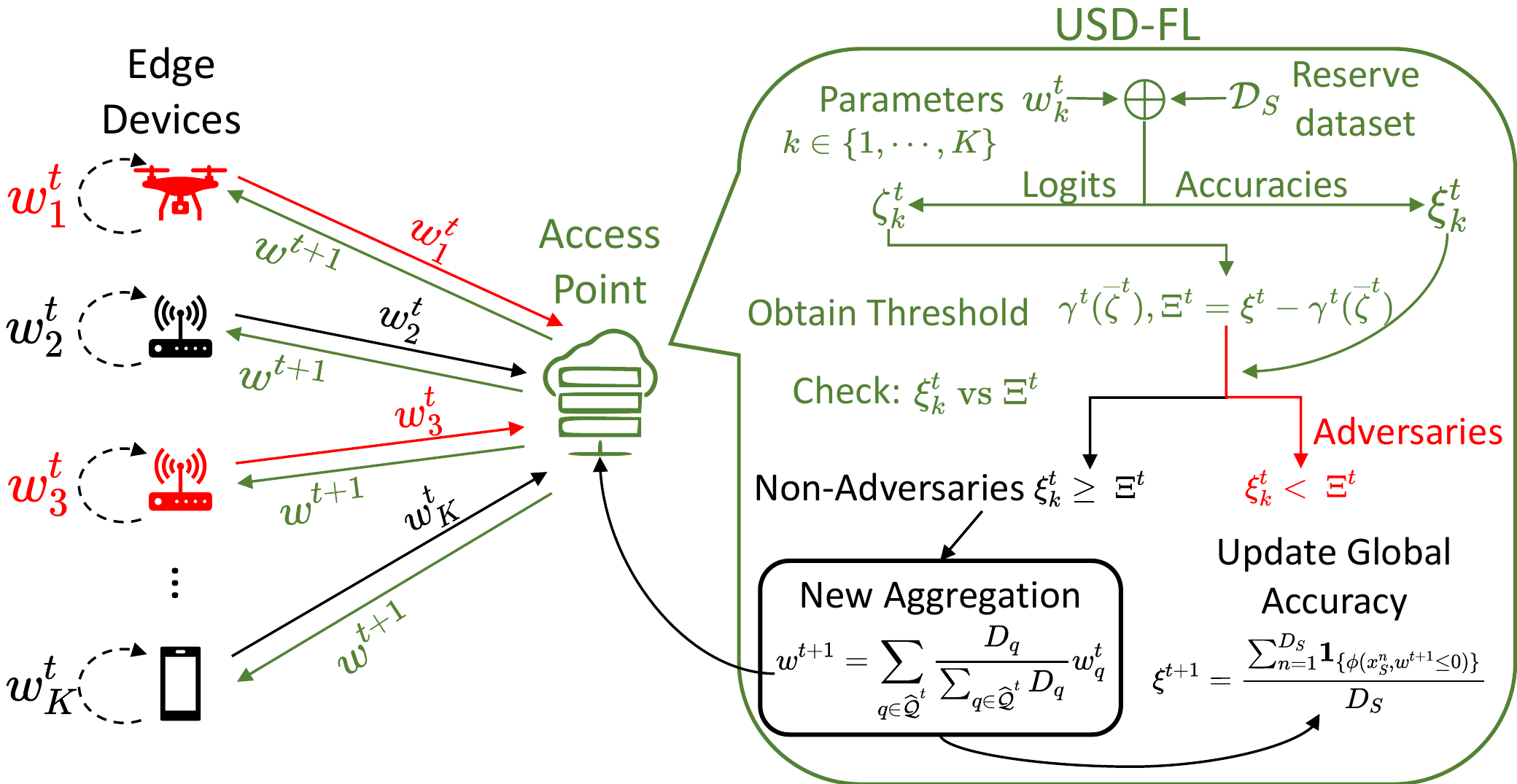}
    \caption{Simplified view of our proposed USD-FL methodology in which the server plays an active role in filtering adversarial vs non-adversarial devices. After determining likely adversaries, the server performs a modified aggregation, using the resulting model parameters to update a performance threshold.}
    \label{fig:def_overview}
    \vspace{-4mm}
\end{figure}

We then estimate the degree of heterogeneity within the network by exploiting the 1-Wasserstein distance, a permutation distance that characterizes the minimum transformation needed to convert one array/distribution into another.  
Subsequently, we design an adaptive threshold function to filter devices based on the average of their pairwise logit distances, which vary over time based on local ML training. 
Using the threshold function, the server finally partitions devices into non-adversarial and likely adversarial groups.
In the following, we describe USD-FL in terms of three components: (a) logit and logit distance computation in Sec.~\ref{ssec:logit_dist_comp}, (b) threshold function development in Sec.~\ref{ssec:threshold_func}, and (c) modification of global aggregations in Sec.~\ref{ssec:mod_agg}.
Fig.~\ref{fig:def_overview} provides a high level overview of our methodology, with focus on the threshold function and subsequent actions on the server-side. 

\subsection{Logits Extraction and Distance Computation} \label{ssec:logit_dist_comp}

We assume the central server $S$ contains an unperturbed dataset, $\mathcal{D}_S$, of size $D_S$ collected from historically received signals, which are distributed in a non-i.i.d. way with respect to the devices’ local datasets. 
{\color{black}At global aggregations, $S$ receives model parameters $w^{t}_{k}$ from devices $k \in \mathcal{K}$ within the network, identically to Sec.~\ref{sec:mod_pois}. 
{\color{black} 
Next, the server $S$ derives a set of logits, i.e., the raw, un-normalized predictions from an ML model, for each device by evaluating each device's set of ML model parameters $w^{t}_k$ on the reserve dataset $\mathcal{D}_S$ and stopping prior to the normalization step.}
In this way, the server obtains a set of logits $\boldsymbol{\zeta}^{t}_k \in \mathbb{R}^{D_S\times C}$ for every $k \in \{1,\cdots,K\}$.}

The server $S$ then computes the pairwise 1-Wasserstein distance among device pairs, which is formally defined as:
\begin{definition} {\color{black}($1$-Wasserstein distance~\cite{villani2009optimal}) \label{def:1-wasserstein_dist}
If $P = \{P_1,\cdots,P_C\}$ and $Q = \{Q_1, \cdots, Q_C\}$, then the $1$-Wasserstein distance can be computed as follows:
\begin{equation} \label{eq:p-wasserstein_dist}
    W_1(P,Q) = \inf_{\pi} \bigg( \sum_{i=1}^{C} \Vert P_i - Q_{\pi(i)} \Vert \bigg) 
\end{equation}
where the infimum is over all permutations $\pi$ of $\{1,\cdots,C\}$.}\footnote{General $p$-Wasserstein distances involve higher powered exponents in~\eqref{eq:p-wasserstein_dist}, and grow rapidly in computational complexity. They are therefore typically reserved for more complicated distributions~\cite{villani2009optimal}.}
\end{definition}
With regards to notation, Definition~\ref{def:1-wasserstein_dist} translates to $W_1(\zeta^{t}_{m}(\mathcal{D}_S,i),\zeta^{t}_{n}(\mathcal{D}_S,i))$, where $m$ and $n \in \{1,\cdots,K \}$, $ i \in \{ 1, \cdots, C\}$, and $\zeta^{t}_{m}(\mathcal{D}_S,i)$ refers to the logits of the $m$-th device extracted from the evaluation of $w^{t}_m$ on $\mathcal{D}_S$ for the $i$-th label. 
Since $\zeta^{t}_{m} \in \mathbb{R}^{D_S \times C}$, as defined earlier, we can express the logits at a device $m$ as $\zeta^{t}_{m} = \{\zeta^{t}_{m}(\mathcal{D}_S,1), \cdots, \zeta^{t}_{m}(\mathcal{D}_S,C) \}$.

Each permutation $\pi$ is a unique arrangement of elements of an array/distribution, and, by measuring the infimum over all permutations $\pi$, the 1-Wasserstein distance is able to provide a quantitative measure of the minimum dissimilarity between two arrays/distributions, regardless of their arrangement. 
Since adversaries in FL-based SC typically have non-i.i.d. underlying data distributions, our intuition is that adversaries are likely to produce logits that share structural similarities but differ in their physical arrangement. 
{\color{black}Hence, we leverage the 1-Wasserstein distance to understand the structural relationships of the logits across pairs of network devices.}

\begin{figure}[t]
    \centering
    \includegraphics[width=0.5\textwidth]{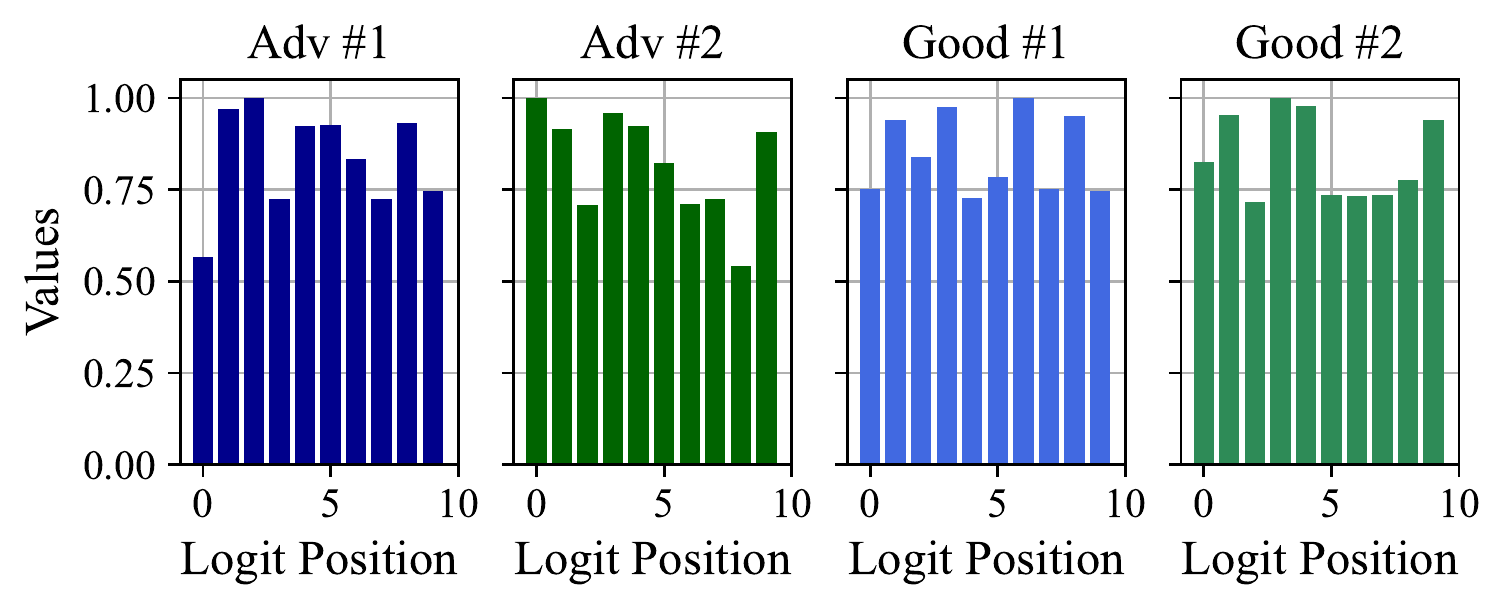}
    \caption{{\color{black}Example logits to highlight the similarity of logits at adversaries as well as the operation of the 1-Wasserstein distance versus other distances in Table~\ref{tab:wasserstein_toy_ex}.}}
    \label{fig:wasserstein_toy_ex}
\end{figure}

We provide an example visualization of devices' logits in Fig.~\ref{fig:wasserstein_toy_ex} along with different distance metrics in Table.~\ref{tab:wasserstein_toy_ex}.
In this example, the network consists of $4$ devices, two adversaries perturbed by the FGSM attack and two good devices, which have data that is distributed in an i.i.d. fashion. 
However, the reserve dataset at the server has data that is distributed in a non-i.i.d. fashion relative to the devices. 
Using the process outlined above, we extract and plot the devices' logits in Fig.~\ref{fig:wasserstein_toy_ex}.

{\color{black} The adversaries' logits, depicted in Fig.~\ref{fig:wasserstein_toy_ex}, share structural similarities, such as comparable minima and maxima and differ structurally from the logits of good devices.}
Intuitively, the distance between Adversary \#1 and Adversary \#2 should be smaller than the distance from any adversary to any good device. 
This is not the case, however, for the $\ell_1$ and $\ell_2$ distances examined in Table~\ref{tab:wasserstein_toy_ex}, which instead suggest that the adversaries are more similar to the good devices than other adversaries. 
{\color{black}Only the 1-Wasserstein distance successfully identifies the structural similarities of the logits at adversaries, consistently providing distances between adversaries that are smaller than those between adversaries and non-adversaries.
This motivates our adoption of the 1-Wasserstein distance over previously proposed logit distances such as the $\ell_1$ or $\ell_2$-norm.}

{\color{black}Moreover, we want to emphasize the significantly larger gaps between the logits of adversaries and good devices in Fig.~\ref{fig:wasserstein_toy_ex} and Table~\ref{tab:wasserstein_toy_ex}. The proposed USD-FL methodology will next leverage this property in Sec.~\ref{ssec:threshold_func} to develop a time-varying threshold function.}

\begin{table}[!t]
\caption{{\color{black}Logit distances from the example logits in Fig.~\ref{fig:wasserstein_toy_ex}. 1-Wasserstein distances yield far larger gaps between adversaries and good devices.}}
\vspace{-1.5mm}
\label{tab:wasserstein_toy_ex} 
\begin{tabularx}{0.48\textwidth}{c| c c c c} 
\toprule[.2em] 
\multicolumn{5}{c}{\textbf{1-Wasserstein Distance}}\\
\centering{Device Num} & Adversary \#1 & Adversary \#2 & Good \#1 & Good \#2 \\
\cmidrule(lr){1-5}
Adversary \#1 & 0 & 0.01367 & 0.03839 & 0.03803 \\
Adversary \#2 & 0.01367 & 0 & 0.04672 & 0.04428 \\
Good \#1 & 0.03839 & 0.04672 & 0 & 0.00796 \\
Good \#2 & 0.03803 & 0.04428 & 0.00796 & 0 \\
\midrule[.2em]
\multicolumn{5}{c}{\textbf{$\ell_1$ Distance}}\\
\centering{Device Num} & Adversary \#1 & Adversary \#2 & Good \#1 & Good \#2 \\
\cmidrule(lr){1-5}
Adversary \#1 & 0 & 2.186 & 1.22 & 1.981 \\
Adversary \#2 & 2.186 & 0 & 1.574 & 0.7043 \\
Good \#1 & 1.22 & 1.574 & 0 & 1.342 \\
Good \#2 & 1.981 & 0.7043 & 1.342 & 0 \\
\midrule[.2em]
\multicolumn{5}{c}{\textbf{$\ell_2$ Distance}}\\
\centering{Device Num} & Adversary \#1 & Adversary \#2 & Good \#1 & Good \#2 \\
\cmidrule(lr){1-5}
Adversary \#1 & 0 & 0.8036 & 0.463 & 0.7126 \\
Adversary \#2 & 0.8036 & 0 & 0.5697 & 0.317 \\
Good \#1 & 0.463 & 0.5697 & 0 & 0.5258 \\
Good \#2 & 0.7126 & 0.317 & 0.5258 & 0 \\
\bottomrule[.2em]
\end{tabularx}
\vspace{-4mm}
\end{table}

\subsection{Threshold Function} \label{ssec:threshold_func}
While the 1-Wasserstein distance can be leveraged to cluster devices into groups based on logit similarities, it does not tell us which groups of devices are adversaries. 
This problem is exacerbated in highly heterogeneous networks, wherein many different groups of non-adversarial devices with structurally similar logits can emerge. 
In response, we leverage the average pairwise logit distance, computed via the 1-Wasserstein distance, to control a threshold function, thereby partitioning devices into adversaries and non-adversaries.


{\color{black}The threshold function is a time-varying accuracy threshold, where if the accuracy of a device's ML model falls beneath the threshold then it is assumed to be an adversary. 
This concept is possible because the server $S$ is assumed to have a small reserve dataset $\mathcal{D}_S$ - similar to~\cite{byz_atk,guo2022deep} - that was earlier used to determine logits.}
Prior to each global aggregation, the server evaluates the previous global ML model $w^{t-1}$ on $\mathcal{D}_S$ to determine its pre-aggregation quality. 
We use the server's classification accuracy $\xi^{t-1}$, which is the percentage of data in $\mathcal{D}_{S}$ that is correctly classified by $w^{t-1}$, as the quality.
Similarly, for each device $k \in \{1,\cdots,K \}$, we can obtain the accuracy $\xi^{t}_k$ of its ML model $w^{t}_k$ evaluated on $\mathcal{D}_S$. 



We next define the average logit distance $\overline{\zeta}^{t}$ as:
\begin{equation} \label{eq:avg_logit_dist}
    \overline{\zeta}^{t} = \sum_{ k_1 \in \mathcal{K} } \frac{\sum_{\substack{k_2 \in \mathcal{K} \\ k_1 \neq k_2}} \sum_{i=1}^{C} W_1( \zeta^{t}_{k_1}(\mathcal{D}_S,i) , \zeta^{t}_{k_2}(\mathcal{D}_S,i) )  }
    {C (K-1)},  
\end{equation}
which enables the formation of an accuracy threshold $\Xi^{t}$ based on the empirical global accuracy $\xi^{t}$ and the average logit distance $\overline{\zeta}^{t}$, thus accounting for extreme network heterogeneity. 
{\color{black}As successful training progresses, devices' ML models will improve in local classification accuracy and confidence (i.e., smaller loss), which translates to logits and logit distances that grow exponentially~\cite{wu2021logit,ishida2020we}.}
In order to linearize the change in average logit distances $\overline{\zeta}^{t}$, the penalty function $\gamma^{t}(\overline{\zeta}^{t})$ was thus chosen to be logarithmic, as follows:
\begin{equation} \label{eq:penalty_logits}
\gamma^{t}(\overline{\zeta}^{t}) = 
\begin{cases} 
    \gamma_{\max} & \text{If } A \log(B\overline{\zeta}^{t}) \geq \gamma_{\max}\\
    \gamma_{\min} & \text{If } A \log(B\overline{\zeta}^{t}) \leq \gamma_{\min} \\
    A \log(B\overline{\zeta}^{t})& \text{Otherwise.}
\end{cases}
\end{equation}
The scaling coefficients $A$ and $B$ can be determined based on the dataset and network conditions under evaluation, and we present our choices of $A$ and $B$ in Sec.~\ref{sec:def_eval}.  


{\color{black}The main idea of the penalty function $\gamma^{t}(\overline{\zeta}^{t})$ is that, in order to distort the global ML model in FL-based SC, adversaries must correspondingly display distorted logits beyond non-adversarial heterogeneity. 
The properties of the 1-Wasserstein distance capture this effect via increases in $\overline{\zeta}^{t}$ when an adversarial attack begins.  
Moreover, assuming effective defense and therefore improvements in ML model quality over time, logits at adversaries become more extreme relative to the non-adversaries in order to effect a change in the global ML model. 
The penalty function in~\eqref{eq:penalty_logits} changes with $\overline{\zeta}^{t}$, integrating in the underlying properties of FL-based SC into $\gamma^{t}(\overline{\zeta}^{t})$.}  

With the penalty function $\gamma^{t}(\overline{\zeta}^{t})$ explained, we now define the accuracy threshold $\Xi^{t}$ as follows:
\begin{equation} \label{eq:thresh_def}
    \Xi^{t} = \xi^{t} - \gamma^{t}(\overline{\zeta}^{t}). 
\end{equation} 
By comparing devices' $\xi^{t}_{k}$ relative to $\Xi^{t}$, the server $S$ can determine if a device $k$ exhibits adversarial characteristics ($\xi^{t}_k < \Xi^{t}$) or not ($\xi^{t}_k \geq \Xi^{t}$). Thereafter, $S$ can partition devices into likely adversaries $\widehat{\mathcal{A}}^t$ and non-adversaries $\widehat{\mathcal{Q}}^t$, analogously to the sets of true adversaries $\mathcal{A}$ and non-adversaries $\mathcal{Q}$. 
{\color{black}These design choices for~\eqref{eq:thresh_def} also enable the time-varying threshold $\Xi^{t}$ to automatically adapt to network changes.}

\begin{algorithm}[h!] 
   \caption{Underlying Server Defense of Federated Learning (\textbf{USD-FL}).}
   \label{alg:usd-fl}
   \begin{algorithmic}[1] 
        \STATE \textbf{input:} $\mathcal{D}_S$: unperturbed dataset at the server $S$\\
        \STATE \textbf{input:} ${w}^{0}$: initial global model parameters \\
        \STATE \textbf{input:} $\eta$: learning rate of model training \\
        \STATE $t=0$ \\
        \WHILE{True}
        \STATE find global ML model accuracy $\xi^{t}$ on $\mathcal{D}_S$ 
            \FOR{$k = 1, \ldots, K$}
                \STATE $w^{t}_k \gets w^{t}$ 
                \STATE device $k$ locally updates $w^{t}_k$
                \STATE $w^{t}_k$ sent to server $S$
            \ENDFOR
            \STATE $\widehat{\mathcal{A}}^t \gets \{\}$, $\widehat{\mathcal{Q}}^t \gets \{\}$
            \STATE Compute $\overline{\zeta}^{t}$, and $\gamma^{t}(\overline{\zeta}^{t})$ \\ 
            \STATE $\Xi^{t} = \xi^{t} - \gamma^{t}(\overline{\zeta}^{t})$ \\
            \FOR{$k = 1, \ldots, K$}
                \STATE find device ML model accuracy $\xi^{t}_k$ on $\mathcal{D}_S$
                \IF {$\xi^{t}_k < \Xi^{t}$}
                    \STATE add $k$ to $\widehat{\mathcal{A}}^t$ 
                \ELSE
                    \STATE add $k$ to $\widehat{\mathcal{Q}}^t$ 
                \ENDIF
            \ENDFOR
            \STATE $w^{t+1} = \sum_{q \in \widehat{\mathcal{Q}}^t} \frac{D_q}{\sum_{q \in \widehat{\mathcal{Q}}^t} D_q} w^{t}_q$
            \STATE $t = t+1$ 
        \ENDWHILE
  \end{algorithmic}
\end{algorithm}

\subsection{Modified Aggregation Rule} \label{ssec:mod_agg}
Post-partition of the $K$ network devices, the server $S$ can modify the global aggregation rule, rather than relying the compromised aggregation procedure in~\eqref{eq:true_avg}, and thereby obtain:
\begin{equation} \label{eq:mod_agg}
    w^{t+1} = \sum_{q \in \widehat{\mathcal{Q}}^t} \frac{D_q}{\sum_{q \in \widehat{\mathcal{Q}}^t} D_q } w^{t}_q \equiv \sum_{k \in \mathcal{K}} \frac{\mathds{1}_{ \{\xi^{t}_k \geq \Xi^{t}\} } w^{t}_k D_k } {\sum_{k \in \mathcal{K}} \mathds{1}_{ \{\xi^{t}_k \geq \Xi^{t} \} } D_k }, \hspace{-2mm}
\end{equation} 
where $\mathds{1}_{ \{\xi^{t}_k \geq \Xi^{t}\} }$ is the indicator function used to check if the $i$-th device accuracy, $\xi^{t}_k$, exceeds the accuracy threshold $\Xi^{t}$. 
Next, $S$ synchronizes model parameters at all devices, including those at devices perceived to be compromised by adversarial poisoning attacks, to the latest global model parameters $w^{t+1}$ from~\eqref{eq:mod_agg}. 
This synchronization is performed in order to minimize the consequences of false-positives in the adversary detection process, as, in highly heterogeneous wireless networks, non-adversarial devices may occasionally fall beneath the server threshold $\Xi^{t}$.

Simultaneously, the server uses the new $w^{t+1}$ to update the empirical accuracy $\xi^{t+1}$, which in turn updates the accuracy threshold $\Xi^{t+1}$. 
It is important to continuously update the accuracy threshold $\Xi^{t}$ because adversarial poisoning attacks may have bounded impacts~\cite{amc_def2,byz_atk,sahay2023defending}. 
In this manner, our proposed USD-FL methodology, summarized in Algorithm~\ref{alg:usd-fl}, is able to continuously defend against adversarial poisoning attacks from true adversaries $a \in \mathcal{A}$. 

{\color{black}
\subsection{Dynamic Network Adjustments} \label{ssec:dyn_net_adj}
In dynamic networks, devices may enter or exit the network, resulting in a time-varying set and number of network devices, $\mathcal{K}_t$ and $K_t$ respectively. 
In such settings, new device arrivals to the network may require a grace period to integrate their non-i.i.d. datasets with the partially trained global ML model.
As such, we augment USD-FL via (i) time-in-network based weighted average logit distances and (ii) personalized accuracy thresholds in order to account for dynamic networks, resulting in dynamic USD-FL (DUSD-FL). 
Formally, we represent time-in-network based weighted average logit distances as follows:
\begin{equation} \label{eq:tin_net_avg_logit}
    \overline{\zeta}^{t} = \sum_{ k_1 \in \mathcal{K}_t } \frac{t_{k_1}}{\overline{t}_{K_t}} \left( \frac{\sum_{\substack{k_2 \in \mathcal{K}_t \\ k_1 \neq k_2}} \sum_{i=1}^{C} W_1( \zeta^{t}_{k_1}(\mathcal{D}_S,i) , \zeta^{t}_{k_2}(\mathcal{D}_S,i) }{C(K_t-1)} \right),  
\end{equation}
where $t_{k_1}$ represents the time-in-network for device $k_1$ at iteration $t$, and $\overline{t}_{K_t}$ denotes the average time-in-network across devices $K_t$. 
The structure of~\eqref{eq:tin_net_avg_logit} biases the average logit distances towards those devices that have been in the network for more training iterations, the reasoning being that such devices have undergone more global aggregations and thus offer greater stability to the ML model training process. 

{\color{black}Moreover,~\eqref{eq:tin_net_avg_logit} downplays the logit distances relative to new arrival devices. The logic is that new network devices may have unique underlying data distributions and untrained local ML models, both of which cause these new devices' logits to be initially larger and divergent from those logits across the rest of the network devices~\cite{wu2021logit}.}

Leveraging~\eqref{eq:tin_net_avg_logit}, personalized accuracy thresholds are then defined as follows:
\begin{equation} \label{eq:dusd_thresh}
    \Xi^{t}_k = \xi^{t} - \gamma^{t}(\overline{\zeta}^{t}) - \left(1 - \frac{t_k}{\widehat{t}_{K_t}} \right) \widehat{\gamma},  
\end{equation} 
where $\widehat{t}_{K_t}$ represents the maximum time-in-network across devices $K_t$, and $\widehat{\gamma}$ indicates the maximum value of accuracy reduction for new devices into the network. 
The scaling $(1-\frac{t_k}{\widehat{t}_{K_t}})$ allows new devices to the network to have lower accuracy thresholds, as such devices may have valuable new information to share but their local ML model performance may initially perform poorly (as they are not yet fully integrated within the global ML model). 
Together~\eqref{eq:tin_net_avg_logit} and~\eqref{eq:dusd_thresh} lead to a modification of the aggregation rule at the server, yielding: 
\begin{equation} \label{eq:dusd_agg}
    w^{t+1} = \sum_{k \in \mathcal{K}_t} \frac{\mathds{1}_{ \{\xi^{t}_k \geq \Xi^{t}_k\} } w^{t}_k D_k } {\sum_{k \in \mathcal{K}_t} \mathds{1}_{ \{\xi^{t}_k \geq \Xi^{t}_k \} } D_k }, 
\end{equation}
which completes the aggregation cycle and begins the subsequent training round. 
Jointly,~\eqref{eq:tin_net_avg_logit}, ~\eqref{eq:dusd_thresh}, and~\eqref{eq:dusd_agg} represent DUSD-FL. 
The key advantage of DUSD-FL over USD-FL is that new arrivals are less penalized and thus less likely to be filtered for having lower performance.
Moreover, in static networks, the values for both average and maximum time-in-network (i.e., $\overline{t}_{K_t}$ and $\widehat{t}_{K_t}$ respectively) are identically $1$ across all training time and all devices.
As a result, DUSD-FL's computations for average logit distances, accuracy thresholds, and modified aggregation rules all reduce to those of USD-FL. 
}

%% file: def_eval.tex
\begin{figure}[t]
    \centering
    \includegraphics[width=0.48\textwidth]{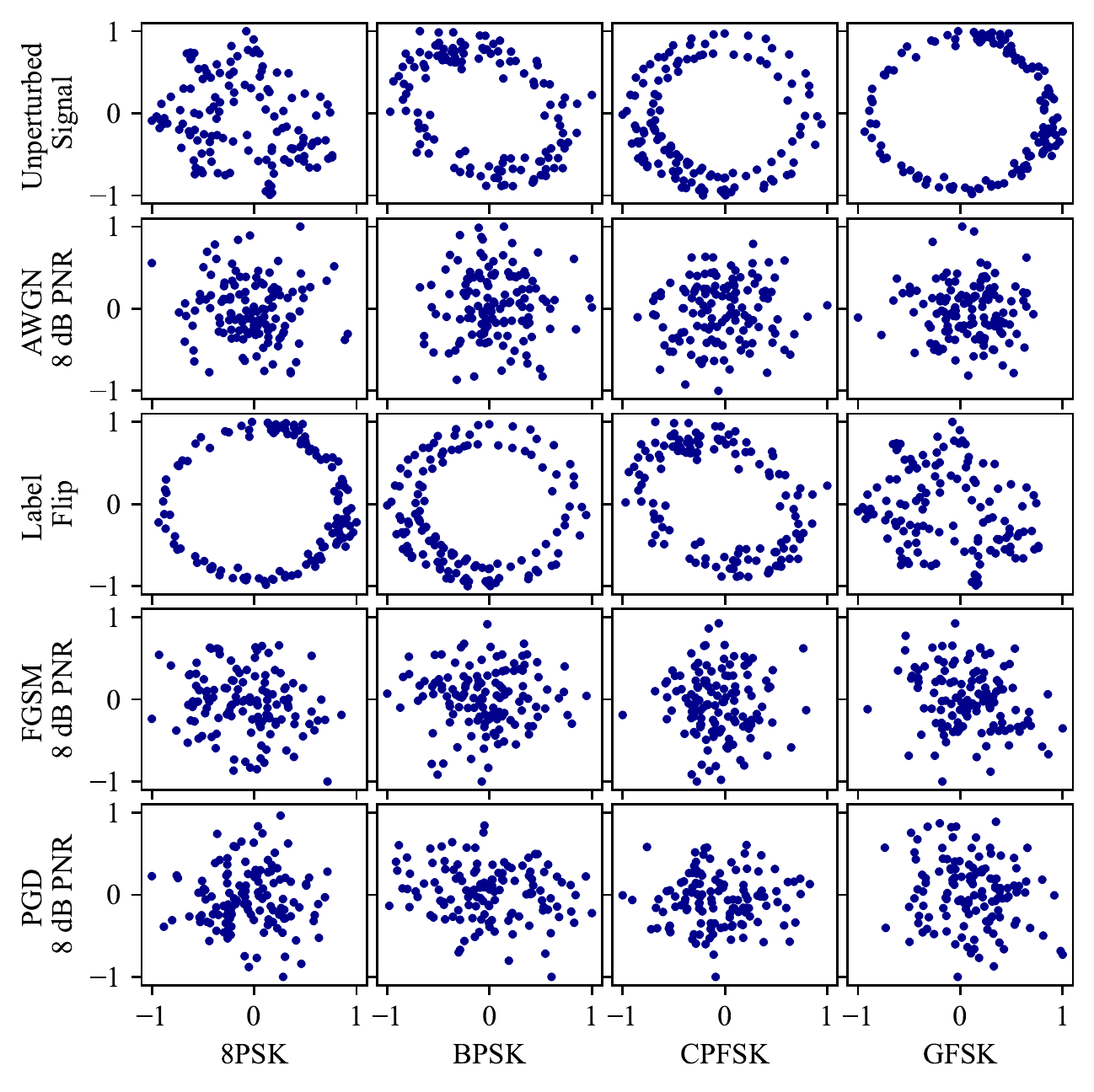}
    \caption{Waveform constellation visualizations for four RML labels: 8PSK, BPSK, CPFSK, and GFSK. The top row depicts received signals with no additive perturbations added at the local device. The AWGN, FGSM, and PGD perturbations are both shown with an average PNR of $8$ dB and visually appear similar. The label flipping attack, shown in the third row, changes the true underlying label, for example the BPSK and CPFSK constellations are flipped.}
    \label{fig:data_const}
    \vspace{-4mm}
\end{figure}

In the following, we first describe the FL and SC architectures in Sec.~\ref{sec:dataset}, present the setup for USD-FL and baseline defenses in Sec.~\ref{ssec:def_setup}, and evaluate USD-FL in Sec.~\ref{ssec:def_eval} by examining accuracy, confusion matrices, and false positive rates relative to baselines. 


\subsection{FL Classification Architecture and Dataset} \label{sec:dataset}
Each device trains a local DL classifier with the VT-CNN2 architecture \cite{dl_amc}. 
Specifically, each local classifier is composed of 2 sequential convolutional layers with 256 and 80 feature maps, consisting of $1 \times 3$ and $2 \times 3$ kernel sizes, respectively, followed by a 256 unit dense layer and a $C$ dimensional output layer. 
Each intermediate layer applies the ReLU activation, and the output layer applies the softmax activation. Thus $\hat{\mathbf{y}}^{n}_{k, j}$ can be interpreted as the probability of the $n^{\text{th}}$ input from the $k^{\text{th}}$ device belonging to the $j^{\text{th}}$ class. 
We use $\eta_k = 0.001 \hspace{1mm} \forall \hspace{1mm} k$, and we set $\alpha_a = 1 \hspace{1mm} \forall \hspace{1mm} a$ to isolate the effect of evasion attacks.

{\color{black}To evaluate our poisoning framework, we employ the RadioML2016.10a dataset (RML)~\cite{o2016radio}, which is an independent AMC dataset commonly used to benchmark the effectiveness of algorithms for radio signal classification. The dataset consists of signals in the following ten modulation constellations stored at 8 and 10 dB SNR: 8PSK, AM-DSB, BPSK, CPFSK, GFSK, PAM4, QAM16, QAM64, QPSK, and WBFM.} 

{\color{black}In total, we apply a $75\%/25\%$ train/test split, resulting in $90$K training samples, split among the participating clients, and $30$K testing samples contained at the global server.
We will also study a variety of adversarial FL contexts, such as evasion attacks of various power levels and architectures as well as time-varying evasion attacks, in the following sections.}

Each RML signal is normalized to unit energy and has observation window of length $\ell = 128$. We depict the RML constellations in the uppermost row of Fig.~\ref{fig:data_const}, and show the signals after perturbing using FGSM as well as after perturbing using AWGN and label flipping in Fig.~\ref{fig:data_const}. 

We measure the potency of the local perturbations in terms of the perturbation to noise ratio (PNR) given by
\begin{equation} \label{eq:pnr}
    \text{PNR [dB]} = \text{PSR [dB]} + \text{SNR [dB]},
\end{equation}
where PSR is the perturbation to signal ratio. 

\begin{figure*}[t]
    \centering
    \includegraphics[width=\textwidth]{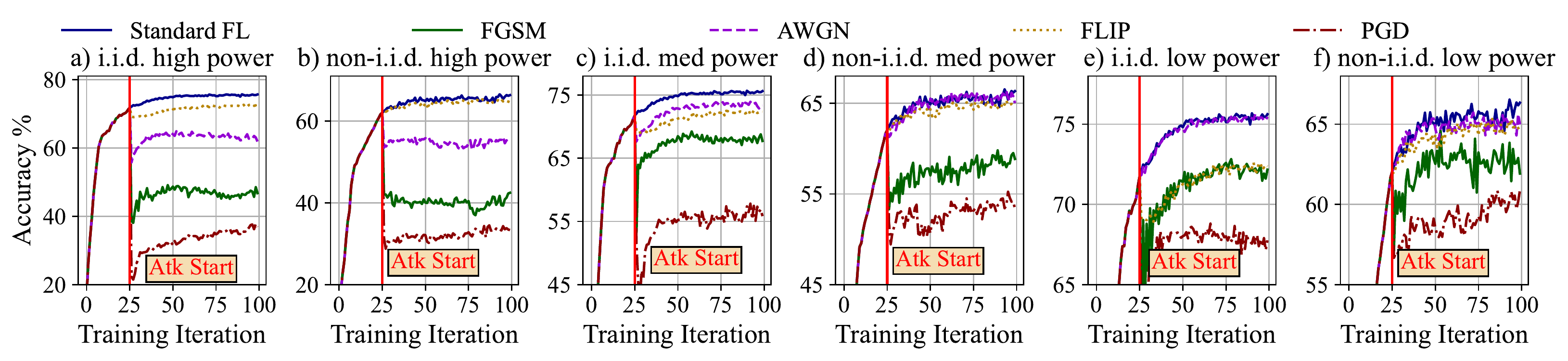}
    \caption{{\color{black}Training performance of higher, medium, and lower power perturbations for a network with $30\%$ adversarial devices. Higher power perturbations at $8$ dB PNR are shown in a) and b), medium power at $4$ dB PNR are in c) and d), and lower power at $0$ dB PNR are shown in e) and f). Lower accuracy indicates higher adversarial impact. All results are averaged over three independent runs, and PGD yields the most potent model poisoning attack across all experiments.} }
    \label{fig:all_ovr}
    \vspace{-4mm}
\end{figure*}

\begin{figure*}[t]
    \centering
    \includegraphics[width=\textwidth]{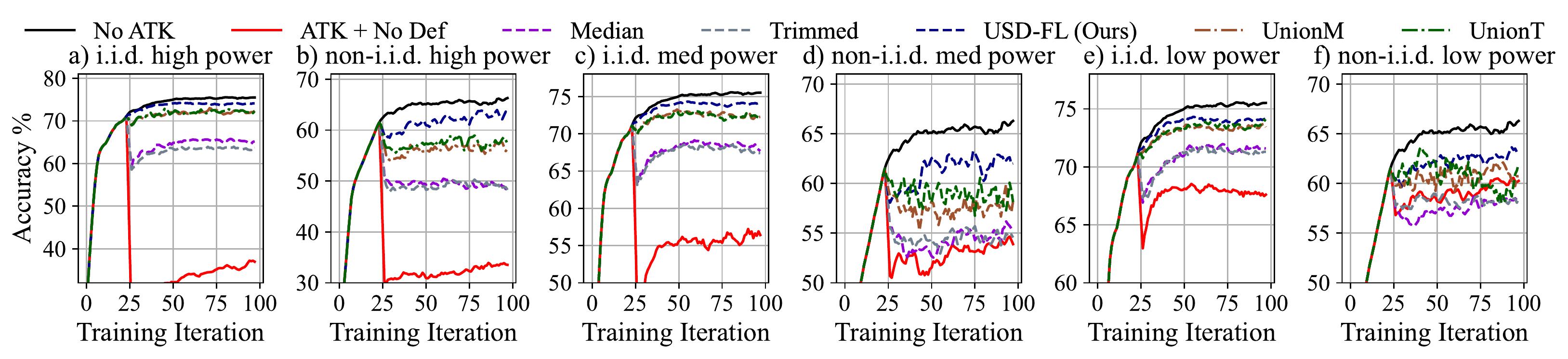} 
    \caption{{\color{black}USD-FL consistently yields the best defense for FL-based SC against PGD-based attacks. 
    For i.i.d. cases, USD-FL mitigates evasion attacks almost entirely, while, for non-i.i.d. cases, USD-FL significantly reduces the impact of evasion attacks.}}
    \label{fig:adv_base_def}
    \vspace{-2mm}
\end{figure*}

\subsection{Defense Configurations for FL-based SC} \label{ssec:def_setup}
The evaluation framework for USD-FL follows the framework presented in Sec.~\ref{sec:dataset}.
{\color{black}For the reserve dataset, the server retains $500$ signals randomly sampled without replacement from $8$ underlying labels, rather than the full $C = 10$ labels in RML.
The remaining data is then partitioned following the $75\%/25\%$ train/test split method presented in Sec.~\ref{sec:dataset}, resulting in roughly 89.6K training samples and 29.9K testing samples. Consequently, the reserve dataset is then $<2\%$ the size of the full testing dataset.}
The server then performs a single round of pre-training on its reserve dataset to obtain starting global ML model paramters $w^{0}$, and follows the steps outlined in Algorithm~\ref{alg:usd-fl}, with coefficients: $A=5$, $B=10^{-7}$ i.i.d. or $B=10^{-5}$ non-i.i.d., $\gamma_{\max} = 0.4$, and $\gamma_{\min} = 0.3$

We evaluate USD-FL relative to four baseline server-driven defenses for FL-based SC: (i) median~\cite{yin2018byzantine,cao2022mpaf}, (ii) trimmed-mean~\cite{yin2018byzantine,cao2022mpaf}, (iii) unionM~\cite{byz_atk}, and (iv) unionT~\cite{byz_atk}. 
{\color{black}These baselines all rely on the server to filter adversaries during the aggregation stage, as in USD-FL.
Both median and trimmed-mean (hereafter, ``trimmed" for conciseness) defenses are conducted element-wise on deivces' ML model parameters. 
Median selects the median element for every ML model parameter, while trimmed first filters away the largest and smallest $z$ parameters then averages the remaining parameters. 
On the other hand, unionM and unionT evaluate devices' ML models $w^{t}_{k}$ on a reserve dataset $\mathcal{D}_S$ (similar to the proposed USD-FL methodology), filter away the $z$ devices with the highest error and loss, and apply a modified aggregation rule (median or trimmed respectively).}

Since the exact quantity of network adversaries is unknown \textit{apriori}, trimmed, unionM, and unionT methodologies must normally estimate a $z$ value, corresponding to the number of adversaries, prior to FL-based SC training. 
In order to have the most potent defense comparison possible, we therefore assume that these baseline defenses have \textit{perfect information}, i.e., the server has exact knowledge of the quantity of adversaries. 
However, USD-FL does \textit{\textbf{not}} know the exact quantity of adversaries in the following evaluation. 

{\color{black}Finally, we note that the initial global ML model is pre-trained for a \textit{single iteration} on the reserve dataset. All experiments, both those for USD-FL as well as those for the server-driven baselines, involve the above identical pre-training process. As such, the following experimental evaluations are based on fair initializations for all server-driven baselines.}

\subsection{Model Poisoning Evaluation} \label{sec:perf_eval}
We now examine the effectiveness of various evasion attacks on FL-based SC.
In our evaluations, we consider a network of $K=10$ devices consisting of classifiers based on the VT-CNN2 architecture described in Sec.~\ref{sec:dataset} in both i.i.d. and non-i.i.d. signal distributions among devices. 
In an i.i.d. environment, all network devices contain the same quantity of local data and have local data sampled uniformly at random from each class of the full training dataset. 
In the non-i.i.d. case, devices have data quantity chosen randomly from $\mathcal{N}(4500,45)$ and data randomly sampled from only five labels as in~\cite{wang2021device}. After training iteration $t_{0} = 25$, $30\%$ of the network is compromised by adversarial attacks similar to~\cite{fl_atk_det}, and begins training on perturbed local datasets. 

We compare the potency of the various evasion attacks outlined in Fig.~\ref{fig:data_const}, namely AWGN, label flipping (FLIP), FGSM, and PGD as these methodologies rely on intentional manipulations of local training data to poison model aggregations and thus the global ML model. 
Specifically, AWGN injects random Gaussian noise into the devices' local training data while FLIP mislabels local training data intentionally. 
In our evaluation, we vary perturbation power to assess attack potency, specifically using $8$ dB PNR for higher power, $4$ dB PNR for medium power, and $0$ dB PNR for lower power in Fig.~\ref{fig:all_ovr}. 
For all cases in Fig.~\ref{fig:all_ovr}, the PGD-based methodology yields the most potent adversarial attack. 
In higher power cases, PGD yields $34\%$, $25\%$, and $8\%$ more accuracy penalty than FLIP, AWGN, and FGSM attacks, respectively, for the i.i.d. case in Fig.~\ref{fig:all_ovr}a), and $28\%$, $20\%$, and $7\%$ more accuracy penalty than  FLIP, AWGN, and FGSM attacks, respectively, for the non-i.i.d. case in Fig.~\ref{fig:all_ovr}b). 
For the medium and lower power scenarios, PGD continues to demonstrate the highest attack potency, though the nominal impact of all adversarial attacks is reduced as compared to higher power attacks. 

The reduction in nominal impact of all evasion attacks in non-i.i.d. cases of Fig.~\ref{fig:all_ovr} is because devices and thus adversaries may not have data from all possible labels. 
As a result, adversaries can only bias the ML model's classification performance on the specific labels that they have corresponding data for. 
Consequently, after model aggregations, the global ML model displays only weaker classification on underlying labels present at the adversaries. 

Owing to the notable reduction in classification accuracies as a result of adversarial evasion attacks (and in particular the FGSM and PGD-based methodologies) in Fig.~\ref{fig:all_ovr}, we next examine the effectiveness of our proposed defense (USD-FL) relative to several baselines from literature. 

\subsection{Defense Performance Evaluation} \label{ssec:def_eval}
In the following experimental results, we investigate three core aspects: (i) the effectiveness of USD-FL versus other defensive baselines, (ii) the confusion matrices of USD-FL and other baselines, and (iii) the false-positive rate of all defense methodologies, all versus the higher power PGD attack. 
{\color{black} Experiments against medium and lower power PGD attacks as well as those versus FGSM-based attacks are left to the supplementary materials due to space limitations.}
All figures and tables are the average of three independent simulations. 


\begin{figure*}[t]
    \centering
    \includegraphics[width=\textwidth]{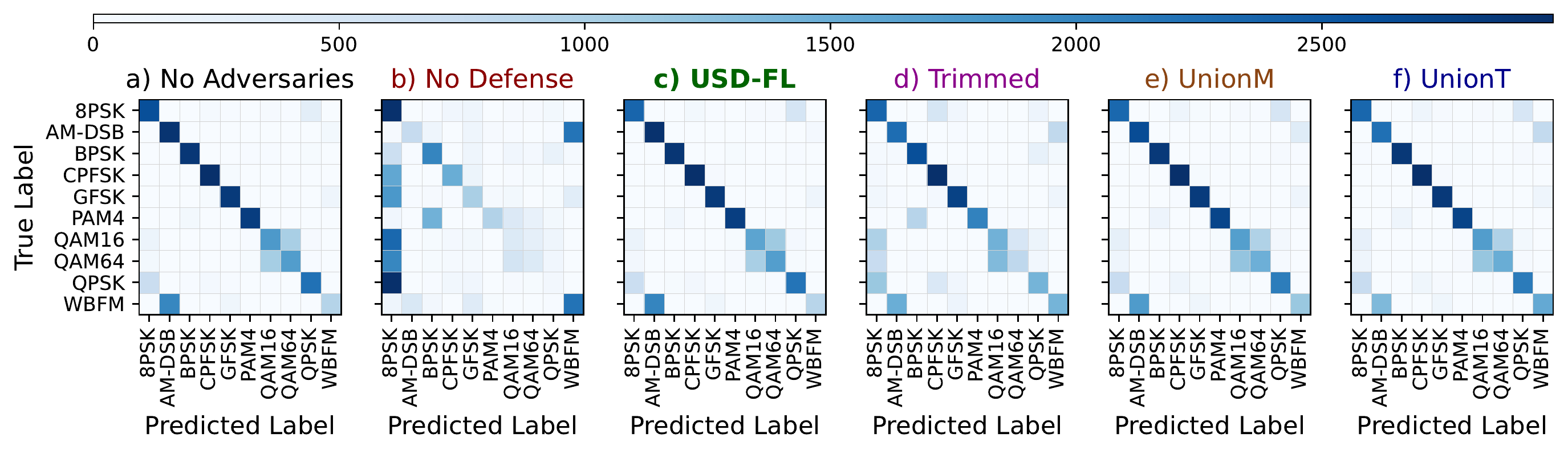}
    \caption{Confusion matrices in an i.i.d. scenario with higher power PGD attacks. USD-FL is nearly identical to the case without adversaries, while baseline defenses, in Fig.~\ref{fig:conf_iid_hp}d), Fig.~\ref{fig:conf_iid_hp}e), and Fig.~\ref{fig:conf_iid_hp}f), misclassify signals more often as 8PSK and WBFM.}
    \label{fig:conf_iid_hp}
    \vspace{-2mm}
\end{figure*}

\begin{figure*}[t]
    \centering
    \includegraphics[width=\textwidth]{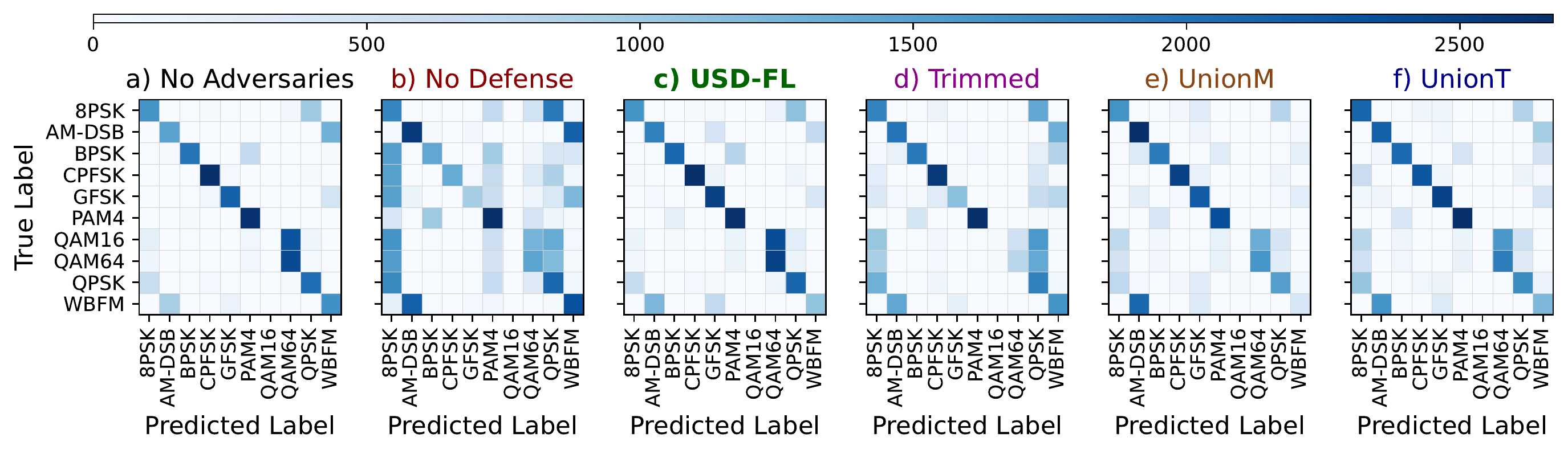}
    \caption{Confusion matrices in a non-i.i.d. scenario versus higher power PGD attacks. 
    This more challenging experimental setting results in all methodologies displaying more frequent misclassification. 
    Nonetheless, USD-FL still demonstrate high similarity to the unperturbed case.} 
    \label{fig:conf_noniid_hp}
    \vspace{-2mm}
\end{figure*}

\subsubsection{Defense methodology comparison} \label{sec:def_comp_exp} 
In Fig.~\ref{fig:adv_base_def}, we compare the global ML model accuracies of various defense methodologies for FL-based SC with the same network setup as that in Sec.~\ref{sec:perf_eval}. 
This experiment examines different defenses for FL-based SC against the PGD-based methodology, which was the most potent evasion attack from Sec.~\ref{sec:perf_eval}.
Due to page limits, another experiment, examining defenses versus the FGSM-based methodology, is left to the supplementary materials. 

In the i.i.d. setting, our proposed USD-FL methodology provides a more robust defense for FL-based SC than the baselines, as measured by ML model accuracy. 
For adversarial perturbations of high, medium, and low power, shown in Fig.~\ref{fig:adv_base_def}a), Fig.~\ref{fig:adv_base_def}c), and Fig.~\ref{fig:adv_base_def}e), respectively, USD-FL yields a final ML model accuracy within 1\% of standard, unperturbed FL-based SC. 
In the scenario involving higher powered attacks in Fig.~\ref{fig:adv_base_def}a), USD-FL significantly mitigates the impact of adversarial perturbations, reducing their potency by approximately 37\%, outperforming the Median and Trimmed baselines by at least 9\% and the unionM and unionT baselines by at least 3\%. 
As the power of the adversarial attack decreases in Fig.~\ref{fig:adv_base_def}c) and Fig.~\ref{fig:adv_base_def}e), the reduction in accuracy drops correspondingly. Therefore, there is less damage to the global ML model to mitigate. 
Nonetheless, USD-FL continues to demonstrate the best performances in the medium and lower power i.i.d. cases.

In the non-i.i.d. scenario, USD-FL also yields the highest accuracies. 
For higher power attacks in Fig.~\ref{fig:adv_base_def}b), USD-FL's performance is within 4\% of unperturbed FL-based SC.
Furthermore, USD-FL reduces the impact of adversaries by over 30\%, simultaneously outperforming the Median and Trimmed defenses by over 13\% and the unionM and unionT baselines by over 6\%.
Similarly, in the medium power case of Fig.~\ref{fig:adv_base_def}d), USD-FL's accuracy is within 3\% of unperturbed FL-based SC, and it reduces the effectiveness of adversarial attacks by roughly 15\%, which is 13\% better than Median and Trimmed as well as 8\% better than unionM and unionT. 
Meanwhile, in the lower power case of Fig.~\ref{fig:adv_base_def}f), the modified aggregation rule-based defenses (i.e., median and trimmed) perform slightly worse than undefended FL. 
This is due to two factors combined: (i) 0 dB PNR perturbations have very small impact to begin with, and (ii) these modified aggregation rule-based defenses may filter non-i.i.d. ML models rather than low power adversaries. 
Thus, it suggests that accidentally filtering a non-adversarial device may have more negative impact than filtering a true adversary compromised by lower power attacks. 


Adversarial attacks can be seen to have a smaller nominal impact in non-i.i.d. settings in Fig.~\ref{fig:adv_base_def} because each adversary only has a subset of all labels. 
Consequently, defense methodologies have less perturbation to mitigate, and thus provide smaller nominal improvements to classification accuracies in the non-i.i.d. scenarios as compared to the i.i.d. cases.

{\color{black}Additionally, we want to emphasize that, in Fig.~\ref{fig:adv_base_def}, USD-FL demonstrates superior performance to the existing server-driven baselines even when the reserve dataset is non-i.i.d. with respect to the datasets across network devices. 
This is because adversarial evasion attacks induce logit distribution shifts for all labels (i.e., types of signals)~\cite{carlini2017towards} when devices' ML model parameters are evaluated on the reserve dataset. 
After extracting devices' logits, our proposed USD-FL methodology leverages the 1-Wasserstein distance (see Sec.~\ref{ssec:logit_dist_comp}) to capture and compare these logit distribution shifts, enabling the server to effectively identify potential adversaries even with non-i.i.d. reserve datasets.}



\subsubsection{Confusion Matrices} \label{sec:conf_mat}
The confusion matrices in Fig.~\ref{fig:conf_iid_hp} and Fig.~\ref{fig:conf_noniid_hp}, for i.i.d. and non-i.i.d. experiments respectively, offer additional insight into the performance of various baselines when subjected to higher power PGD-based attacks. Medium power ($4$ dB PNR) and lower power ($0$ dB PNR) PGD-based attacks are presented in the supplementary materials. 
Moreover, we also examine the confusion matrices as a result of FGSM-based attacks of varying power levels in the supplementary materials.
For these experiments, we examine (a) unperturbed FL-based SC, (b) undefended FL-based SC in the presence of adversaries, (c) USD-FL, (d) Trimmed-Mean, (e) UnionM, and (f) UnionT, obtaining confusion matrices via evaluation of the ML model, post-training, on the validation dataset. 
We omit the Median defense here as its confusion matrices are nearly identical to those of the Trimmed-Mean results. 

For the i.i.d. scenario in Fig.~\ref{fig:conf_iid_hp}, Fig.~\ref{fig:conf_iid_hp}b) shows the confusion matrix of undefended FL-based SC, revealing that adversarial attacks cause the ML model to falsely classify many signals as 8PSK. 
While the other defense methodologies, presented in Fig.~\ref{fig:conf_iid_hp}d)-\ref{fig:conf_iid_hp}f), struggle to mitigate this damage, USD-FL, by contrast, yields a confusion matrix in Fig.~\ref{fig:conf_iid_hp}c) nearly identical to the unperturbed FL-based SC scenario depicted in Fig.~\ref{fig:conf_iid_hp}a).
This confirms that USD-FL successfully overcame the adversarial perturbations, hence it's robust performance in Fig.~\ref{fig:adv_base_def}a). 
In all cases, we notice that the QAM16 and QAM64 signals often get confused for each other, regardless of defense. 
This is due to the fact that QAM16 and QAM64 belong to the same family of modulated signals (i.e., Quadrature Amplitude Modulation), and thus have highly similar waveforms relative to other modulation schemes, such as AM-DSB.

For the non-i.i.d. experiment in Fig.~\ref{fig:conf_noniid_hp}, confusion matrices are more varied than those for the i.i.d. experiment, confirming that non-i.i.d. scenarios result in lower global ML model accuracy (which can also be seen via the classification accuracies in Fig.~\ref{fig:adv_base_def}).  
Many more signals are incorrectly classified by the ``No Defense" scenario in Fig.~\ref{fig:conf_noniid_hp}b) than in the i.i.d. case of Fig.~\ref{fig:conf_iid_hp}b). 
Even in this more challenging scenario, USD-FL, shown in Fig.~\ref{fig:conf_noniid_hp}c), continues to mitigate the bulk of the damage of adversarial attacks, yielding a confusion matrix that is again nearly identical to the case with no adversaries, i.e., Fig.~\ref{fig:conf_noniid_hp}a). 

{\color{black} To summarize the key insights of Fig.~\ref{fig:conf_iid_hp} and~\ref{fig:conf_noniid_hp} numerically, USD-FL correctly classifies at least $300$ and $1400$ more signals than the server-driven baselines for i.i.d. and non-i.i.d. settings, respectively.}

\subsubsection{False positive adversary detection} \label{sec:false_positive} 
Next, we investigate the false positive detection rates for the different defenses, which explain how USD-FL outperformed the baselines in Fig.~\ref{fig:adv_base_def}. Table~\ref{tab:fp_pgd_hp} shows the false positive rates for various defenses versus higher power PGD attacks. 
Additional false positive tables for medium and lower power perturbations (as well as those for FGSM attacks) are left to the supplementary materials. 
The measurements in both tables are the result of averaging the false positive rates, computed every $5$ global aggregations.



{\color{black}In the i.i.d. setting, USD-FL exhibits no false positives (i.e., an average false positive rate of $0\%$) and thus achieves the high accuracies shown in Fig.~\ref{fig:adv_base_def}a). By contrast, the baseline defenses exhibit non-zero false positive rates, allowing undetected adversaries to degrade classification performance, as seen in Fig.~\ref{fig:adv_base_def}a).}


\begin{figure*}[t!]
\centering
    \centering
    \includegraphics[width=0.8\textwidth]{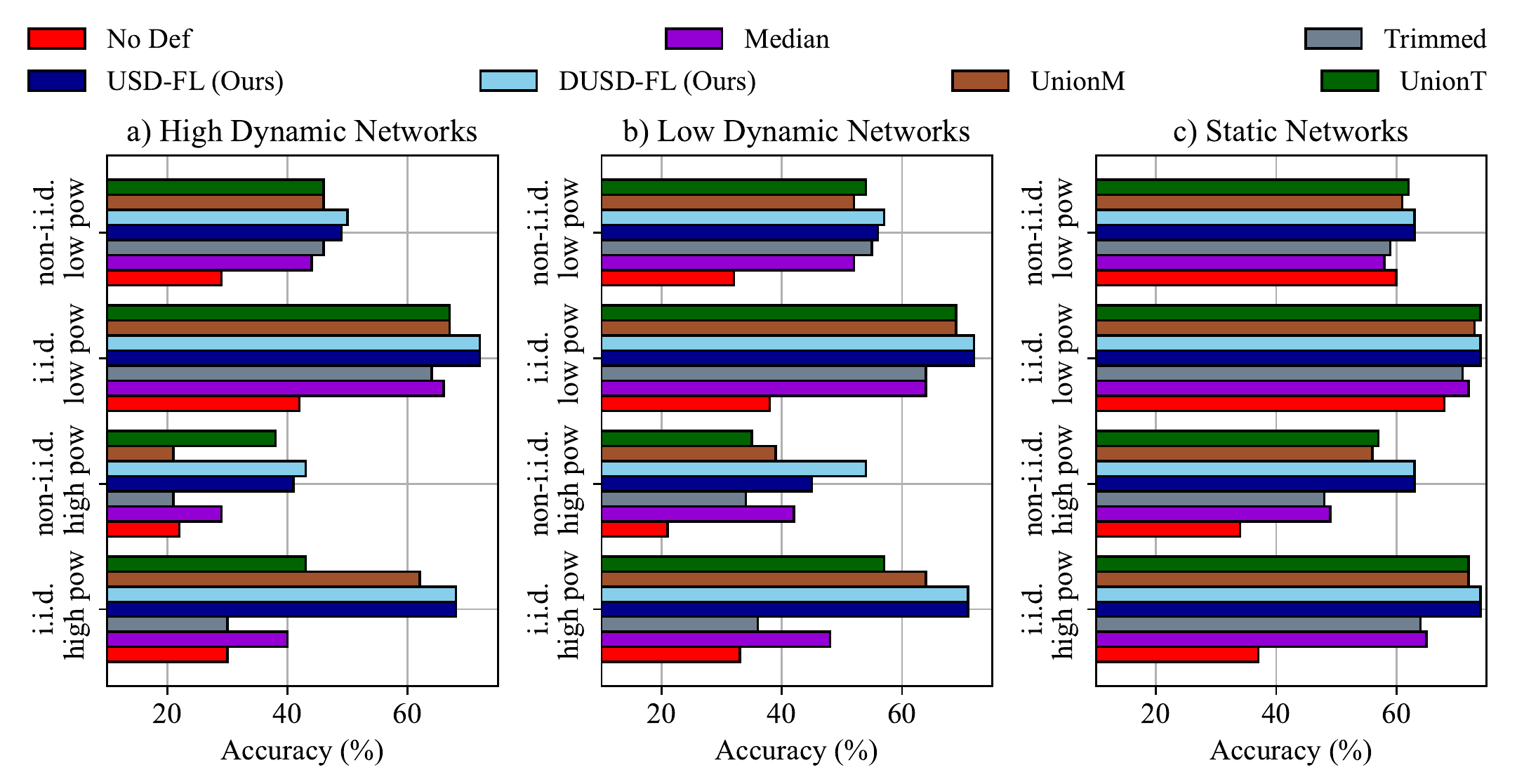}
    \caption{{\color{black}Examining USD-FL and DUSD-FL in dynamic network settings. For high dynamic networks, we assume that, every $10$ training iterations, $30\%$ of the devices exits the network while, simultaneously, an equal number of devices enters the network. The same process occurs for low dynamic networks, except that the percentage decreases to $10\%$ of the network. Finally, we show that DUSD-FL reduces to USD-FL in static network settings in Fig.~\ref{fig:dyn_nets}c).}}
  \label{fig:dyn_nets}
  \vspace{-2mm}
\end{figure*}

Meanwhile, for the non-i.i.d. setting, USD-FL again demonstrates the lowest false positive rates, which explains its superior performance relative to the baseline defenses in Fig.~\ref{fig:adv_base_def}b). 
That being said, USD-FL does yield a non-zero false positive rate in the non-i.i.d. setting.
In non-i.i.d. scenarios, devices will exhibit a larger degree of performance variability~\cite{chen2020convergence}, which can lead non-adversarial devices to be overly biased to unique wireless signal data distributions.
Therefore, adversaries are harder to detect in non-i.i.d. settings, and USD-FL does yield the occasional misstep, albeit less often than the baselines. 


\begin{table}[!t]
\caption{{\color{black}Average Defense False Positive Rates Versus High Power PGD Attacks}}
\vspace{-1.5mm}
\label{tab:fp_pgd_hp} 
{\color{black}
\begin{tabularx}{0.48\textwidth}{c c c c c c}
\toprule[.2em] 
& & \multicolumn{4}{c}{\textbf{Experiment Condition}} \\
\cmidrule(lr){3-6}
& & \multicolumn{2}{c}{\textbf{i.i.d.}} & \multicolumn{2}{c}{\textbf{non-i.i.d.}}\\
\cmidrule(lr){3-4} \cmidrule{5-6}
\multicolumn{2}{c}{\textbf{Defense Method}} & \textbf{Nominal} & \textbf{Rate (\%)} &\textbf{Nominal} & \textbf{Rate (\%)} \\
\midrule
\multicolumn{2}{c}{\textbf{UnionM}} & 0.86 & 12.3 & 0.74 & 10.6 \\
\multicolumn{2}{c}{\textbf{UnionT}} & 0.88 & 12.6 & 0.79 & 11.3 \\
\multicolumn{2}{c}{\textbf{Median}} & 2.41 & 34.4 & 2.37 & 33.9\\
\multicolumn{2}{c}{\textbf{Trimmed}} & 2.69 & 38.4 & 2.65 & 37.9 \\
\multicolumn{2}{c}{\textbf{USD-FL}} & 0 & 0 & 0.02  & 0.3 \\
\bottomrule
\end{tabularx} }
\vspace{-4mm}
\end{table}

\subsection{USD-FL and DUSD-FL in dynamic networks}
{\color{black}To understand the defensive capability of both DUSD-FL and USD-FL in dynamic networks, we perform additional experiments, measuring defensive efficiency under high and low dynamic network conditions in Fig.~\ref{fig:dyn_nets}. To model dynamic networks, we assume that, every 10 training iterations, a percentage of devices exit the network and an equivalent quantity of devices enters the network. For high and low dynamic network settings in Fig.~\ref{fig:dyn_nets}, we use 30\% and 10\% respectively. Firstly, in Fig.~\ref{fig:dyn_nets}, we see that both USD-FL and DUSD-FL consistently outperform the baseline server-driven defenses for both high and low dynamic networks. While the performances for USD-FL and DUSD-FL are nearly identical for i.i.d. settings, DUSD-FL obtains a consistent advantage over USD-FL in non-i.i.d.settings, specifically at least 3\% for both high and low dynamic networks respectively.}

{\color{black}In non-i.i.d. settings, devices need a grace period for their local ML model training to integrate their non-i.i.d. datasets with the partially trained global ML model, until then they may yield lower accuracies on the reserve dataset. 
As a result, the information at new devices, although non-adversarial, may be filtered by USD-FL, whereas, owing to the reduced accuracy threshold for new device entries (i.e., $\widehat{\gamma}$ in~\eqref{eq:dusd_thresh}), DUSD-FL is able to include such devices within global ML model aggregations, thus enabling DUSD-FL to achieve better performance.}


%% file: conclusion.tex
The growing adoption of FL based methodologies to improve wireless signal classification has many potential benefits. However, there are specific challenges within wireless environments that can impede the performance and training of such methodologies. 
In the first part, we examined the potency of various evasion attacks in compromising FL-based SC, showing that specific attacks can achieve over $40\%$ reduction in classification accuracy. 

In the second part, we proposed USD-FL, a server-driven defense for FL-based SC. The USD-FL algorithm relies on a server-side reserve dataset, which is smaller and non-i.i.d. distributed relative to the devices' local datasets, to compare and contrast the logits of devices' ML models.
Subsequently, USD-FL checks the classification accuracy of devices' SC models, then partitions devices into adversaries and non-adversaries based on a threshold function controlled by the average logit distance (as computed by the 1-Wasserstein distance). 
Finally, USD-FL performs an aggregation, similar to FedAvg~\cite{fl}, based entirely on the non-adversaries. 

{\color{black}
The key steps of USD-FL summarized above are all applied at the server during the ML model aggregation, and, thus, implementation of USD-FL follows in a plug-and-play fashion from that of FedAvg. As such, USD-FL requires little additional overhead costs to FedAvg yet offers significant benefits to resilience against adversarial evasion attacks.
In future work, we plan on further investigating fully decentralized FL-based SC, and methodologies to mitigate adversarial attacks therein.} 

{\color{black}
Additional topics for future study include defending FL in scenarios with extreme device-level heterogeneity, such as heterogeneous data structures commonly seen in federated domain adaptation problems~\cite{peng2020federated}, heterogeneous channel distributions across network devices~\cite{uzlaner2025asynchronous}, or distinct learning tasks (with some degree of commonality) explored in federated multi-task learning scenarios~\cite{smith2017federated}.}

%% file: more_exps/conf_mats.tex
\section{Additional Experiments}
\label{sec:app_more_def_exps}
Within this appendix, we present and discuss additional experiments for medium and lower power PGD attacks as well as FGSM attacks of varying power levels. 
The performance of various defenses versus the FGSM attacks is in Sec.~\ref{sec:app_alld_vs_fgsm}. 
Then, experiments examining the defenses versus medium and lower power PGD attacks in discussed in Sec.~\ref{sec:app_pgd_other_pow}.
Finally, we examine the confusion matrices and false positive rates for the FGSM attacks for varying power levels in Sec.~\ref{sec:app_fgsm_other_pow}.


\subsection{Baseline defenses versus the FGSM-based methodology} \label{sec:app_alld_vs_fgsm}
\begin{figure*}[h]
    \centering
    \includegraphics[width=\textwidth]{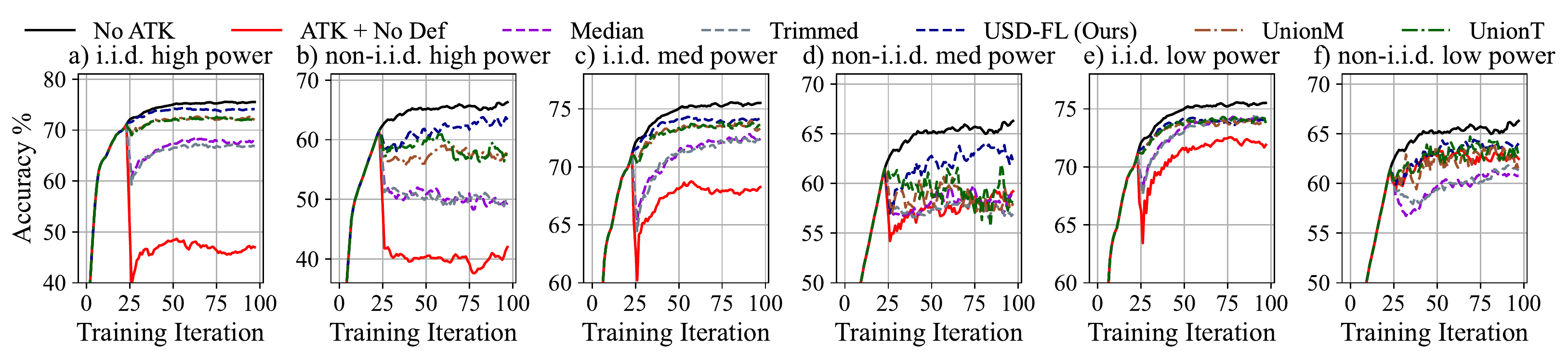} 
    \caption{{\color{black}Examining the efficacy of various defenses versus FGSM attacks of varying power. 
    For all cases, USD-FL demonstrates the best defense for FL-based SC. Higher power attacks represent $8$ dB PNR, medium attacks use $4$ dB PNR, and lower power attacks have on $0$ dB PNR.}}
    \label{fig:adv_base_def_fgsm}
    \vspace{-2mm}
\end{figure*}

Similar to the result in Fig.~\ref{fig:adv_base_def}, we first examine the defensive efficacy of various baseline methodologies relative to the proposed USD-FL defense versus FGSM attacks of varying power levels in Fig.~\ref{fig:adv_base_def_fgsm}.
In this experiment, we also rely on higher power attacks of $8$ dB PNR, medium power attacks of $4$ dB PNR, and lower power attacks of $0$ dB PNR. 
From the undefended cases (the red lines in Fig.~\ref{fig:adv_base_def_fgsm}), we can see that the FGSM attacks have much lower attack potency compared to the PGD attacks of the same power - see Fig.~\ref{fig:all_ovr} for more precision. 
Correspondingly, all defenses appear to have nominally less impact, as there is less adversarial damage to mitigate. 
Nonetheless, the key takeaways are qualitatively the same in Fig.~\ref{fig:adv_base_def_fgsm} as those in Fig.~\ref{fig:adv_base_def}, with USD-FL demonstrating the best defensive performance in all of the cases. 

\subsection{Medium and lower power PGD attacks} \label{sec:app_pgd_other_pow}
We next examine the confusion matrices and false positive rates of various defenses for FL-based SC versus medium and lower power PGD attacks. 
First, we show the confusion matrices for medium power PGD attacks in Fig.~\ref{fig:conf_mp_iid} and Fig.~\ref{fig:conf_mp_noniid} and then discuss the confusion matrices for lower power PGD attacks in Fig.~\ref{fig:conf_lp_iid} and Fig.~\ref{fig:conf_lp_noniid}.  
Finally, we present the false positive rates of various defenses versus medium and lower power PGD attacks in Table.~\ref{tab:fp_pgd_mp} and Table.~\ref{tab:fp_pgd_lp}, respectively. 
\\

\subsubsection{Confusion matrices}
\begin{figure*}[h!]
    \centering
    \includegraphics[width=\textwidth]{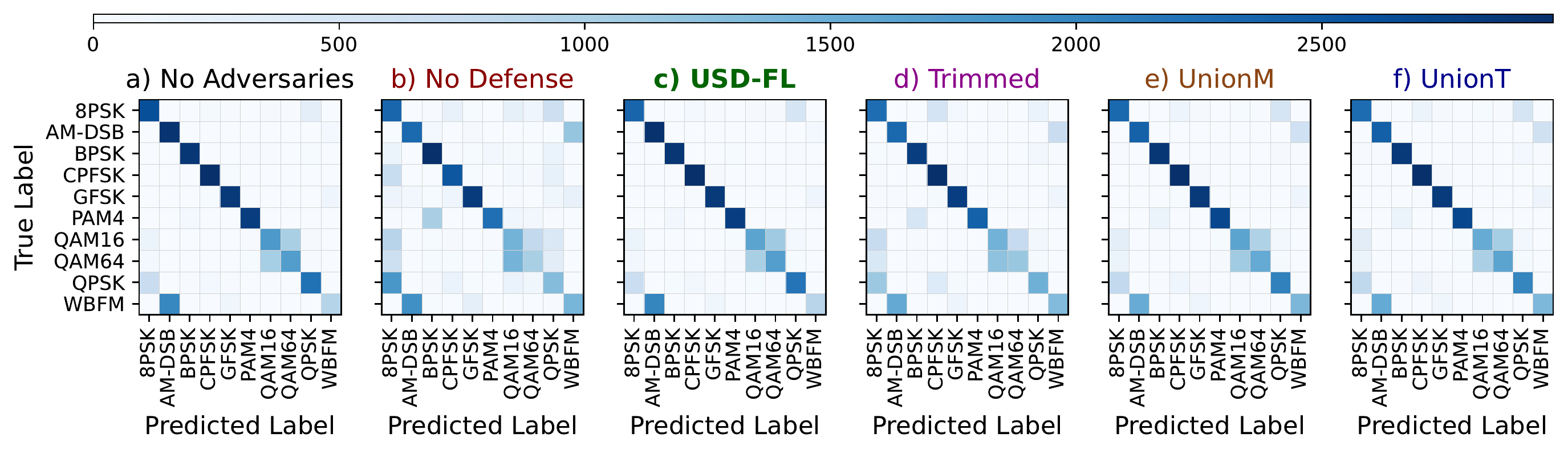}
    \caption{{\color{black}Confusion matrices in an i.i.d. scenario versus medium power PGD attacks. While all defenses demonstrate a significant improvement over the undefended FL-based SC in Fig.~\ref{fig:conf_mp_iid}b), USD-FL yields the greatest similar to the unperturbed FL-based in Fig.~\ref{fig:conf_mp_noniid}a).}}
    \label{fig:conf_mp_iid}
    \vspace{-3mm}
\end{figure*}

\begin{figure*}[h!]
    \centering
    \includegraphics[width=\textwidth]{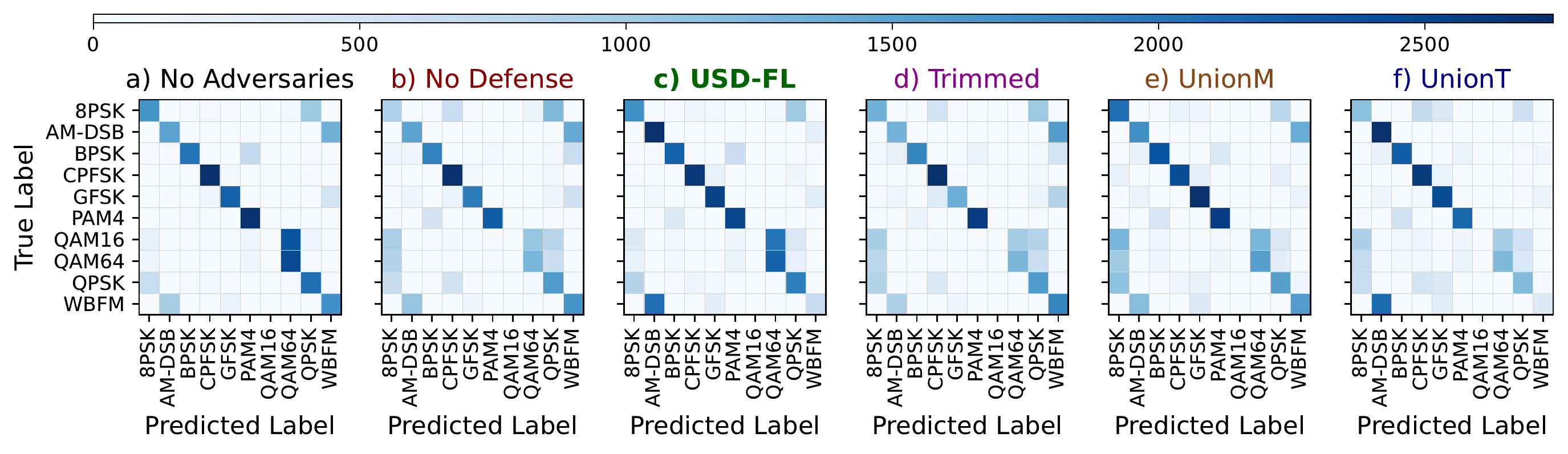}
    \caption{{\color{black}Confusion matrices in a non-i.i.d. scenario versus medium power PGD attacks. USD-FL continues to demonstrate high similarly to the case without adversaries. }}
    \label{fig:conf_mp_noniid}
    \vspace{-3mm}
\end{figure*}

For the medium power PGD attacks, we can clearly see that the undefended cases, in Fig.~\ref{fig:conf_mp_iid}b) and Fig.~\ref{fig:conf_mp_noniid}b), show fewer misclassifications than in the case of higher power PGD attacks, which were discussed in the main manuscript as Fig.~\ref{fig:conf_iid_hp}b) and Fig.~\ref{fig:conf_noniid_hp}b) 
Specifically, in the i.i.d. case, the medium power attacks in Fig.~\ref{fig:conf_mp_iid}b) result in fewer predictions as 8PSK and WBFM and better classification results for the QAM signals relative to the higher power attacks in Fig.~\ref{fig:conf_iid_hp}b). 
A similar takeaway is apparent for the non-i.i.d. experiments. 
\\

Even with the less potent attack, the baseline defenses continue to demonstrate significantly more misclassifications than the proposed USD-FL methodology. 
For instance, the trimmed, unionM, and unionT defenses continue to be more biased to towards the WBFM and 8PSK signals than USD-FL. 
As an example, in the i.i.d. case, we can see that USD-FL has much fewer shading (i.e., fewer classifications) of signals as 8PSK in Fig.~\ref{fig:conf_mp_iid}c) relative to Fig.~\ref{fig:conf_mp_iid}d)-f).
Meanwhile, in the non-i.i.d. case, the baseline defenses, in Fig.~\ref{fig:conf_mp_noniid}d)-f), demonstrate significantly more scattered misclassifications (i.e., many non-diagonal boxes are shaded) than USD-FL in Fig.~\ref{fig:conf_mp_noniid}c). 
\\

\begin{figure*}[h!]
    \centering
    \includegraphics[width=\textwidth]{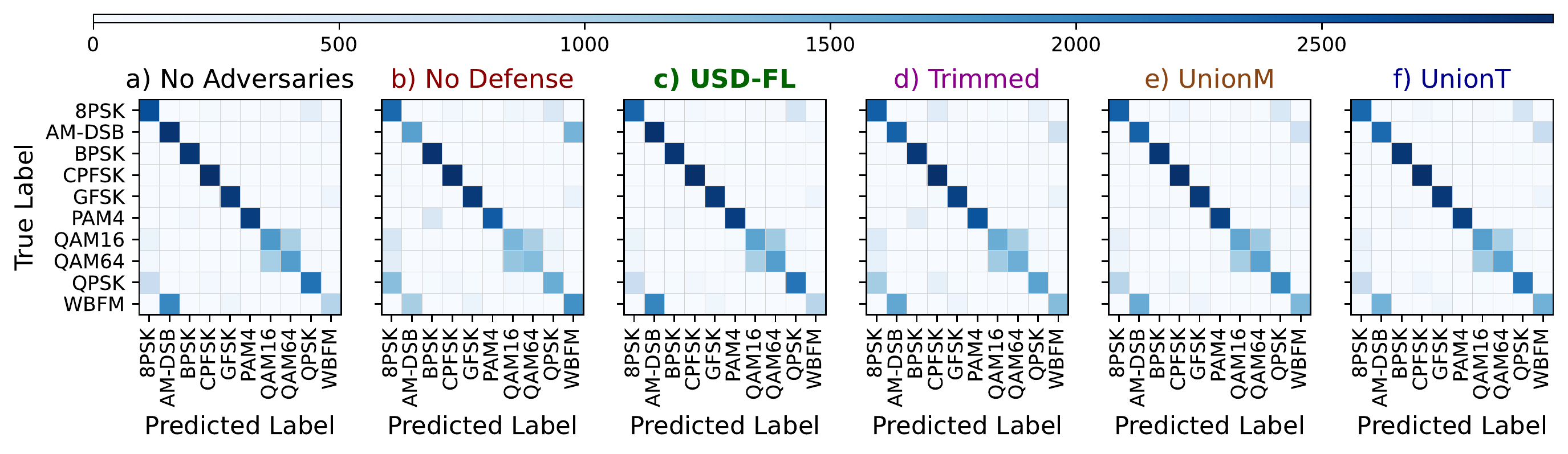}
    \caption{{\color{black}Confusion matrices in an i.i.d. scenario versus lower power PGD attacks. The perturbation appears to be very small, as the undefended FL-based SC in Fig.~\ref{fig:conf_lp_iid}b) appears very similar to the unperturbed FL-based SC in Fig.~\ref{fig:conf_lp_iid}a). }}
    \label{fig:conf_lp_iid}
    \vspace{-3mm}
\end{figure*}

\begin{figure*}[h!]
    \centering
    \includegraphics[width=\textwidth]{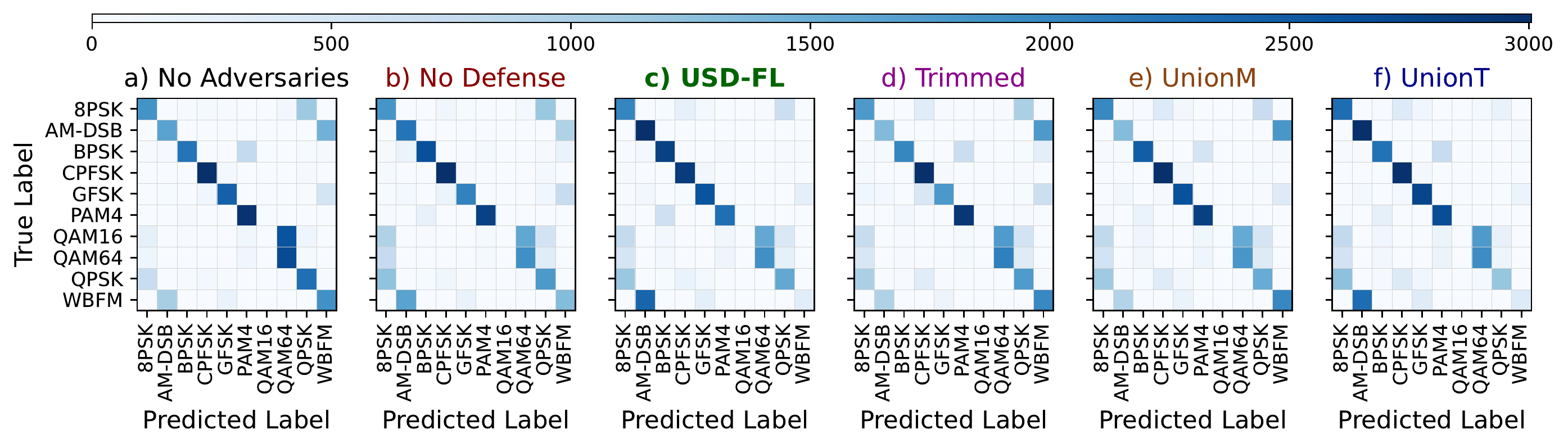}
    \caption{{\color{black}Confusion matrices in a non-i.i.d. scenario versus lower power PGD attacks. The perturbation impact is quite small, and most defenses appear to provide similar performances.}}
    \label{fig:conf_lp_noniid}
    \vspace{-3mm}
\end{figure*}

For the experiments involving lower power PGD attacks in Fig.~\ref{fig:conf_lp_iid} and Fig.~\ref{fig:conf_lp_noniid}, we can see that the confusion matrix of \textit{undefended} FL-based SC, in Fig.~\ref{fig:conf_lp_iid}b) and Fig.~\ref{fig:conf_lp_noniid}b), becomes more similar to \textit{unperturbed} FL-based SC, in Fig.~\ref{fig:conf_lp_iid}a) and Fig.~\ref{fig:conf_lp_noniid}a). 
That being said, the key takeaways remain similar to those confusion matrices from the higher and medium power PGD attacks, with USD-FL again demonstrating the greatest similarity to unperturbed FL-based SC. 
\\

\clearpage
\newpage 
\clearpage 

\subsubsection{False positive rates}


The false positive rates for medium power PGD attacks and lower power PGD attacks are presented in Table.~\ref{tab:fp_pgd_mp} and Table.~\ref{tab:fp_pgd_lp} respectively. 
Comparison with the higher power PGD attacks in Table.~\ref{tab:fp_pgd_hp} yields three main takeaways.
{\color{black} Firstly, USD-FL yields the smallest false positive rates across all experiments. 
This was shown to be true for the higher power PGD-based methodology in Table.~\ref{tab:fp_pgd_hp}, and continues to hold in Table.~\ref{tab:fp_pgd_mp} and Table.~\ref{tab:fp_pgd_lp}. 
Secondly, false positive rates generally increase as adversarial attacks decrease in power. 
For instance, USD-FL demonstrates an increased false positive rate for non-i.i.d. settings, increasing from $0.3\%$ versus higher power PGD attacks in Table.~\ref{tab:fp_pgd_hp} to $12.6\%$ versus lower power PGD attacks in Table.~\ref{tab:fp_pgd_lp}. 
By contrast, the next best/smallest false positive rate defense (i.e., UnionM) for non-i.i.d. in Table.~\ref{tab:fp_pgd_lp} yields a false positive rate of $26.1\%$, which is $13.5\%$ larger than that offered by USD-FL.} 
Finally, we noticed that the pure aggregation rule defenses, i.e., no pre-filtering prior to an aggregation rule, demonstrate relatively stable false positive rates. For example, median and trimmed continue display roughly $34\%$ and $37\%$ false positive rates, respectively, across higher, medium, and lower PGD attacks. 
This final point highlights that adversarial evasion attacks are quite capable at evading detection by traditional defenses. 
\\

\begin{table}[h!]
\caption{{\color{black}Average Defense False Positive Rates Versus Medium Power PGD Attacks}}
\label{tab:fp_pgd_mp} 
\begin{tabularx}{0.96\textwidth}{>{\centering\arraybackslash}m{10em} *{5}{>{\centering\arraybackslash}X}}
\toprule[.2em] 
& & \multicolumn{4}{c}{\textbf{Experiment Condition}} \\
\cmidrule(lr){3-6}
& & \multicolumn{2}{c}{\textbf{i.i.d.}} & \multicolumn{2}{c}{\textbf{non-i.i.d.}}\\
\cmidrule(lr){3-4} \cmidrule{5-6}
\multicolumn{2}{c}{\textbf{Defense Method}} &  \textbf{Nominal} & \textbf{Rate (\%)} & \textbf{Nominal} &  \textbf{Rate (\%)} \\
\midrule
\multicolumn{2}{c}{\textbf{UnionM}} & 0.76 & 10.9 & 1.4 & 20 \\
\multicolumn{2}{c}{\textbf{UnionT}} & 0.83 & 11.9 & 1.52 & 21.7\\
\multicolumn{2}{c}{\textbf{Median}} & 2.43 & 34.7 & 2.39 & 34.1\\
\multicolumn{2}{c}{\textbf{Trimmed}} & 2.69 & 38.4 & 2.61 & 37.3 \\
\multicolumn{2}{c}{\textbf{USD-FL}} & 0 & 0 & 0.55 & 7.9\\
\bottomrule
\end{tabularx}
\end{table}


\begin{table}[h!]
\caption{{\color{black}Average Defense False Positive Rates Versus Low Power PGD Attacks}}
\label{tab:fp_pgd_lp} 
\begin{tabularx}{0.96\textwidth}{>{\centering\arraybackslash}m{10em} *{5}{>{\centering\arraybackslash}X}}
\toprule[.2em] 
& & \multicolumn{4}{c}{\textbf{Experiment Condition}} \\
\cmidrule(lr){3-6}
& & \multicolumn{2}{c}{\textbf{i.i.d.}} & \multicolumn{2}{c}{\textbf{non-i.i.d.}}\\
\cmidrule(lr){3-4} \cmidrule{5-6}
\multicolumn{2}{c}{\textbf{Defense Method}} &  \textbf{Nominal} & \textbf{Rate (\%)} & \textbf{Nominal} &  \textbf{Rate (\%)} \\
\midrule
\multicolumn{2}{c}{\textbf{UnionM}} & 0.81 & 11.6 & 1.83 & 26.1 \\
\multicolumn{2}{c}{\textbf{UnionT}} & 0.88 & 12.6 & 1.9 & 27.1\\
\multicolumn{2}{c}{\textbf{Median}} & 2.4 & 34.3 & 2.39 & 34.1 \\
\multicolumn{2}{c}{\textbf{Trimmed}} & 2.61 & 37.1 & 2.57 & 36.7 \\
\multicolumn{2}{c}{\textbf{USD-FL}} & 0 & 0 & 0.88 & 12.6 \\
\bottomrule
\end{tabularx}
\end{table}

\newpage
\clearpage
\newpage 

\subsection{Qualitative results for defenses versus FGSM attacks of varying power}
\label{sec:app_fgsm_other_pow}

\subsubsection{Confusion matrices for FGSM-based adversarial attacks}
\begin{figure*}[h!]
    \centering
    \includegraphics[width=\textwidth]{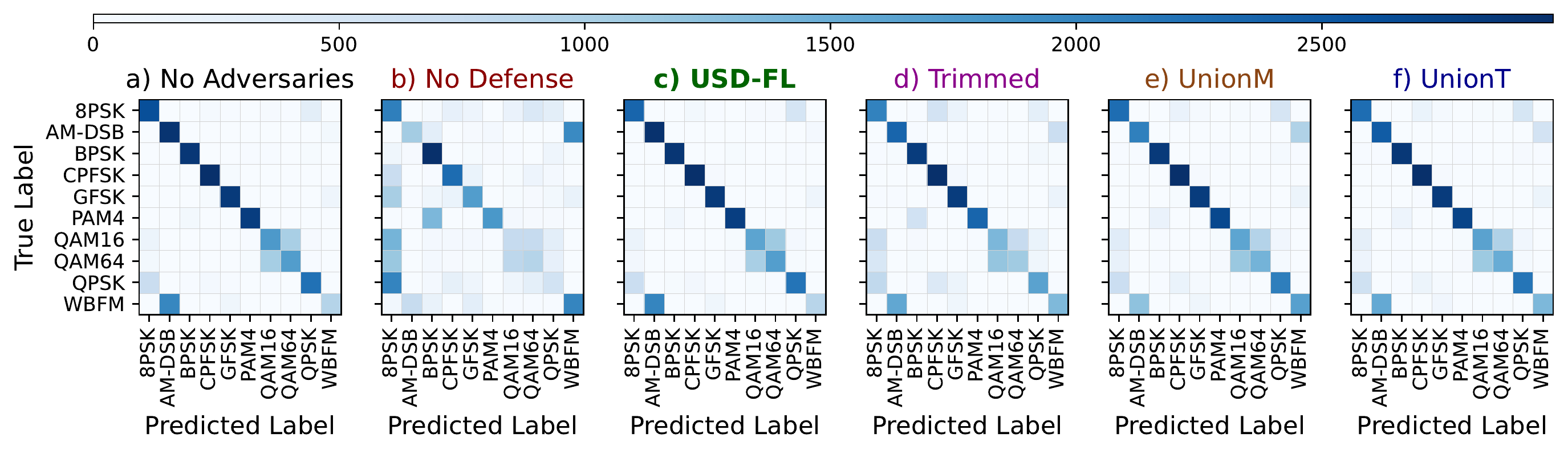}
    \caption{{\color{black}Confusion matrices in an i.i.d. scenario versus higher power FGSM attacks. USD-FL's confusion matrix is nearly identical to the case without adversaries. Most confusion seems to be from misclassifying signal modulations as 8PSK.}}
    \label{fig:fgsm_conf_hp_iid}
    \vspace{-3mm}
\end{figure*}

\begin{figure*}[h!]
    \centering
    \includegraphics[width=\textwidth]{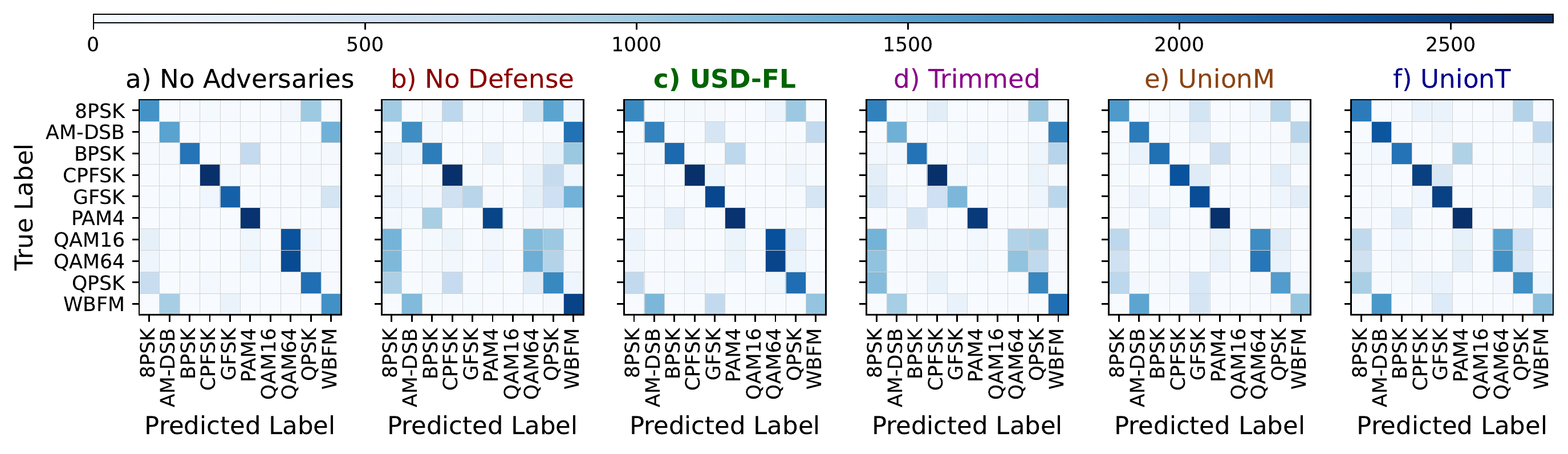}
    \caption{{\color{black}Confusion matrices in a non-i.i.d. scenario versus higher power FGSM attacks. Baseline defenses demonstrate more diverse types of confusion/misclassification than the i.i.d. case depicted in Fig.~\ref{fig:fgsm_conf_hp_iid}. Nonetheless, USD-FL continues to demonstrate high similarly to the case without adversaries. }}
    \label{fig:fgsm_conf_hp_noniid}
    \vspace{-3mm}
\end{figure*}

\begin{figure*}[h!]
    \centering
    \includegraphics[width=\textwidth]{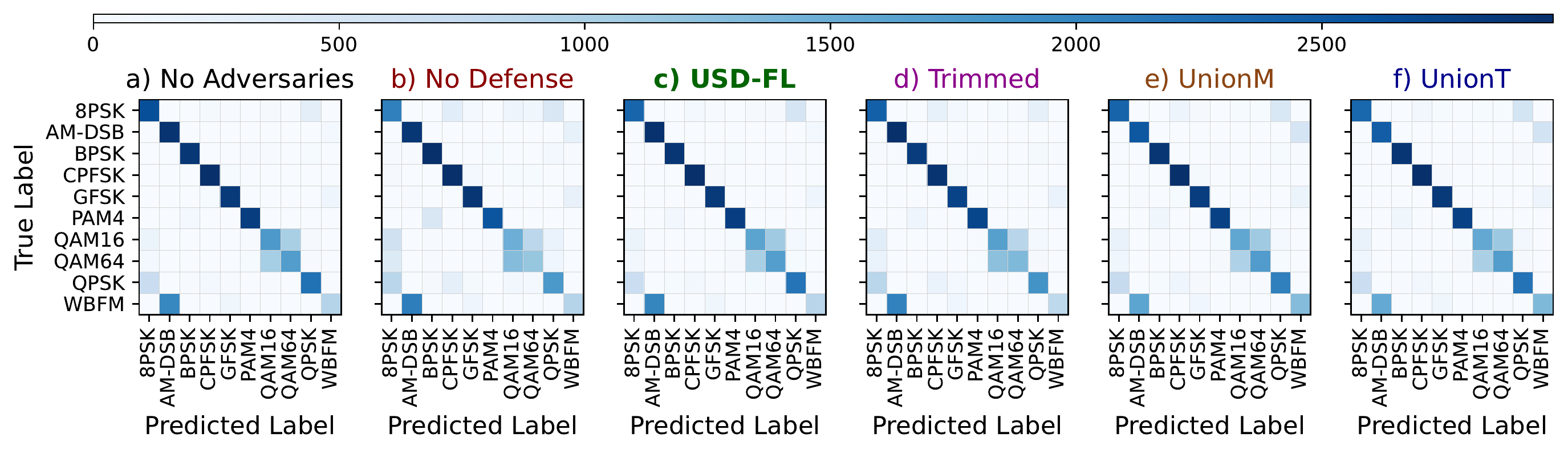}
    \caption{{\color{black}Confusion matrices in an i.i.d. scenario versus medium power FGSM attacks. Comparing the undefended case in Fig.~\ref{fig:fgsm_conf_mp_iid}b) with the higher power FGSM attack in Fig.~\ref{fig:fgsm_conf_hp_iid}b) shows much fewer confusion for the medium power perturbation scenario. }}
    \label{fig:fgsm_conf_mp_iid}
    \vspace{-3mm}
\end{figure*}

\begin{figure*}[h!]
    \centering
    \includegraphics[width=\textwidth]{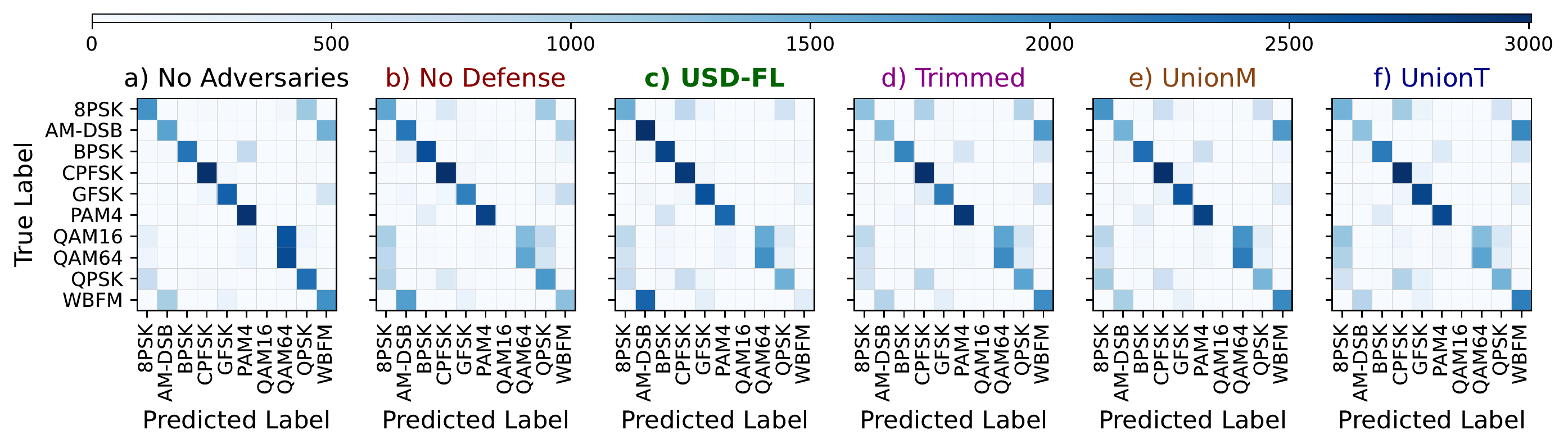}
    \caption{{\color{black}Confusion matrices in a non-i.i.d. scenario versus medium power FGSM attacks. By examining the undefended case in Fig.~\ref{fig:fgsm_conf_mp_noniid}b), we can see that the attack still results in considerable confusion, especially as many signals continue to be misclassified as 8PSK. }}
    \label{fig:fgsm_conf_mp_noniid}
    \vspace{-3mm}
\end{figure*}

\begin{figure*}[h!]
    \centering
    \includegraphics[width=\textwidth]{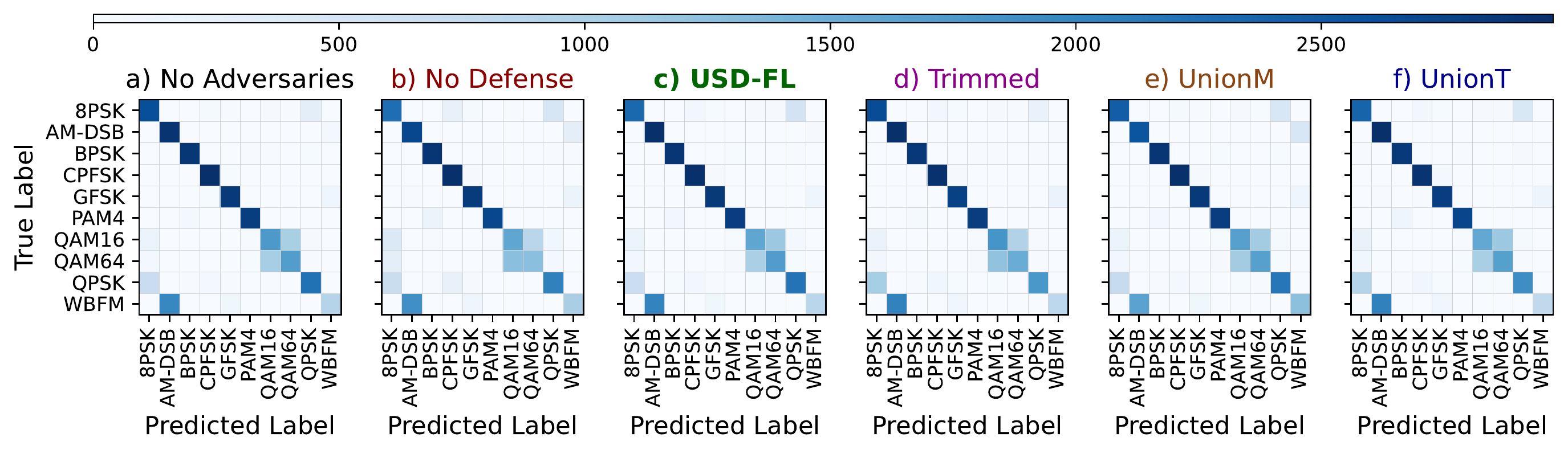}
    \caption{{\color{black}Confusion matrices in an i.i.d. scenario versus lower power FGSM attacks. At $0$ dB PNR, the potency of the adversarial attack becomes quite diminished. Most defenses appear highly similar to the undefended as well as the unperturbed scenarios.} }
    \label{fig:fgsm_conf_lp_iid}
    \vspace{-3mm}
\end{figure*}

\begin{figure*}[h!]
    \centering
    \includegraphics[width=\textwidth]{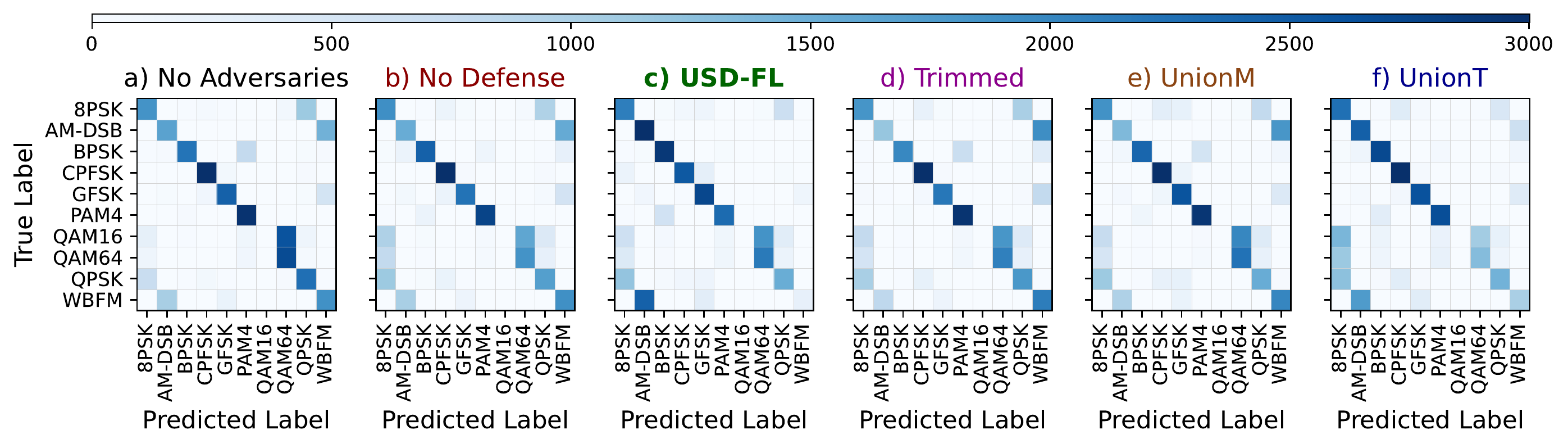}
    \caption{{\color{black}Confusion matrices in a non-i.i.d. scenario versus lower power FGSM attacks. Similarly to Fig.~\ref{fig:fgsm_conf_lp_iid}, the confusion matrices begin to appear highly similar when subject to $0$ dB PNR adversarial perturbations.}}
    \label{fig:fgsm_conf_lp_noniid}
    \vspace{-3mm}
\end{figure*}

In the following, we present the confusion matrices for various methodologies versus FGSM-based adversarial attacks of varying power. 
The performance of higher power perturbations is in Fig.~\ref{fig:fgsm_conf_hp_iid} and Fig.~\ref{fig:fgsm_conf_hp_noniid}, while the experiment involving medium power attacks is in Fig.~\ref{fig:fgsm_conf_mp_iid} and Fig.~\ref{fig:fgsm_conf_mp_noniid}. Finally, we show the behavior of lower power perturbations in Fig.~\ref{fig:fgsm_conf_lp_iid} and Fig.~\ref{fig:fgsm_conf_lp_noniid}. 
The key takeaways remain qualitatively the same as those for PGD attacks, with the main difference being that FGSM perturbations are less potent than PGD ones. 
\\

\newpage 
\clearpage 
\newpage 

\subsubsection{False positive rates versus FGSM-based attacks}
In the following, we present the false positive rates versus higher power FGSM-based attacks in Table.~\ref{tab:fp_fgsm_hp}, medium power FGSM-based attacks in Table.~\ref{tab:fp_fgsm_mp}, and lower power FGSM-based attacks in Table.~\ref{tab:fp_fgsm_lp}. 
We note that the key takeaways remain the same as that discussed for the PGD-based attacks.

\begin{table}[h!]
\caption{Average False Positive Rates - High Power FGSM}
\label{tab:fp_fgsm_hp} 
\begin{tabularx}{0.96\textwidth}{>{\centering\arraybackslash}m{10em} *{5}{>{\centering\arraybackslash}X}}
\toprule[.2em] 
& & \multicolumn{4}{c}{\textbf{Experiment Condition}} \\
\cmidrule(lr){3-6}
& & \multicolumn{2}{c}{\textbf{i.i.d.}} & \multicolumn{2}{c}{\textbf{non-i.i.d.}}\\
\cmidrule(lr){3-4} \cmidrule{5-6}
\multicolumn{2}{c}{\textbf{Defense Method}} &  \textbf{Nominal} & \textbf{Rate (\%)} & \textbf{Nominal} &  \textbf{Rate (\%)} \\
\midrule
\multicolumn{2}{c}{\textbf{UnionM}} & 0.9 & 12.9 & 1.1 & 15.7 \\
\multicolumn{2}{c}{\textbf{UnionT}} & 0.88 & 12.6 & 0.88 & 12.6\\
\multicolumn{2}{c}{\textbf{Median}} & 2.42 & 34.6 & 2.39 & 34.1 \\
\multicolumn{2}{c}{\textbf{Trimmed}} & 2.68 & 38.3 & 2.63 & 37.6 \\
\multicolumn{2}{c}{\textbf{USD-FL}} & 0 & 0 & 0.02 & 0.3\\
\bottomrule
\end{tabularx}
\end{table}

\begin{table}[h!]
\caption{Average False Positive Rates - Medium Power FGSM}
\label{tab:fp_fgsm_mp} 
\begin{tabularx}{0.96\textwidth}{>{\centering\arraybackslash}m{10em} *{5}{>{\centering\arraybackslash}X}}
\toprule[.2em] 
& & \multicolumn{4}{c}{\textbf{Experiment Condition}} \\
\cmidrule(lr){3-6}
& & \multicolumn{2}{c}{\textbf{i.i.d.}} & \multicolumn{2}{c}{\textbf{non-i.i.d.}}\\
\cmidrule(lr){3-4} \cmidrule{5-6}
\multicolumn{2}{c}{\textbf{Defense Method}} &  \textbf{Nominal} & \textbf{Rate (\%)} & \textbf{Nominal} &  \textbf{Rate (\%)} \\
\midrule
\multicolumn{2}{c}{\textbf{UnionM}} & 0.83 & 11.9 & 1.9 & 27.1 \\
\multicolumn{2}{c}{\textbf{UnionT}} & 0.76 & 10.9 & 1.83 & 26.1\\
\multicolumn{2}{c}{\textbf{Median}} & 2.39 & 34.1 & 2.38 & 34 \\
\multicolumn{2}{c}{\textbf{Trimmed}} & 2.6 & 37.1 & 2.57 & 36.7 \\
\multicolumn{2}{c}{\textbf{USD-FL}} & 0 & 0 & 0.88  & 12.6 \\
\bottomrule
\end{tabularx}
\end{table}

\begin{table}[h!]
\caption{Average False Positive Rates - Low Power FGSM}
\label{tab:fp_fgsm_lp} 
\begin{tabularx}{0.96\textwidth}{>{\centering\arraybackslash}m{10em} *{5}{>{\centering\arraybackslash}X}}
\toprule[.2em] 
& & \multicolumn{4}{c}{\textbf{Experiment Condition}} \\
\cmidrule(lr){3-6}
& & \multicolumn{2}{c}{\textbf{i.i.d.}} & \multicolumn{2}{c}{\textbf{non-i.i.d.}}\\
\cmidrule(lr){3-4} \cmidrule{5-6}
\multicolumn{2}{c}{\textbf{Defense Method}} &  \textbf{Nominal} & \textbf{Rate (\%)} & \textbf{Nominal} &  \textbf{Rate (\%)} \\
\midrule
\multicolumn{2}{c}{\textbf{UnionM}} & 0.95 & 13.6 & 2.21 & 31.6 \\
\multicolumn{2}{c}{\textbf{UnionT}} & 0.95 & 13.6 & 2.02 & 28.9 \\
\multicolumn{2}{c}{\textbf{Median}} & 2.36 & 33.7 & 2.37 & 33.9 \\
\multicolumn{2}{c}{\textbf{Trimmed}} & 2.55 & 36.4 & 2.54 & 36.3 \\
\multicolumn{2}{c}{\textbf{USD-FL}} & 0.02 & 0.3 & 0.98 & 14 \\
\bottomrule
\end{tabularx}
\end{table}

\newpage 
\clearpage 
\newpage 

{\color{black}
\subsection{Statistical significance of USD-FL vs baselines}
We further verify the statistical significance of USD-FL’s results relative to server-driven baselines by measuring p-values derived from the paired t-test in Table~\ref{tab:stat_sig} . 
P-values smaller than $0.05$ indicate statistical significance, meaning that USD-FL’s performance improvements are structural rather than due to random variation. 
In this regard, Table~\ref{tab:stat_sig} confirms that USD-FL’s superiority over server-driven baselines (versus PGD attacks of varying power levels) are statistically significant as all entries are below $0.05$, with the vast majority being at least an order of magnitude smaller.
}

\begin{table}[h!]
\caption{{\color{black} Statistical significance for classification accuracies of USD-FL vs server-driven baselines against PGD-attacks of varying power levels. P-values are obtained via paired t-tests and those values smaller than $0.05$ indicate statistical significance, meaning that USD-FL's performance improvements are structural rather than due to random variations.}}
{\color{black}
\begin{tabularx}{0.99\textwidth}
{>{\centering\arraybackslash}m{10em} *{6}{>{\centering\arraybackslash}X}}
\toprule[.2em]
& \multicolumn{2}{c}{\textbf{High Power}} & \multicolumn{2}{c}{\textbf{Medium Power}} & \multicolumn{2}{c}{\textbf{Low Power}} \\
\cmidrule(lr{0.25em}){2-3} \cmidrule(lr{0.25em}){4-5} \cmidrule(lr{0.25em}){6-7} 
\textbf{USD-FL versus} & \textbf{i.i.d.} & \textbf{non-i.i.d.} & \textbf{i.i.d.} & \textbf{non-i.i.d.} & \textbf{i.i.d.} & \textbf{non-i.i.d.} \\
\midrule
\textbf{Undefended} & $8.24\textrm{e}\textnormal{-}8$ & $3.76\textrm{e}\textnormal{-}6$ & $2.60\textrm{e}\textnormal{-}8$ & $9.86\textrm{e}\textnormal{-}8$ & $2.25\textrm{e}\textnormal{-}10$ & $8.20\textrm{e}\textnormal{-}8$ \\
\textbf{Median} & $2.79\textrm{e}\textnormal{-}9$ & $6.35\textrm{e}\textnormal{-}7$ & $3.52\textrm{e}\textnormal{-}6$ & $9.14\textrm{e}\textnormal{-}7$ & $6.38\textrm{e}\textnormal{-}8$ & $6.33\textrm{e}\textnormal{-}6$ \\
\textbf{Trimmed} & $4.35\textrm{e}\textnormal{-}7$ & $2.25\textrm{e}\textnormal{-}8$ & $2.50\textrm{e}\textnormal{-}8$ & $4.94\textrm{e}\textnormal{-}7$ & $5.44\textrm{e}\textnormal{-}8$ & $1.20\textrm{e}\textnormal{-}6$ \\
\textbf{UnionM} & $1.43\textrm{e}\textnormal{-}3$ & $2.26\textrm{e}\textnormal{-}4$ & $7.67\textrm{e}\textnormal{-}7$ & $1.03\textrm{e}\textnormal{-}3$ & $1.83\textrm{e}\textnormal{-}5$ & $6.50\textrm{e}\textnormal{-}4$ \\
\textbf{UnionT} & $3.00\textrm{e}\textnormal{-}2$ & $1.06\textrm{e}\textnormal{-}2$ & $1.40\textrm{e}\textnormal{-}5$ & $1.36\textrm{e}\textnormal{-}3$ & $1.24\textrm{e}\textnormal{-}5$ & $2.50\textrm{e}\textnormal{-}4$ \\
\bottomrule
\end{tabularx}}
\parbox{0.95\textwidth}{\footnotesize {\color{black}*All numerical values represent p-values obtained via paired t-tests.}}
\label{tab:stat_sig}
\end{table}

\begin{figure}[h!]
    \centering
    \includegraphics[width=0.85\linewidth]{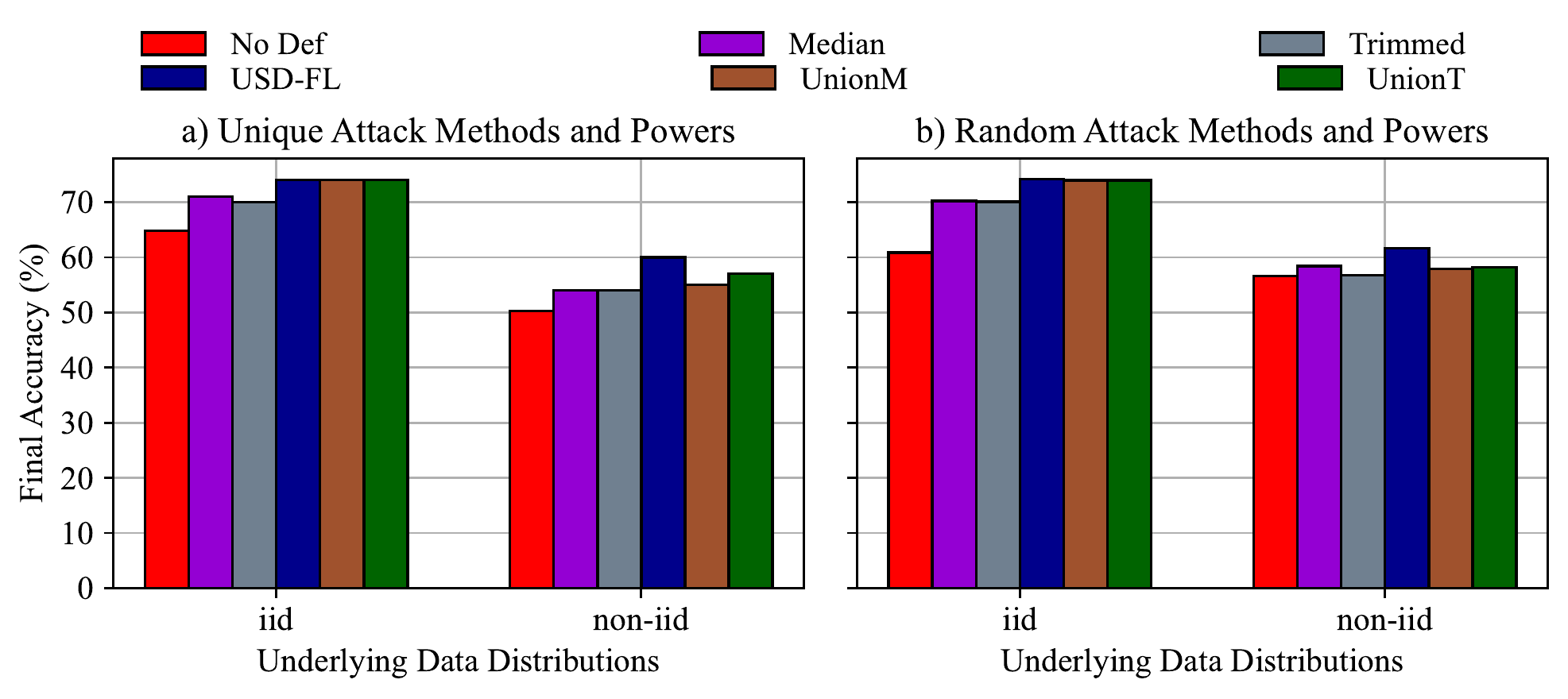}
    \caption{{\color{black}Final accuracies of server-driven defenses in response to unique and random attack architectures and powers.}}
    \label{fig:unique_random_atks}
    \vspace{-2mm}
\end{figure}

{\color{black}
\subsection{Unique and random adversarial attack architecture and power}
We further investigate the performances of USD-FL versus the baseline server-driven defenses when (i) different evasion attack architectures and attack powers happen simultaneously (i.e., PGD, FGSM, and AWGN attacks occur jointly on FL-based SC with randomly assigned high, medium, and low power) and (ii) attack architectures and attack powers are time-varying at adversaries, meaning that, at each training iteration, adversaries will randomly choose among PGD, FGSM, and AWGN attack architectures and high, medium, and low attack power.  
We refer to the first and second scenarios as having unique and random attacks, respectively, and show the resulting final classification accuracies in Fig.~\ref{fig:unique_random_atks}a) and~\ref{fig:unique_random_atks}b), respectively. 
For the i.i.d. scenarios in Fig.~\ref{fig:unique_random_atks}, we can see that USD-FL maintains similar performance to UnionM and UnionT, with all three reaching approximately $73\%$ accuracy. 
Meanwhile, in the non-i.i.d. scenario, USD-FL maintains an advantage of at least $3\%$ over the existing server-driven baselines across both unique and random adversarial attack methods. 
{\color{black}USD-FL is able to consistently outperform these existing server-driven baselines because it is able to group similarities across adversaries together (even when attack architectures differ in Fig.~\ref{fig:unique_random_atks}) and subsequently filter likely adversaries away from the global aggregations.}
}

{\color{black}
\subsection{Defense effectiveness with limited network information}
We next investigate how different server-driven baselines perform under scenarios with limited information regarding adversarial attack start time and the quantity of adversaries in Table~\ref{tab:perfect_knowledge}. Here, we examine four categories: (i) \textit{All} which means that the server knows both the adversarial attack start time as well as the number of adversaries, (ii) \textit{Adversaries} which indicates that the server only knows the exact number of adversaries, (iii) \textit{Attack Time} which denotes that the server only knows the starting time of the adversarial attacks, and (iv) \textit{Nothing} which means that the server has no knowledge.

From Table~\ref{tab:perfect_knowledge}, we see that, aside from USD-FL, all of the baselines suffer major reductions to their defensive effectiveness when they lose knowledge of the adversarial attack start time, i.e., when the network only knows the number of adversaries. 
In such settings, median and trimmed baselines both experience over $18\%$ declines in their final classification accuracies (for both i.i.d. and non-i.i.d. settings) while UnionM and UnionT baselines experience over $7\%$ and $5\%$ drops in accuracies for i.i.d. and non-i.i.d. settings respectively. 
By contrast, when the baseline server-driven defenses know the attack start time but not the number of adversaries, they experience smaller declines in performance. 
Here, the performance of median is identical to that of ``All" as the median defense is independent of the quantity of adversaries. 
On the other hand, the trimmed, UnionM, and UnionT baselines experience at most a $5\%$ decline in classification accuracies for both i.i.d. and non-i.i.d. settings. 
Finally, for the no knowledge scenario (i.e., the ``Nothing" column in Table~\ref{tab:perfect_knowledge}), we see the worst performance for trimmed, UnionM, and UnionT baselines, with $\%$ declines relative to the perfect network knowledge scenario (the ``All" column in Table~\ref{tab:perfect_knowledge}). 
As USD-FL does not rely on the knowing the exact number of adversaries or the exact attack start time, its final classification accuracies are identical to that of the perfect knowledge case for all of these experiments, meaning that USD-FL yields bigger performance improvements (e.g., at least $1\%$ and $5\%$ improvements in perfect knowledge to at least $12\%$ and $23\%$ improvements in no knowledge over all baselines) in more challenging settings. 
As such, we can see that USD-FL can indeed prevent significant disruption to FL systems even with limited network information at the server.

USD-FL does require some network information, namely knowledge of the ML model architecture, dataset structure, and the possible set of labels as these factors influence accuracy and total logit value limits. Fortunately, however, these three factors are givens in supervised learning and thus FL problems, and, moreover, baseline server-driven defenses also require such network information.
}

\begin{table}[h]
\caption{
{\color{black} USD-FL yields even larger performance improvements relative to baseline server-driven defenses in networks without perfect information. \textit{All} indicates that the server knows both the number of adversaries and the attack start time, \textit{Adversaries} denotes that the server only knows the number of adversaries, \textit{Attack Time} means that the server knows the attack time only, and \textit{None} indicates that the server has no knowledge of either.}}
{\footnotesize
{\color{black}
\begin{tabularx}{0.99\textwidth}
{>{\centering\arraybackslash}m{10em} *{8}{>{\centering\arraybackslash}X}}
\toprule[.2em]
& \multicolumn{8}{c}{Network Knowledge} \\
\cmidrule(l{1.5em}r{1.5em}){2-9} 
& \multicolumn{2}{c}{All} & \multicolumn{2}{c}{Adversaries} & \multicolumn{2}{c}{Attack Time} & \multicolumn{2}{c}{Nothing} \\
\cmidrule(l{1.5em}r{1.5em}){2-3} \cmidrule(l{1.5em}r{1.5em}){4-5} \cmidrule(l{1.5em}r{1.5em}){6-7} \cmidrule(l{1.5em}r{1.5em}){8-9}
Defense Method & i.i.d. & non-i.i.d. & i.i.d. & non-i.i.d. & i.i.d. & non-i.i.d. & i.i.d. & non-i.i.d. \\ 
\midrule
Median & 65.1 & 48.1 & 40.1 & 25.3 & 65.1 & 48.1 & 40.1 & 25.3 \\
Trimmed & 63.8 & 48.0 & 41.5 & 30.8 & 62.0 & 47.0 & 40.6 & 27.7 \\ 
\textbf{USD-FL} & 74.1 & 62.5 & 74.1 & 62.5 & 74.1 & 62.5 & 74.1 & 62.5 \\ 
UnionM & 73.0 & 57.7 & 65.6 & 41.2 & 71.3 & 55.5 & 51.6 & 33.0 \\
UnionT & 73.1 & 56.9 & 65.0 & 51.8 & 68.1 & 55.2 & 52.1 & 39.2 \\
\bottomrule
\end{tabularx}}}
\label{tab:perfect_knowledge}
\end{table}

\begin{figure}[t!]
\centering
    \centering
    \includegraphics[width=0.95\linewidth]{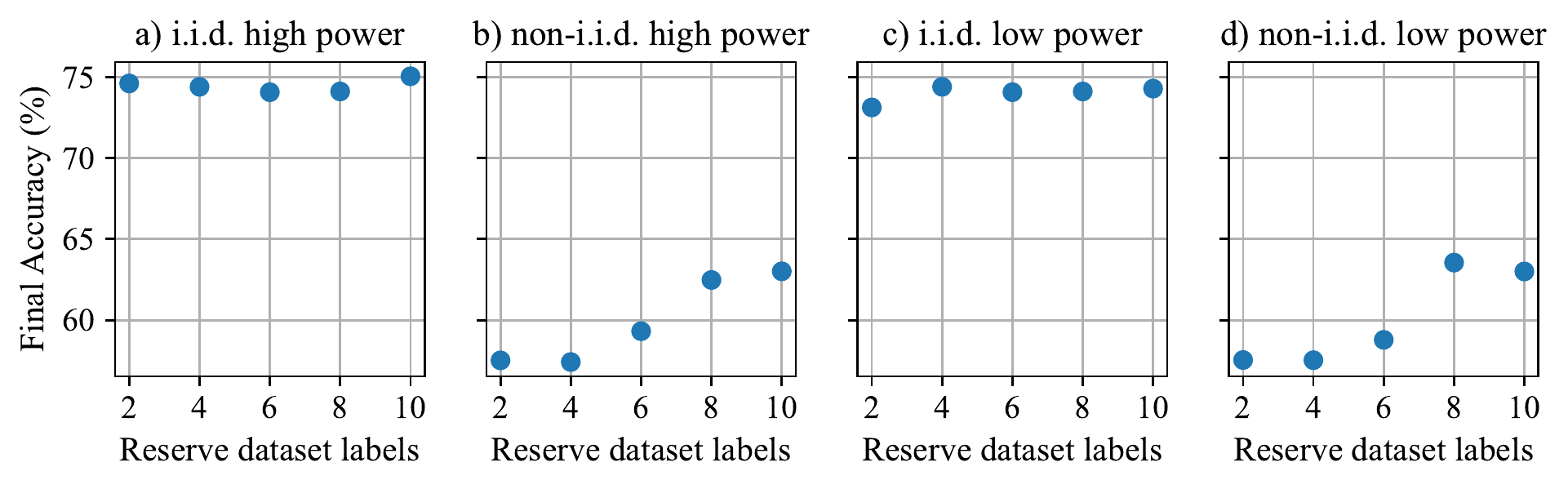}
    \caption{{\color{black}USD-FL's final accuracies against PGD attacks as a function of the number of unique labels in the reserve dataset. In i.i.d. scenarios, USD-FL achieves similar accuracies regardless of the number of labels while, in non-i.i.d. scenarios, the final accuracies generally increase as the reserve dataset holds more unique labels.}}
  \label{fig:vary_labs}
  \vspace{-2mm}
\end{figure}

\begin{figure}[t!]
\centering
    \centering
    \includegraphics[width=0.95\linewidth]{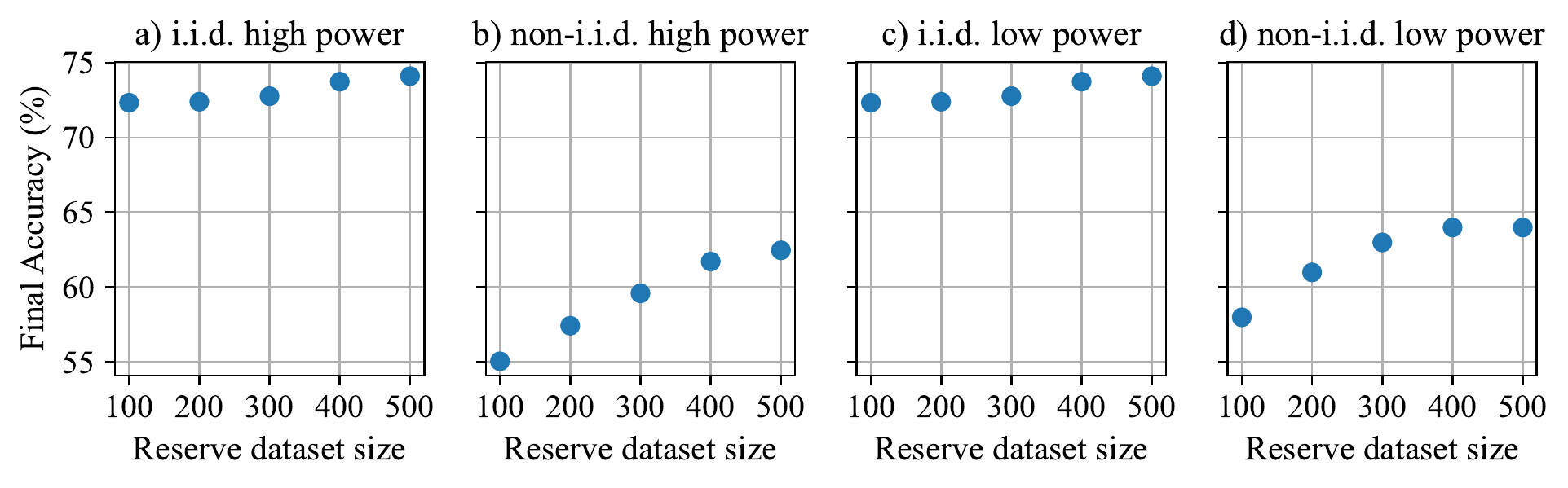}
    \caption{{\color{black}USD-FL's final accuracies against PGD attacks as a function of the reserve dataset size. For both i.i.d. and non-i.i.d. settings, the final accuracies increases as the reserve dataset grows in size.}} 
  \label{fig:vary_size}
  \vspace{-2mm}
\end{figure}

{\color{black}
\subsection{Impact of reserve dataset size and heterogeneity}
We further examine the impact of the reserve dataset by (i) varying the degree of heterogeneity of the server's reserve dataset via the number of labels present within it and (ii) varying the size of the reserve dataset. 
For the first experiment involving number of labels in the reserve dataset in Fig.~\ref{fig:vary_labs}, the reserve dataset has $500$ signals.
Meanwhile the second experiment for reserve dataset size in Fig.~\ref{fig:vary_size} has $8$ labels.

{\color{black} First, we examine the impact of the reserve dataset's distribution in Fig.~\ref{fig:vary_labs}.
For the i.i.d. settings in Fig.~\ref{fig:vary_labs}a) and~\ref{fig:vary_labs}c), we can see that the final accuracies are all very similar, regardless of the number of unique labels at the reserve dataset. 
This is because, in these scenarios, devices' local datasets are drawn i.i.d. from the training dataset, RadioML2016~\cite{o2016radio}, which contains $10$ unique labels, and their local ML models are not biased to favor any particular labels. 
Therefore, the devices' ML models, when handling the server's reserve dataset, are able to return unbiased outputs/accuracies, enabling USD-FL to filter adversaries appropriately.
By contrast, in the non-i.i.d. experiments in Fig.~\ref{fig:vary_labs}b) and~\ref{fig:vary_labs}d), devices' datasets consist of data from only $5$ random labels, and thus their ML models are locally biased to fit those specific labels. 
When evaluated on the reserve dataset, the devices' ML models yield biased outputs, which favor classification towards the labels within their local datasets. 
Thus, USD-FL generally has decreasing final accuracies as the number of labels within the reserve dataset decreases in non-i.i.d. settings.}

Next, we investigate the impact of the size of the reserve dataset in Fig.~\ref{fig:vary_size}. 
Here, both i.i.d. and non-i.i.d. scenarios depict the same trend - as the reserve dataset grows in size from $100$ signals to $500$ signals, the final accuracy increases correspondingly. 
For both i.i.d. settings in Fig.~\ref{fig:vary_size}a) and~\ref{fig:vary_size}c), the change from $100$ to $500$ signals only results in less than $3\%$ change in final accuracies, demonstrating USD-FL's resilience to reserve dataset size in i.i.d. settings. 
Meanwhile, in the non-i.i.d. settings of Fig.~\ref{fig:vary_size}b) and~\ref{fig:vary_size}d), the final accuracies change by roughly $12.5\%$ and $11\%$, respectively. 
The final accuracy changes in non-i.i.d. settings are substantial because smaller reserve datasets result in noisier classification (and thus lower accuracies), especially as the devices' ML models themselves are also trained on biased datasets. 
}

{\color{black} We further study the corresponding false positive rates as the reserve dataset varies with respect to unique number of labels. Here, we examine the case with high and low power attacks in Table~\ref{tab:fp_pgd_hp_vsl} and ~\ref{tab:fp_pgd_lp_vsl} respectively.
These false positive rates further confirm the insights derived from Fig.~\ref{fig:vary_labs}. 
For the high power attacks in Table~\ref{tab:fp_pgd_hp_vsl}, we can see that i.i.d. settings all have near $0$ nominal false positive rates.
Meanwhile, for the non-i.i.d. setting, we see a clear trend in which USD-FL yields smaller false positive rates as the reserve dataset grows in the number of labels, confirming the key takeaway from Fig.~\ref{fig:vary_labs}. 
Simultaneously, the lower power attacks in Table~\ref{tab:fp_pgd_lp_vsl} yield the same general trends as that for the higher power attacks, with a few differences in nominal values as a result of higher power attacks generally being more detectable.}

\begin{table}[h!]
\caption{{\color{black}Average False Positive Rates With Respect to Reserve Dataset Labels - USD-FL Versus High Power PGD Attacks.}}
\vspace{-1.5mm}
\label{tab:fp_pgd_hp_vsl} 
{\footnotesize
{\color{black}
\begin{tabularx}{0.96\textwidth}{>{\centering\arraybackslash}m{10em} *{5}{>{\centering\arraybackslash}X}}
\toprule[.2em] 
& & \multicolumn{4}{c}{\textbf{Experiment Condition}} \\
\cmidrule(lr){3-6}
& & \multicolumn{2}{c}{\textbf{i.i.d.}} & \multicolumn{2}{c}{\textbf{non-i.i.d.}}\\
\cmidrule(lr){3-4} \cmidrule{5-6}
\multicolumn{2}{c}{\textbf{Num Labels}} & \textbf{Nominal} & \textbf{Rate (\%)} &\textbf{Nominal} & \textbf{Rate (\%)} \\
\midrule
\multicolumn{2}{c}{\textbf{2}} & 0.02 & 0.3 & 2.45 & 35.0 \\
\multicolumn{2}{c}{\textbf{4}} & 0.00 & 0 & 2.38 & 34.0 \\
\multicolumn{2}{c}{\textbf{6}} & 0.00 & 0 & 1.24 & 17.7 \\
\multicolumn{2}{c}{\textbf{8}} & 0.00 & 0 & 0.02 & 0.3 \\
\multicolumn{2}{c}{\textbf{10}} & 0.00 & 0 & 0.02 & 0.3 \\
\bottomrule
\end{tabularx} } }
\end{table}

\begin{table}[h!]
\caption{{\color{black}Average False Positive Rates With Respect to Reserve Dataset Labels - USD-FL Versus Low Power PGD Attacks.}}
\vspace{-1.5mm}
\label{tab:fp_pgd_lp_vsl} 
{\footnotesize
{\color{black}
\begin{tabularx}{0.96\textwidth}{>{\centering\arraybackslash}m{10em} *{5}{>{\centering\arraybackslash}X}}
\toprule[.2em] 
& & \multicolumn{4}{c}{\textbf{Experiment Condition}} \\
\cmidrule(lr){3-6}
& & \multicolumn{2}{c}{\textbf{i.i.d.}} & \multicolumn{2}{c}{\textbf{non-i.i.d.}}\\
\cmidrule(lr){3-4} \cmidrule{5-6}
\multicolumn{2}{c}{\textbf{Num Labels}} & \textbf{Nominal} & \textbf{Rate (\%)} &\textbf{Nominal} & \textbf{Rate (\%)} \\
\midrule
\multicolumn{2}{c}{\textbf{2}} & 0.90 & 13.9 & 1.67 & 23.9 \\
\multicolumn{2}{c}{\textbf{4}} & 0.00 & 0 & 1.36 & 19.4 \\
\multicolumn{2}{c}{\textbf{6}} & 0.00 & 0 & 1.10 & 15.7 \\
\multicolumn{2}{c}{\textbf{8}} & 0.00 & 0 & 0.88 & 12.6 \\
\multicolumn{2}{c}{\textbf{10}} & 0.00 & 0 & 0.74 & 10.6 \\
\bottomrule
\end{tabularx} } }
\end{table}

\newpage
\clearpage
\newpage 




%% file: main.bbl
\begin{thebibliography}{10}
\providecommand{\url}[1]{#1}
\csname url@samestyle\endcsname
\providecommand{\newblock}{\relax}
\providecommand{\bibinfo}[2]{#2}
\providecommand{\BIBentrySTDinterwordspacing}{\spaceskip=0pt\relax}
\providecommand{\BIBentryALTinterwordstretchfactor}{4}
\providecommand{\BIBentryALTinterwordspacing}{\spaceskip=\fontdimen2\font plus
\BIBentryALTinterwordstretchfactor\fontdimen3\font minus \fontdimen4\font\relax}
\providecommand{\BIBforeignlanguage}[2]{{%
\expandafter\ifx\csname l@#1\endcsname\relax
\typeout{** WARNING: IEEEtran.bst: No hyphenation pattern has been}%
\typeout{** loaded for the language `#1'. Using the pattern for}%
\typeout{** the default language instead.}%
\else
\language=\csname l@#1\endcsname
\fi
#2}}
\providecommand{\BIBdecl}{\relax}
\BIBdecl

\bibitem{wang2023potent}
S.~Wang, R.~Sahay, and C.~G. Brinton, ``How potent are evasion attacks for poisoning federated learning-based signal classifiers?'' \emph{arXiv:2301.08866}, 2023.

\bibitem{dl_amc}
T.~J. O’Shea, T.~Roy, and T.~C. Clancy, ``Over-the-air deep learning based radio signal classification,'' \emph{IEEE J. Sel. Topics Signal Process.}, 2018.

\bibitem{fl}
B.~McMahan, E.~Moore, D.~Ramage, S.~Hampson, and B.~A. y~Arcas, ``Communication-efficient learning of deep networks from decentralized data,'' in \emph{AISTATS}, 2017.

\bibitem{wang2023towards}
S.~Wang, S.~Hosseinalipour, V.~Aggarwal, C.~G. Brinton, D.~J. Love, W.~Su, and M.~Chiang, ``Towards cooperative federated learning over heterogeneous edge/fog networks,'' \emph{arXiv:2303.08361}, 2023.

\bibitem{dong2022federated}
J.~Dong, L.~Wang, Z.~Fang, G.~Sun, S.~Xu, X.~Wang, and Q.~Zhu, ``Federated class-incremental learning,'' in \emph{Proceedings of the IEEE/CVF conference on computer vision and pattern recognition}, 2022.

\bibitem{dong2023federated}
J.~Dong, D.~Zhang, Y.~Cong, W.~Cong, H.~Ding, and D.~Dai, ``Federated incremental semantic segmentation,'' in \emph{Proceedings of the IEEE/CVF Conference on Computer Vision and Pattern Recognition}, 2023.

\bibitem{hu2019decentralized}
C.~Hu, J.~Jiang, and Z.~Wang, ``Decentralized federated learning: A segmented gossip approach,'' \emph{arXiv preprint arXiv:1908.07782}, 2019.

\bibitem{hegedHus2019gossip}
I.~Heged{\H{u}}s, G.~Danner, and M.~Jelasity, ``Gossip learning as a decentralized alternative to federated learning,'' in \emph{DAIS 2019}.\hskip 1em plus 0.5em minus 0.4em\relax Springer, 2019, pp. 74--90.

\bibitem{wang2023multi}
S.~Wang, S.~Hosseinalipour, and C.~G. Brinton, ``Multi-source to multi-target decentralized federated domain adaptation,'' \emph{arXiv preprint arXiv:2304.12422}, 2023.

\bibitem{amc_fl1}
Y.~Wang, G.~Gui, H.~Gacanin, B.~Adebisi, H.~Sari, and F.~Adachi, ``Federated learning for automatic modulation classification under class imbalance and varying noise condition,'' \emph{IEEE Trans. Cogn. Commun. Netw.}, 2022.

\bibitem{evasion_atks}
A.~Chakraborty, M.~Alam, V.~Dey, A.~Chattopadhyay, and D.~Mukhopadhyay, ``Adversarial attacks and defences: A survey,'' \emph{arXiv:1810.00069}, 2018.

\bibitem{model_poison}
A.~N. Bhagoji, S.~Chakraborty, P.~Mittal, and S.~Calo, ``Analyzing federated learning through an adversarial lens,'' in \emph{Proc. of the 36th ICML}, 2019.

\bibitem{cao2022mpaf}
X.~Cao and N.~Z. Gong, ``Mpaf: Model poisoning attacks to federated learning based on fake clients,'' in \emph{Proc. IEEE/CVF Conf. Comput. Vision Pattern Recognit.}, 2022.

\bibitem{yin2018byzantine}
D.~Yin, Y.~Chen, R.~Kannan, and P.~Bartlett, ``Byzantine-robust distributed learning: Towards optimal statistical rates,'' in \emph{Int. Conf. Mach. Learn.}\hskip 1em plus 0.5em minus 0.4em\relax PMLR, 2018.

\bibitem{blanchard2017machine}
P.~Blanchard, E.~M. El~Mhamdi, R.~Guerraoui, and J.~Stainer, ``Machine learning with adversaries: Byzantine tolerant gradient descent,'' \emph{Advances in neural information processing systems}, vol.~30, 2017.

\bibitem{byz_atk}
M.~Fang, X.~Cao, J.~Jia, and N.~Gong, ``Local model poisoning attacks to byzantine-robust federated learning,'' in \emph{29th USENIX Secur.}, 2020, pp. 1605--1622.

\bibitem{carlini2017towards}
N.~Carlini and D.~Wagner, ``Towards evaluating the robustness of neural networks,'' in \emph{2017 Proc. {IEEE} Symp. Secur. Privacy}.\hskip 1em plus 0.5em minus 0.4em\relax IEEE, 2017, pp. 39--57.

\bibitem{kannan2018adversarial}
H.~Kannan, A.~Kurakin, and I.~Goodfellow, ``Adversarial logit pairing,'' \emph{\textnormal{arXiv:1803.06373}}, 2018.

\bibitem{adv_in_rf}
D.~Adesina, C.-C. Hsieh, Y.~E. Sagduyu, and L.~Qian, ``Adversarial machine learning in wireless communications using rf data: A review,'' \emph{IEEE Commun. Surv. Tut.}, 2022.

\bibitem{amc_adv_atk1}
M.~Sadeghi and E.~G. Larsson, ``Adversarial attacks on deep-learning based radio signal classification,'' \emph{IEEE Wireless Commun. Letters}, 2018.

\bibitem{amc_adv_atk2}
Y.~{Lin}, H.~{Zhao}, Y.~{Tu}, S.~{Mao}, and Z.~{Dou}, ``Threats of adversarial attacks in dnn-based modulation recognition,'' in \emph{Proc. of IEEE INFOCOM}, 2020.

\bibitem{amc_adv_atk3}
B.~{Kim}, Y.~E. {Sagduyu}, K.~{Davaslioglu}, T.~{Erpek}, and S.~{Ulukus}, ``Over-the-air adversarial attacks on deep learning based modulation classifier over wireless channels,'' in \emph{Proc. of 54th Annual CISS}, 2020, pp. 1--6.

\bibitem{amc_def1}
R.~Sahay, C.~G. Brinton, and D.~J. Love, ``A deep ensemble-based wireless receiver architecture for mitigating adversarial attacks in automatic modulation classification,'' \emph{IEEE Trans. Cogn. Commun. Netw.}, 2022.

\bibitem{amc_def2}
S.~{Kokalj-Filipovic}, R.~{Miller}, N.~{Chang}, and C.~L. {Lau}, ``Mitigation of adversarial examples in rf deep classifiers utilizing autoencoder pre-training,'' in \emph{Proc. of ICMCIS}, 2019, pp. 1--6.

\bibitem{adv_trn1}
R.~Sahay, D.~J. Love, and C.~G. Brinton, ``Robust automatic modulation classification in the presence of adversarial attacks,'' in \emph{Proc. of 55th Annual CISS}, 2021.

\bibitem{adv_trn2}
L.~Zhang, S.~Lambotharan, G.~Zheng, G.~Liao, A.~Demontis, and F.~Roli, ``A hybrid training-time and run-time defense against adversarial attacks in modulation classification,'' \emph{IEEE Wireless Commun. Letters}, 2022.

\bibitem{tian2022exploring}
J.~Tian, B.~Wang, J.~Li, Z.~Wang, B.~Ma, and M.~Ozay, ``Exploring targeted and stealthy false data injection attacks via adversarial machine learning,'' \emph{IEEE Internet Things J.}, 2022.

\bibitem{adv_trn_overfitting}
L.~Rice, E.~Wong, and Z.~Kolter, ``Overfitting in adversarially robust deep learning,'' in \emph{Int. Conf. Mach. Learn.}\hskip 1em plus 0.5em minus 0.4em\relax PMLR, 2020.

\bibitem{atks_in_fl}
M.~S. Jere, T.~Farnan, and F.~Koushanfar, ``A taxonomy of attacks on federated learning,'' \emph{IEEE Secur. Privacy}, 2021.

\bibitem{label_flip}
V.~Tolpegin, S.~Truex, M.~E. Gursoy, and L.~Liu, ``Data poisoning attacks against federated learning systems,'' in \emph{Eur. Symp. Res. Comput. Secur.}\hskip 1em plus 0.5em minus 0.4em\relax Springer, 2020, pp. 480--501.

\bibitem{fl_atk_det}
Z.~Liu, J.~Mu, W.~Lv, Z.~Jing, Q.~Zhou, and X.~Jing, ``A distributed attack-resistant trust model for automatic modulation classification,'' \emph{IEEE Commun. Letters}, 2022.

\bibitem{wu2024model}
M.~Wu, B.~Zhao, Y.~Xiao, C.~Deng, Y.~Liu, and X.~Liu, ``Model: A model poisoning defense framework for federated learning via truth discovery,'' \emph{IEEE Trans. Inf. Forensics Security}, 2024.

\bibitem{xu2024dual}
R.~Xu, S.~Gao, C.~Li, J.~Joshi, and J.~Li, ``Dual defense: Enhancing privacy and mitigating poisoning attacks in federated learning,'' \emph{Adv. Neural Inf. Process. Syst. (NeurIPS)}, vol.~37, pp. 70\,476--70\,498, 2024.

\bibitem{kumar2024revamping}
K.~N. Kumar, R.~Mitra, and C.~K. Mohan, ``Revamping federated learning security from a defender's perspective: A unified defense with homomorphic encrypted data space,'' in \emph{Proc. IEEE/CVF Conf. Comput. Vis. Pattern Recognit. (CVPR)}, 2024, pp. 24\,387--24\,397.

\bibitem{lyu2024adversarial}
X.~Lyu, S.~Li, N.~Wang, T.~Li, D.~Chen, and Y.~Chen, ``Adversarial attacks on federated learning revisited: a client-selection perspective,'' in \emph{2024 Proc. IEEE Conf. Commun. Netw. Secur. (CNS)}.\hskip 1em plus 0.5em minus 0.4em\relax IEEE, 2024, pp. 1--9.

\bibitem{kumar2023impact}
K.~N. Kumar, C.~K. Mohan, and L.~R. Cenkeramaddi, ``The impact of adversarial attacks on federated learning: A survey,'' \emph{IEEE Trans. Pattern Anal. Mach. Intell.}, vol.~46, no.~5, pp. 2672--2691, 2023.

\bibitem{shen2016auror}
S.~Shen, S.~Tople, and P.~Saxena, ``Auror: Defending against poisoning attacks in collaborative deep learning systems,'' in \emph{Proc. 32nd Annu. Conf. Comput. Secur. Appl.}, 2016, pp. 508--519.

\bibitem{li2021lomar}
X.~Li, Z.~Qu, S.~Zhao, B.~Tang, Z.~Lu, and Y.~Liu, ``Lomar: A local defense against poisoning attack on federated learning,'' \emph{IEEE Trans. Dependable Secure Comput.}, 2021.

\bibitem{zhang2024flpurifier}
J.~Zhang, C.~Zhu, X.~Sun, C.~Ge, B.~Chen, W.~Susilo, and S.~Yu, ``Flpurifier: backdoor defense in federated learning via decoupled contrastive training,'' \emph{IEEE Trans. Inf. Forensics Security}, 2024.

\bibitem{tao2023byzantine}
Y.~Tao, S.~Cui, W.~Xu, H.~Yin, D.~Yu, W.~Liang, and X.~Cheng, ``Byzantine-resilient federated learning at edge,'' \emph{IEEE Trans. Comput.}, 2023.

\bibitem{jiang2023data}
Y.~Jiang, W.~Zhang, and Y.~Chen, ``Data quality detection mechanism against label flipping attacks in federated learning,'' \emph{IEEE Trans. Inf. Forensics Secur.}, 2023.

\bibitem{xie2019zeno}
C.~Xie, S.~Koyejo, and I.~Gupta, ``Zeno: Distributed stochastic gradient descent with suspicion-based fault-tolerance,'' in \emph{Int. Conf. Mach. Learn.}\hskip 1em plus 0.5em minus 0.4em\relax PMLR, 2019, pp. 6893--6901.

\bibitem{yang2024roseagg}
H.~Yang, W.~Xi, Y.~Shen, C.~Wu, and J.~Zhao, ``Roseagg: Robust defense against targeted collusion attacks in federated learning,'' \emph{IEEE Trans Inf. Forensics Security}, vol.~19, pp. 2951--2966, 2024.

\bibitem{chen2022uav}
J.~Chen and J.~Tang, ``Uav-assisted data collection for dynamic and heterogeneous wireless sensor networks,'' \emph{IEEE Wireless Commun. Letters}, 2022.

\bibitem{jia2022energy}
R.~Jia, J.~Wu, J.~Lu, M.~Li, F.~Lin, and Z.~Zheng, ``Energy saving in heterogeneous wireless rechargeable sensor networks,'' in \emph{Proc. of IEEE INFOCOM 2022}.\hskip 1em plus 0.5em minus 0.4em\relax IEEE, 2022.

\bibitem{wang2022uav}
S.~Wang, S.~Hosseinalipour, M.~Gorlatova, C.~G. Brinton, and M.~Chiang, ``Uav-assisted online machine learning over multi-tiered networks: A hierarchical nested personalized federated learning approach,'' \emph{IEEE Trans. Netw. Service Manage.}, 2022.

\bibitem{vilajosana2023challenges}
X.~Vilajosana, G.~Boquet, J.~Melia-Segui, P.~Tuset-Peiro, B.~Martinez, and F.~Adelantado, ``Challenges and opportunities for simultaneous multifunctional wireless networks in the uhf band,'' \emph{IEEE Commun. Mag.}, 2023.

\bibitem{lin20215g}
X.~Lin, S.~Rommer, S.~Euler, E.~A. Yavuz, and R.~S. Karlsson, ``5g from space: An overview of 3gpp non-terrestrial networks,'' \emph{IEEE Communications Standards Magazine}, 2021.

\bibitem{wild2021joint}
T.~Wild, V.~Braun, and H.~Viswanathan, ``Joint design of communication and sensing for beyond 5g and 6g systems,'' \emph{IEEE Access}, 2021.

\bibitem{mohamed2020strategies}
M.~Mohamed, S.~Handagala, J.~Xu, M.~Leeser, and M.~Onabajo, ``Strategies and demonstration to support multiple wireless protocols with a single rf front-end,'' \emph{IEEE Wireless Commun.}, 2020.

\bibitem{cai2017modulation}
Y.~Cai, Z.~Qin, F.~Cui, G.~Y. Li, and J.~A. McCann, ``Modulation and multiple access for 5g networks,'' \emph{IEEE Commun. Surv. Tut.}, 2017.

\bibitem{liu2022integrated}
F.~Liu, Y.~Cui, C.~Masouros, J.~Xu, T.~X. Han, Y.~C. Eldar, and S.~Buzzi, ``Integrated sensing and communications: Towards dual-functional wireless networks for 6g and beyond,'' \emph{IEEE J. Sel. Areas Commun.}, 2022.

\bibitem{alharbi2024collusive}
S.~Alharbi, Y.~Guo, and W.~Yu, ``Collusive backdoor attacks in federated learning frameworks for iot systems,'' \emph{IEEE Internet Things J.}, 2024.

\bibitem{fgsm}
I.~J. Goodfellow, J.~Shlens, and C.~Szegedy, ``Explaining and harnessing adversarial examples,'' \emph{arXiv:1412.6572}, 2014.

\bibitem{madry2018towards}
A.~Madry, A.~Makelov, L.~Schmidt, D.~Tsipras, and A.~Vladu, ``Towards deep learning models resistant to adversarial attacks,'' in \emph{Int. Conf. Learn. Representations}, 2018.

\bibitem{biggio2013evasion}
B.~Biggio, I.~Corona, D.~Maiorca, B.~Nelson, N.~{\v{S}}rndi{\'c}, P.~Laskov, G.~Giacinto, and F.~Roli, ``Evasion attacks against machine learning at test time,'' in \emph{Proc. Mach. Learn. Knowl. Discov. Databases}.\hskip 1em plus 0.5em minus 0.4em\relax Springer, 2013, pp. 387--402.

\bibitem{zhang2020voiceprint}
L.~Zhang, Y.~Meng, J.~Yu, C.~Xiang, B.~Falk, and H.~Zhu, ``Voiceprint mimicry attack towards speaker verification system in smart home,'' in \emph{Proc. IEEE INFOCOM, 2020}.\hskip 1em plus 0.5em minus 0.4em\relax IEEE, 2020, pp. 377--386.

\bibitem{yan2023rethinking}
Y.~Yan, C.-M. Feng, M.~Ye, W.~Zuo, P.~Li, R.~S.~M. Goh, L.~Zhu, and C.~Chen, ``Rethinking client drift in federated learning: A logit perspective,'' \emph{arXiv preprint arXiv:2308.10162}, 2023.

\bibitem{zhang2022federated}
J.~Zhang, Z.~Li, B.~Li, J.~Xu, S.~Wu, S.~Ding, and C.~Wu, ``Federated learning with label distribution skew via logits calibration,'' in \emph{Int. Conf. Mach. Learn.}\hskip 1em plus 0.5em minus 0.4em\relax PMLR, 2022, pp. 26\,311--26\,329.

\bibitem{itahara2021distillation}
S.~Itahara, T.~Nishio, Y.~Koda, M.~Morikura, and K.~Yamamoto, ``Distillation-based semi-supervised federated learning for communication-efficient collaborative training with non-iid private data,'' \emph{IEEE Trans. Mobile Comput.}, 2021.

\bibitem{chen2023importance}
H.-Y. Chen, C.-H. Tu, Z.~Li, H.~W. Shen, and W.-L. Chao, ``On the importance and applicability of pre-training for federated learning,'' in \emph{Eleventh Int. Conf. Learn. Representations}, 2023.

\bibitem{tian2022fedbert}
Y.~Tian, Y.~Wan, L.~Lyu, D.~Yao, H.~Jin, and L.~Sun, ``Fedbert: when federated learning meets pre-training,'' \emph{ACM Trans. Intell. Syst. Technol. (TIST)}, vol.~13, no.~4, pp. 1--26, 2022.

\bibitem{chen2022pre}
H.-Y. Chen, C.-H. Tu, Z.~Li, H.-W. Shen, and W.-L. Chao, ``On pre-training for federated learning,'' \emph{arXiv:2206.11488}, 2022.

\bibitem{guo2022deep}
Z.~Guo, K.~Yu, Z.~Lv, K.-K.~R. Choo, P.~Shi, and J.~J. Rodrigues, ``Deep federated learning enhanced secure poi microservices for cyber-physical systems,'' \emph{IEEE Wireless Commun.}, 2022.

\bibitem{villani2009optimal}
C.~Villani, \emph{Optimal transport: old and new}.\hskip 1em plus 0.5em minus 0.4em\relax Springer, 2009, vol. 338.

\bibitem{wu2021logit}
H.~Wu and D.~Klabjan, ``Logit-based uncertainty measure in classification,'' in \emph{2021 IEEE Int. Conf. Big Data}.\hskip 1em plus 0.5em minus 0.4em\relax IEEE, 2021, pp. 948--956.

\bibitem{ishida2020we}
T.~Ishida, I.~Yamane, T.~Sakai, G.~Niu, and M.~Sugiyama, ``Do we need zero training loss after achieving zero training error?'' in \emph{Int. Conf. Mach. Learn.}\hskip 1em plus 0.5em minus 0.4em\relax PMLR, 2020, pp. 4604--4614.

\bibitem{sahay2023defending}
R.~Sahay, M.~Zhang, D.~J. Love, and C.~G. Brinton, ``Defending adversarial attacks on deep learning-based power allocation in massive mimo using denoising autoencoders,'' \emph{IEEE Trans. Cogn. Commun. Netw.}, 2023.

\bibitem{o2016radio}
T.~J. O'shea and N.~West, ``Radio machine learning dataset generation with gnu radio,'' in \emph{Proc. 6th GNU Radio Conf.}, vol.~1, no.~1, 2016.

\bibitem{wang2021device}
S.~Wang, M.~Lee, S.~Hosseinalipour, R.~Morabito, M.~Chiang, and C.~G. Brinton, ``Device sampling for heterogeneous federated learning: Theory, algorithms, and implementation,'' in \emph{Proc. of IEEE INFOCOM}, 2021.

\bibitem{chen2020convergence}
M.~Chen, H.~V. Poor, W.~Saad, and S.~Cui, ``Convergence time optimization for federated learning over wireless networks,'' \emph{IEEE Trans. Wireless Commun.}, 2020.

\bibitem{peng2020federated}
X.~Peng, Z.~Huang, Y.~Zhu, and K.~Saenko, ``Federated adversarial domain adaptation,'' in \emph{Proc. 8th Int. Conf. Learn. Representations}, 2020.

\bibitem{uzlaner2025asynchronous}
N.~Uzlaner, T.~Raviv, N.~Shlezinger, and K.~Todros, ``Asynchronous online adaptation via modular drift detection for deep receivers,'' \emph{IEEE Trans. Wireless Commun.}, 2025.

\bibitem{smith2017federated}
V.~Smith, C.-K. Chiang, M.~Sanjabi, and A.~S. Talwalkar, ``Federated multi-task learning,'' \emph{Advances Neural Inf. Process. Syst.}, vol.~30, 2017.

\end{thebibliography}
